\documentclass[aps,pra,reprint,twocolumn,superscriptaddress,showpacs,showkeys,a4paper,english]{revtex4-1}
\usepackage{amsmath,amssymb,amstext}
\usepackage[usenames,dvipsnames]{color}
\usepackage{graphicx,subfigure,relsize}
\usepackage{bm,braket}
\usepackage{natbib}
\usepackage{umoline}
\usepackage{hyperref}
\usepackage[english]{babel}
\usepackage[OT1]{fontenc}
\definecolor{mygreen}{rgb}{0,0.5,0}
\definecolor{myblue}{rgb}{0,0,0.75}
\definecolor{mymagenta}{cmyk}{0,1,0,0.12}

\newcommand{\proj}[1]{\ket{ #1}\bra{ #1}}

\begin{document}

\title{Tensor Networks for Lattice Gauge Theories with Continuous Groups}

\author{L.~Tagliacozzo}
    \email{luca.tagliacozzo@icfo.es}
    \affiliation{ICFO-Institut de Ciencies Fotoniques, Av. Carl Friedrich Gauss, 3, 08860 Castelldefels, Barcelona, Spain.}

\author{A.~Celi}
    \email{alessio.celi@icfo.es}
    \affiliation{ICFO-Institut de Ciencies Fotoniques, Av. Carl Friedrich Gauss, 3, 08860 Castelldefels, Barcelona, Spain.}
\author{M.~Lewenstein}
    \email{maciej.lewenstein@icfo.es}
    \affiliation{ICFO-Institut de Ciencies Fotoniques, Av. Carl Friedrich Gauss, 3, 08860 Castelldefels, Barcelona, Spain.}

\date{\today}

\begin{abstract}
We discuss how to formulate lattice gauge theories in the Tensor Network language. In this way we obtain  both a consistent-truncation scheme  of the Kogut-Susskind lattice gauge theories and a Tensor Network variational ansatz for gauge invariant states that can be used in actual numerical computations. Our construction is also applied to the simplest realization of the quantum link models/gauge magnets,  and provides a clear way to understand their microscopic relation with the Kogut-Susskind lattice gauge theories. We also introduce a new set of gauge invariant operators that modify continuously  Rokshar-Kivelson wave functions, and can be used to extend the phase diagrams of known models.  As an example we characterize the transition between the deconfined phase of the $Z_2$ lattice gauge theory and  the Rokshar-Kivelson point of  the U(1) gauge magnet in 2D in terms of entanglement entropy. The topological entropy serves as an order parameter for the transition, but not the Schmidt gap.

\end{abstract}



\maketitle

Tensor Network (TN) techniques are starting to play an important role in our understanding of many-body quantum systems, both on the lattice and in the continuum.  They can be used as a framework to classify the phases of quantum matter \cite{chen_classification_2011,turner_topological_2011,schuch_classifying_2011}, or as powerful  numerical ansatz in actual computations of 1D \cite{white_real-space_1992,white_density_1992} and 2D strongly correlated quantum magnets \cite{tagliacozzo_simulation_2009,stoudenmire_studying_2012,xie_tensor_2014},  fermionic systems \cite{corboz_simulation_2010,kraus_fermionic_2010}, or anyonic systems \cite{konig_anyonic_2010,pfeifer_simulation_2010}. They  have also recently made their way into  quantum chemistry as computational tool to study the structure of molecules from the first principles \cite{boguslawski_accurate_2012,nakatani_efficient_2013,nakatani_efficient_2013}. 

While numerical simulations based on Monte Carlo (MC)  are still the most successful  techniques in some of these fields,  TNs start to provide  viable alternatives to them, particularly in those contexts where MC has troubles, such as the physics of frustrated anti-ferromagnets \cite{yan_spin-liquid_2011,jiang_identifying_2012,depenbrock_nature_2012}, and the real time evolution of out of equilibrium systems \cite{vidal_efficient_2004,white_real_2004,daley_time-dependent_2004}.  

At present, the main limitation of numerical  TN techniques is that the cost of the simulations increases rapidly with the amount of correlations in the system (which is encoded in the bond dimension $D$ of the elementary tensors), and thus TNs tend to be biased towards weakly correlated phases. 

However, the steady improvement of the TN algorithms \cite{lubasch_unifying_2014,lubasch_algorithms_2014} makes us confident that these limitations will soon be overcome, and as a consequence TN will become more and more useful  in the physics of quantum many-body systems.
Among interesting  quantum many-body systems, we focus here on gauge theories, a context in which TN have recently made a spectacular debut  \cite{tagliacozzo_entanglement_2011,banuls_mass_2013,banuls_matrix_2013,buyens_matrix_2013,liu_exact_2013}.

Gauge theories (GT) \cite{oraifeartaigh_gauge_2000} 
describe three of the four fundamental interactions (electromagnetic, weak, and strong interactions). In particular, strong interactions, are described by an $SU(3)$ gauge theory, called Quantum Chromo-Dynamics (QCD)
\cite{smilga_lectures_2001}). GT also allow to understand  emergent phenomena at low energies in condensed matter systems, { e.g.}
anti-ferromagnets \cite{balents_spin_2010} and high-temperature
superconductors \cite{anderson_resonating_1987,lee_doping_2006,mann_high-temperature_2011}.

The phase diagrams of  GT, similarly to those of most strongly correlated many-body quantum systems, are still debated. Still, there are exactly solvable GT  that  display topological phases.
Recently, topological states have been proposed  as possible  hardware for quantum computers, and thus there is an urgent need for clarification of generic
GT phase diagrams where these states could appear \cite{kitaev_fault-tolerant_2003,trebst_breakdown_2007,tupitsyn_topological_2010, tagliacozzo_entanglement_2011,dusuel_robustness_2011}.

Wilson's 
formulation of lattice gauge theories 
(LGT) \cite{wilson_confinement_1974}  was obtained by substituting the continuous space-time with  a discrete set
of points (the lattice). It provided the  breakthrough that has allowed  to develop numerical tools based on MC, able to address the
strong coupling regime of GT.  These tools are,  as today, the main resource to compare various aspects of QCD at strong-coupling  with experiments \cite{bazavov_nonperturbative_2010}. Those aspects of QCD that are hard or
impossible to address with MC are indeed in most of the cases still unclear. For example, the mechanism of charge confinement
\cite{polyakov_compact_1975}, invented to explain the absence of
isolated quarks \cite{kim_search_2007}, still stands as a conjecture in full QCD,
four decades since it was first understood in  Abelian models. 
Furthermore, MC simulations struggle to  address hot and dense nuclear matter \cite{satz_sps_2004,gupta_scale_2011}, probed
by heavy nuclei collisions at CERN  and RHIC \cite{alice_collaboration_j/_2012,star_collaboration_directed_2012}. In their current formulation MC simulations of LGT cannot  be used to characterize  the real time out of equilibrium dynamics of GT.

The aim of this paper is two-fold. On one side we develop the theory of TNs for LGT with arbitrary groups. On the other, we  provide a constructive approach to LGT using the TN formalism (reviewed in Sec. \ref{sec:tn}). In this framework, TN are used as a model-building tool that, given a group $G$,  allows to design the most general gauge invariant theory out of the simple  knowledge of the group representation matrices.  The approach is based on reformulating very simple results about the theory of group representations (reviewed in Sec. \ref{sec:gt}) in the TN formalism. 

In particular, we use the approach in which  LGT differ from standard  many-body quantum systems due to the presence  of a large amount of local symmetry constraints (Sec. \ref{sec:lgt}), which arise as a consequence of generalizations of the Gauss law. One of the guiding principles in designing LGT will thus be the possibility  of  defining such local symmetry constraints  (Sec. \ref{sec:lgth} and Sec. \ref{sec:lgt_gt}). In particular,  we identify the ``physical Hilbert space'', ${\cal H}_P$, as the space of states that fulfill those constraints (Sec. \ref{sec:lgt_ph}). We show how these constraints can be naturally embedded in a TN. We thus construct an exact projector onto ${\cal H}_P$ as a TN (Sec. \ref{sec:proj_tn}).  We  re-derive, with our formalism, the elementary  gauge invariant operators (Sec. \ref{sec:g_inv_op})  necessary to describe the dynamics  inside  ${\cal H}_P$.

In the course of our discussion we explain that continuous groups are associated to infinite dimensional local Hilbert spaces, and thus the TN network construction for them  has infinite bond dimension $D$, and thus is computationally intractable.

In order to cure this, we introduce a scheme that allows to truncate, in a gauge invariant way, the infinite dimensional local Hilbert spaces.
In this way we obtain a version of the KS LGT for (compact) continuous groups defined on finite dimensional Hilbert spaces (Sec. \ref{sec:t_ks}). The projector onto ${\cal H}_P$ for these models can be expressed as a TN with finite bond dimension, and can thus be used in practical computations.

We review the alternative constructions of LGT with  continuous gauge symmetry and discrete local Hilbert spaces, called gauge magnets (or link models) \cite{horn_finite_1981,orland_lattice_1990,orland_exact_1992,chandrasekharan_quantum_1996,brower_qcd_1999}  (Sec. \ref{sec:gm}). We  generalize  it to arbitrary groups and we show that in the non-Abelian case, the gauge magnets are not equivalent to a  local gauge invariant  truncation of the KS LGT. 

\begin{figure}
 \includegraphics[width=\columnwidth]{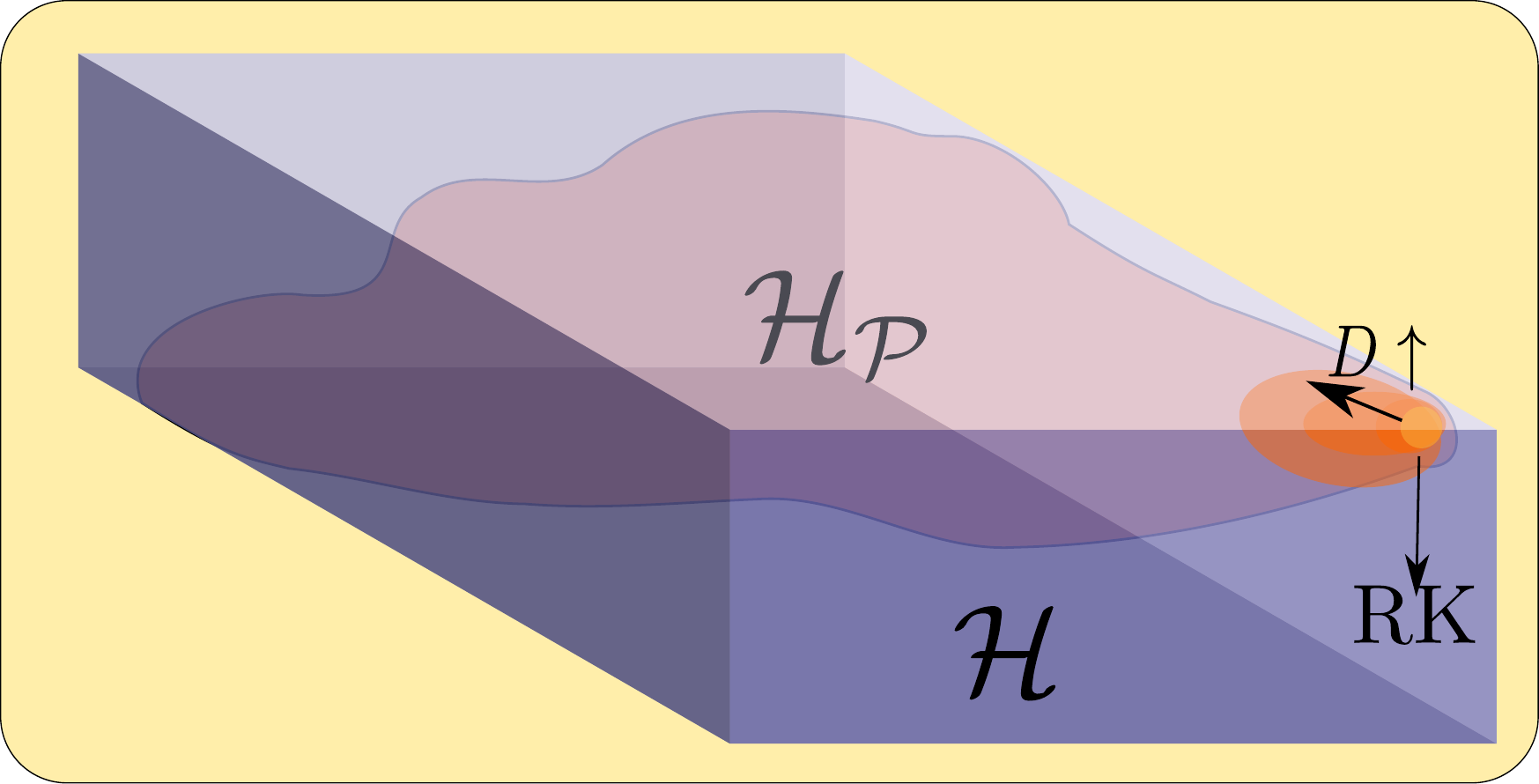}
 \caption{The Hilbert space ${\cal H}$ of a quantum many-body system (represented here by a 3D box) is exponentially large, since it is  the tensor product of the Hilbert spaces of the constituents. Gauge symmetry allows to identify a smaller space that we call the \emph{physical Hilbert space} ${\cal H }_P$. This  is the subspace spanned  by those states that fulfill all the local constraints imposed by the gauge symmetry and  is represented by a membrane inside ${\cal H}$.  ${\cal H }_P$ is smaller than the full ${\cal H}$ but it is still exponentially large. Low-energy states of local gauge invariant Hamiltonians, however, are expected to live in a small corner of ${\cal H }_P$, in the same way as low-energy states of generic local Hamiltonians live in a small corner of ${\cal H}$ \cite{verstraete_criticality_2006,hastings_area_2007,wolf_area_2008,masanes_area_2009}.
 For this reason  we design a variational ansatz based on TN that allows to explore this small corner of ${\cal H }_P$ (orange oval). By increasing the bond dimension of the elementary tensors in the TN ($D\uparrow$) we can explore increasingly large regions of ${\cal H }_P$, and eventually for $D\to \infty$ we can cover the whole   ${\cal H }_P$.  The projector on ${\cal H}_P$, and a family of interesting Rokshar-Kivelson (RK) states, are obtained exactly with  TN  with minimal bond dimension that scales as $D\simeq\sqrt{d}$ where $d$ is the dimension of the local Hilbert space. \label{fig:hilb_emb} }
\end{figure}

At this stage we are able to introduce a general TN variational ansatz for LGT, with both discrete and continuous groups, that again automatically embeds all local constraints dictated by the gauge symmetry. The states described by this ansatz are indeed gauge symmetric by construction.
Gauge-symmetry constraints, indeed, allow to restrict the attention to ${\cal H}_P$, which is still, however, exponentially large as shown in Fig. \ref{fig:hilb_emb}. Low-energy states of local gauge invariant Hamiltonians are  expected to live only on a small region of ${\cal H}_P$, in the same way as low-energy states of generic local Hamiltonians live in a small region of the unconstrained Hilbert space, since they fulfill the ``area-law'' for the entanglement \cite{verstraete_criticality_2006,hastings_area_2007,wolf_area_2008,masanes_area_2009}.

We explicitly construct a TN ansatz that allows  to explore this small corner of  ${\cal H }_P$ (Sec. \ref{sec:var_anst}). In its simplest form, the TN ansatz requires the same bond dimension as the projector onto ${\cal H}_P$. In this case $D\simeq\sqrt{d}$, with $d$ the dimension of the local Hilbert space, and it allows to characterize the physics of generalized Rokshar-Kivelson states (RK) \cite{rokhsar_superconductivity_1988}. By increasing the bond dimension, one can gradually explore all the space of gauge invariant states ${\cal H}_P$, as represented in Fig. \ref{fig:hilb_emb}, by increasingly large orange circles.

The TN ansatz depends on several elementary  tensors, each made by two distinguished parts. One part is  completely determined  by the gauge-symmetry constraints, while the other contains the free parameters to be used in variational calculations. As in the case of globally invariant TN \cite{mcculloch_non-abelian_2002,singh_tensor_2011,singh_tensor_2012}, our formalism allows to unveil interesting connections between gauge symmetric quantum states and spin networks \cite{rovelli_spin_1995}.  

As a further application we define  gauge invariant vertex operators \cite{ardonne_topological_2004} for arbitrary  gauge theories (Sec. \ref{sec:vertex_op}). In this way we open new possibilities to use them  as extensions of the standard Hamiltonians in order to explore extended phase diagrams of the known models. 

We benchmark these new tools in the context of RK states. In particular we focus on  the recent proposals about the characterization of  quantum phases based on the  analysis of  the entanglement scaling of the ground state wave function \cite{kitaev_topological_2006,levin_detecting_2006,li_entanglement_2008}. We analyze the well known transition between the eight-vertex and the six-vertex models. In the gauge theory language this transition is induced by applying the vertex operators \cite{ardonne_topological_2004} onto the RK $\mathbb{Z}_2$ wavefunctions. In this way we  provide an example of phase transition between a $\mathbb{Z}_2$ gapped spin liquid and a $U(1)$ algebraic spin liquid that is detected by the topological entropy, but elusive for the lowest part of the entanglement spectrum (Sec. \ref{sec:num_res}).  

All the discussion about connections of our results with other works in the literature is postponed to Sec. \ref{sec:about}, and we conclude with a summary of our results and an outlook on future developments in Sec. \ref{sec:conc}.

\section{Summary of the results}
Here we briefly summarize the most important results of our paper so that the reader interested in applying our formalism to specific models will easily find the relevant material.

- We derive the TN representation of the standard KS LGT Hamiltonian for arbitrary compact groups in Eq. \ref{eq:one_site_op}, \ref{eq:u_ks}, \ref{eq:h_ks}.

-  We present an extra gauge invariant operator that can be added to the KS Hamiltonian to explore generalized KS LGT for arbitrary compact groups in Eq. \ref{eq:v_op}.

- We provide the exact TN representation for the projector onto ${\cal H}_P$  of the KS LGT  (represented as an  hyperplane in Fig. \ref{fig:hilb_emb}) in Fig. \ref{fig:proj_gauge}. It  is the contraction of several copies of elementary  tensors $\cal{C}$ and $\cal{G}$ defined in Fig. \ref{fig:proj_k_s} and Eq. \ref{eq:dis}. 

- We describe a truncation of the Hilbert space of the KS LGT that is consistent with gauge symmetry in Sect. \ref{sec:t_ks}. This allows to study LGT with continuous groups (i.e. $U(1)$ and $SU(N)$) with constituents leaving in finite dimensional Hilbert spaces.

- In this way we introduce an exact TN representation with finite bond dimension of the projector onto  the ${\cal H}_P$ of an arbitrary  LGT. It is encoded in the TN of Fig. \ref{fig:proj_gauge} with the $\cal{C}$ and $\cal{G}$ tensors defined in Fig. \ref{fig:proj_k_s_t}.
  
- We discuss the relation of this truncation scheme  with  gauge magnets or quantum link models. We build the exact  TN representation of the projector onto  ${\cal H}_P$ also for gauge magnets. It is the TN of  Fig. \ref{fig:proj_gauge} with the $\cal{C}$ and $\cal{G}$ tensors defined in Fig. \ref{fig:proj_gm}.

- We provide a variational ansatz for generic states of ${\cal H}_P$  of all the LGT discussed as the TN of Fig. \ref{fig:gen_ans}. The ansatz is the contraction of several copies of  sparse tensors that unveils a connection between the LGT gauge invariant Hilbert space and spin networks.

- We confirm that the topological entropy  detects  phase transitions elusive to standard local order parameters. In particular, we characterize the transition between two different topological phases the 2D 8-vertex  and the 6-vertex topological phases. These results are presented  in Fig. \ref{fig:topo_ent}.  In Fig. \ref{fig:num_sp} we show that the same phase transition   does not affect the  behavior of the lowest part of the entanglement spectrum. 

\section{Tensor Networks}
\label{sec:tn}
Tensors are multi-linear maps $\bar{X}^a_{bc}$ acting among different Hilbert spaces. In particular the coefficients of a state of a quantum many-body systems are encoded in the element of a very large tensor  $T^{i_1 \dots i_N}$, 
\begin{equation}
\ket{\psi} = \sum_{i_1 \dots i_N }T^{i_1 \dots i_N }\ket{i_1 \dots i_N}. \label{eq:qstate}
\end{equation}
In general, the tensor $T$ is too large to be stored on a computer and thus it is useful to express it as the contraction of smaller elementary tensors. These contracted tensors are called TN. When dealing with large TN, the formulas become easily large and complex and it is simpler to resort to a  graphical notation.  The graphical notation is explained  already  in the literature \cite{tagliacozzo_simulation_2009,evenbly_tensor_2011,eisert_entanglement_2013,orus_practical_2013}, but we also shortly  review it here in order to fix the notation we use in the paper. 

\begin{figure}
 \includegraphics[width=\columnwidth]{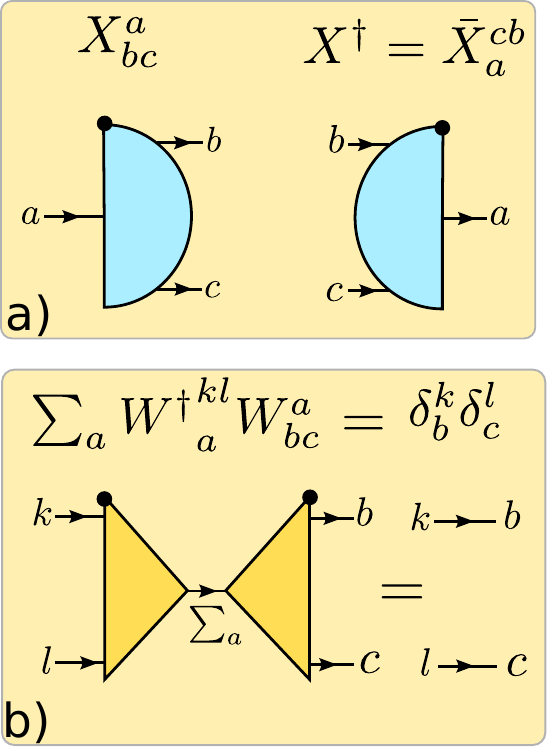}
 \caption{Graphic representation of tensors and their contractions. Tensors are generalized vertexes (geometric shapes) whose legs are represented by dangling arrows. {In panel a)} we represent a tensor with three legs $X^a_{b c}$. Upper indexes are  incoming legs while lower indexes are outgoing legs. Complex conjugation is denoted by mirror reflection of the vertex. Different colors denote different tensors. Indexes are ordered clockwise starting from the solid dot on the vertex. The contraction of two tensors is denoted by an arrow joining them. The dagger operation $X^{\dagger}$ involves both the complex conjugation of the elements of the tensor, and the inversion of the arrows attached to its legs. As a consequence  the order of the legs also changes, $X^{\dagger}=\bar{{X}^{c b}_a}$ \cite{cvitanovic_group_2008}. {Panel b) }An  isometric tensor (or simply isometry)  is always represented   by a triangular vertex.  A tensor $W$ is isometric if there is a specific choice of legs, which identifies two vector spaces $U$ and $V$, such that, when  $W$ is contracted with $W^{\dagger}$ through the legs in $U$,  the result is the identity tensor in $V$. In the figure,  $U$ is spanned by the leg $a$, while $V$ is spanned by the two legs $b,c$. \label{fig:tens}}
\end{figure}

In the graphical notation,  geometric shapes are associated to tensors and lines or ``legs'' attached to them  represent their indexes. As an example,  the upper panel of Fig. \ref{fig:tens} represents a tensor with three indexes $X^a_{bc}$. A small dot on the shape allows to keep track of the index ordering that is assumed  clockwise starting from the dot. Incoming legs \emph{in-legs} (upper indexes) are drawn with entering arrows while outgoing legs \emph{out-legs} are drawn as outgoing arrows. The Hermitian conjugate of a tensor, $X^{\dagger}$,  involves taking the complex conjugate of the elements and exchanging the in-, with the out-legs,  $X^{\dagger}=\bar{X}^{cb}_a$. This is represented  graphically  by mirroring the tensor and inverting the arrows on the lines (left of the upper panel of Fig. \ref{fig:tens}). 

%

A leg connecting two tensors represents their multiplication through the contraction of the corresponding  indexes (summation over all the values of that indexes). An example is represented on the left of the lower panel of Fig. \ref{fig:tens}, where the tensor  $W^{\dagger}$ is contracted with the tensor $W$.

In  our formulas we sometime omit explicit summation and  use the Einstein notation, where the summation is intended over repeated indexes. 

A tensor can always be interpreted as a matrix by dividing its legs into two groups, one group identifies  a  vectors spaces $U$ and the other $V$. In this way the tensor becomes a map from $U$ to $V$. A natural choice is to interpret $U$ as the collection of in-legs and $V$ as the collection of out legs, but this is not the only possibility.  

An important class of tensors are isometric tensors. For a specific choice of $U$ and $V$, they fulfill
\begin{equation}
 W W^ {\dagger} = 1_{V}. \label{eq:iso}
\end{equation}

In our drawings triangular shapes always represent isometric tensors \footnote{Reader acquainted with TN drawings should not assume that in this paper  the triangles provide the orientation  of the isometry, that is that the triangle can be interpreted as the tip of an arrows that points from  $U$ to $V$.  Here we use an arbitrary orientation for the triangle and explicitly write the isometric property so to identify  $U$ and $V$}.

The lower panel of Fig. \ref{fig:tens} illustrates a specific case of  a three-leg  \emph{isometric} tensors $W$, where $U$ is spanned by the leg $a$ and $V$ is spanned by the two legs $b,c$.
This implies that $W^{\dagger}$ is defined as $W^{\dagger} \equiv \bar{W}^{cb}_a$ and equation \eqref{eq:iso} reads $\bar{W}^{kl}_a W^{a}_{bc} = \delta^k_{b}\delta^l_{c}$ (remember that there is a sum over $a$). Graphically the contraction  is represented by the  line connecting the two tensors. The result of the contraction is a four-leg tensor, explicitly written as the tensor product of two identity tensors represented by straight lines. 

When TNs represent quantum states of many-body systems, the legs related to the constituents ($i_1 \to i_N$ in Eq. \eqref{eq:qstate}) are called \emph{physical} legs and are typically represented by Latin letters,  while all the others legs (those that are contracted) are legs called \emph{auxiliary} legs, and represented by Greek letters.

\section{Group theory in the tensor network language}
\label{sec:gt}
 Here we assume that the reader is familiar with  basic concepts of the representation theory of both finite and continuous groups, and we list the relevant results for our paper in order to  express them  in the TN language.
 In particular, while this paper deals with gauge theories with continuous groups, we develop the formalism  by using  discrete groups $G$.
 The underlining ideas, are indeed completely independent from the fact whether the group is discrete or continuous,  and we feel that discrete groups allow to present these ideas in a simpler way. 
 
 Concretely, all the results we present, which involve the summation over group elements,  can be rewritten for  continuous compact groups by substituting the sums  with integrals over the group, defined through the appropriate invariant measure (see i.e. chapter 4 of \cite{serre_linear_1977}).
 
Here, we remind that a collection of elements $\set{g}$ closed under multiplication forms a group $G$. We are mostly interested in  matrix representations of the group $G$ that are obtained by associating to each group element $g$ a unitary  matrix, $\Gamma(g)$, acting on a given vector space.
In this way the group multiplication table is rephrased  into specific relations between the representing matrices. In general, given a matrix representation of a group, there is a well defined procedure to reduce it to a block diagonal form, where each of the blocks constitute a ``smaller'', independent representation of the same group $G$. If those blocks are not further reducible into smaller blocks they define an irreducible representation 
of the group  $G$ that in this paper will be labeled by $r$. 
The dimension and number of the irreducible representations depend on the group $G$, and their study is the subject of the theory of group representations. In the following we denote by  $\Gamma_r(g)$ the matrix representation of $g$ in  the irreducible representation  $r$. 

One of the most important results of the theory of group representations is, what is typically called, the ``great orthogonality theorem'' \cite{got}. It states that given a group $G$ with elements $g$, and given any pair of irreducible representations  $r,r'$ the following relation holds 
\begin{equation}
 \frac{\sqrt{n_r n_{r'}}}{|G|}\sum_g {\Gamma_r(g^{-1})}^{i}_{j} \Gamma_{r'}(g)^{l}_{k} = \delta^{i}_{k} \delta^{j}_{l} \delta(r,r'), \label{eq:orth_rel}
 \end{equation}
where $\Gamma_r(g)^{l}_{k}$ are the $l,k$ matrix elements of the irreducible representation (irrep) $r$ of $g$, $|G|$ is the number of elements of $G$, and $n_r$ and $n_{r'}$ are  the dimensions of the irrep's $r$ and $r'$.
We now want to reinterpret this relation in terms of TN diagrams. 
In order to do this  we need to  identify two vector spaces $U$ and $V$. $U$ is  spanned by the group elements $g$.
This vector space   is called group algebra $\mathbb{C}(g)$, and has dimension equal to $|G|$. $V$  is the vector space spanned by the direct sum $V= \oplus_r (V_r \otimes V_{\bar{r}})$.  Each of $V_r$ is the defining space of the irrep $r$. $V_{\bar{r}}$ is the defining space of its conjugate representation, obtained by taking the Hermitian conjugation of the matrices, $\Gamma^{\dagger}_r(g)\equiv \Gamma_r(g^{-1})$, where we have used the property that we are dealing with unitary representations. The fact that the vector space $V$ has a direct sum structure is encoded in the  $\delta(r,r')$ factor in the right hand side of Eq. \eqref{eq:orth_rel}. 

\begin{figure}
 \includegraphics[width=\columnwidth]{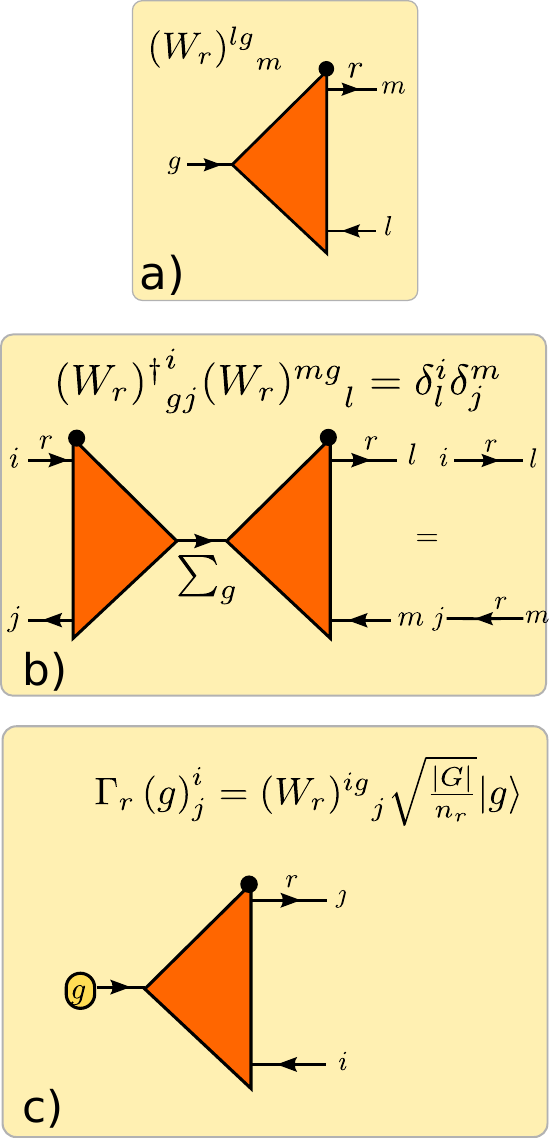}
 \caption{a) The orthogonality relation in \eqref{eq:orth_rel} allows to identify a set of isometries $W_r$ projecting  $\mathbb{C}(g)$ onto $V_r\otimes \bar{V}_r$. If one knows the matrices $\Gamma_r(g)$ for all $g \in G$,  $W_r$ is  defined by collecting them inside a three-index tensor. The index $r$ identifying the irrep is written on the top of one of the two  legs. b) The isometric property of the tensor $W_r$. 
 c) Alternatively, if we know the $W_r$ isometry, we can obtain the matrices in the $r$ representation by acting with $W_r$ on a vector proportional to $\ket{g}$ (yellow circle in the figure). The resulting tensor  has two indexes $i,j$, and is the matrix $\Gamma_r(g)^i_j$.
 \label{fig:ortho}}
\end{figure}

We start by focusing on the above formula in the case $r = r'$.
In this case  Eq. \eqref{eq:orth_rel} tells that the tensor   $W_r$, 
\begin{equation}
 W_r =\sqrt{\frac{{ n_r}}{|G|}}\Gamma_r(g)^{l}_{k}, \label{eq:wr}
\end{equation}
is an isometry. The tensor $W_r$ is represented in the panel a) of Fig. \ref{fig:ortho}. 
The fact that it is an isometry means that
\begin{equation}
 W_r ^{\dagger} W_r = 1_{V_r \otimes V_{\bar{r}}},
\end{equation}
as represented in the panel b) of Fig. \ref{fig:ortho}.

Each of the $W_r$ thus allows to project a vector $\ket{g}$ in the group algebra $\mathbb{C}(G)$ onto a vector $\ket{(e_r)^m_l}$ of the tensor product $V_r \otimes {V}_{\bar{r}}$.
This  can be used in two ways. 
Reading Eq. \eqref{eq:wr} from left to right, it  tells us that if we know  all the $\Gamma_r(g)$ for all the elements $g \in G$,  we can collect them in a three-leg tensor and  obtain  an isometry that projects ${\cal C}(G)$ onto $V_r \otimes {V}_{\bar{r}}$.
From right to left,  we can obtain the irrep's matrices. If we are given the isometry $W_r$, by acting onto the vector $\ket{\tilde{g}}=\frac{|G|}{\sqrt{ n_r}}\ket{g}$ of $\mathbb{C}(g)$ we  obtain  $\Gamma_r(g)$, 
\begin{equation}
 \Gamma_r(g) = W_r \ket{\tilde{g}}, \label{eq:gamma_g} 
\end{equation}
as we represent graphically in the panel c) of Fig. \ref{fig:ortho}.
 
\begin{figure}
\includegraphics[width=\columnwidth]{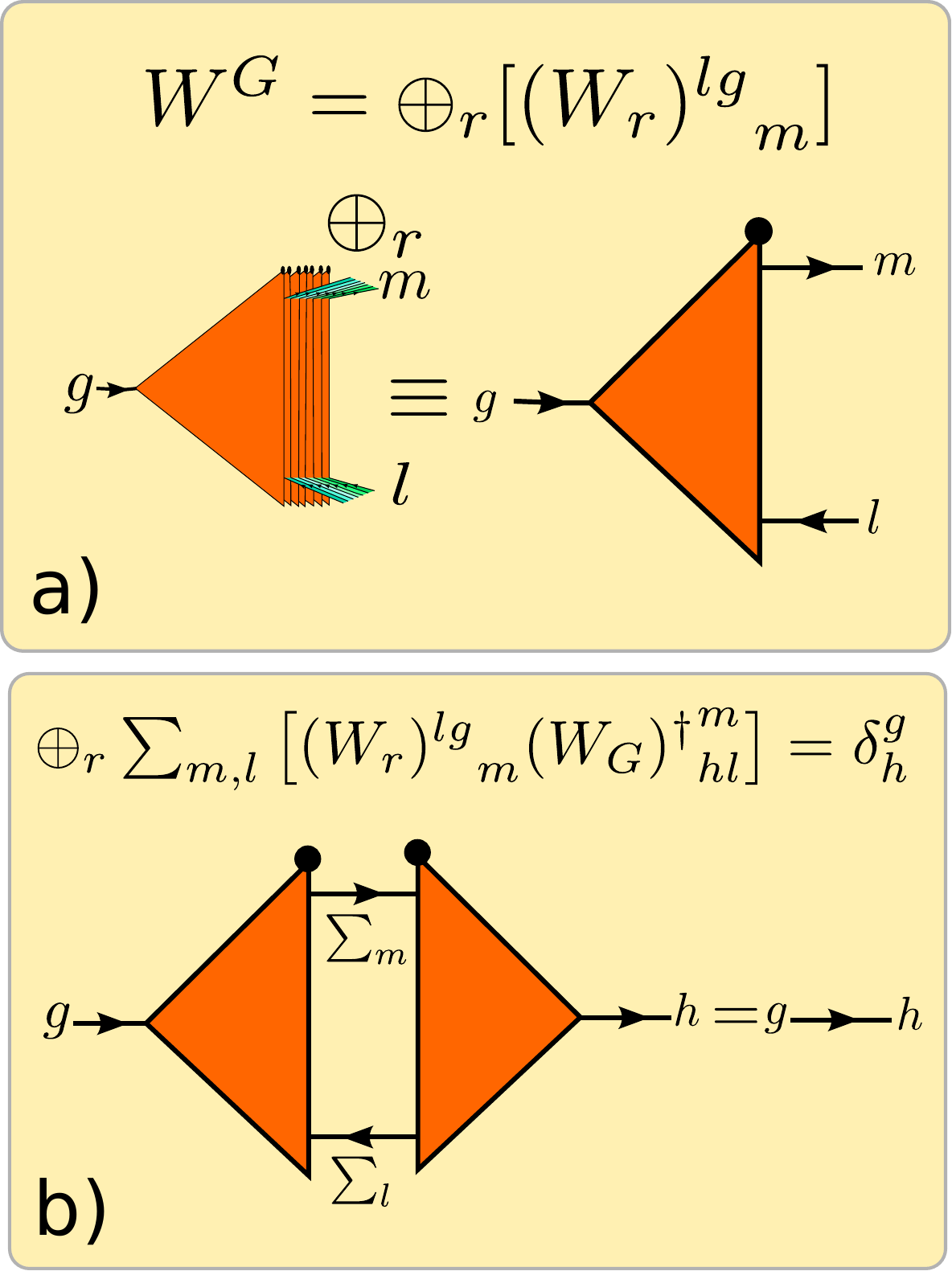}
\caption{ a) The direct sum of $W_r$ is a unitary tensor. It allows to  change basis from ${\mathbb C}(G)$ to the direct sum of all the irrep's times their conjugate, $\oplus_r (V_r \otimes V_{\bar{r}})$. On the upper part of the figure we explicitly draw  the collection of all the $W_r$'s. A simplified picture is shown on the lower part, where the direct sum is implicit in the absence of the label $r$ attached to the legs of the tensor. This is the notation we will use in the following. When we want to represent  $W_r$ we will attach $r$ to the leg, while we represent $W_G$ by exactly the same drawing but without the $r$. From now on, for simplicity, we will also omit to label the legs with letters.
b) The unitarity of $W_G$ is encoded in the fact that the contraction over the two legs  acting on $\oplus (V_r \otimes V_{\bar{r}})$ gives a delta function in ${\mathbb C}(G)$.
\label{fig:W_G}}
\end{figure}

A peculiarity of our notation, that we inherit from spin networks, is that each leg of the tensor also carries a representation index $r$ that is typically superimposed to the line. In order to avoid confusion, after  Fig. \ref{fig:W_G} we will drop all the letters labeling the indexes of the tensors (unless really needed), and keep only the letters related to the irrep $r$.

A second result of the theory of group representations is that $\sum_r n_r^2 = |G|$,  that is the direct sum of all the $W_r$ is a unitary transformation as encoded in following relations,
 \begin{equation}
  W_G = \oplus W_r ,\quad   W_G W_G^ {\dagger} = 1_{\mathbb{C}(G)}, \quad W_G^ {\dagger} W_G = 1_{\oplus_r (V_r\otimes {V}_{\bar{r}})}, \label{eq:wg}
 \end{equation}

A graphical representation of the direct sum of $W_r$ leading to $W_G$ is presented in Fig. \ref{fig:W_G}.

%
\subsection{Symmetric tensors}
\label{sec:symm}
\begin{figure}
 \includegraphics[width=\columnwidth]{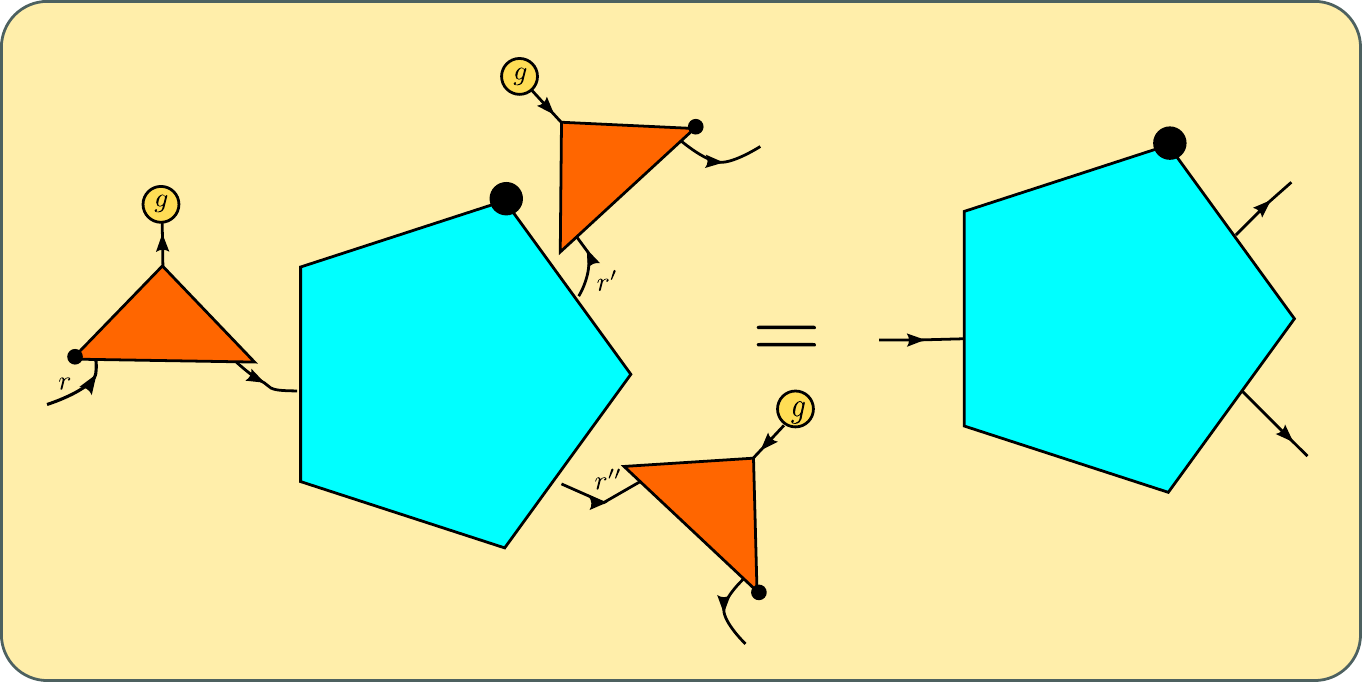}
 \caption{A symmetric tensor is left invariant by the simultaneous rotation of all incoming legs by $\Gamma^{\dagger}(g)$, and its out-going legs by $\Gamma(g)$ in the appropriate representation. Here we exemplify the case of a three-leg tensor with an incoming leg that transforms in the irrep $r$, and two out-legs that transform under irrep's $r'$ and $r''$. \label{fig:symm}}
\end{figure}

In the context of many-body quantum systems, symmetries play a fundamental role in the classification of phases. Recently, in the context of TN states, symmetries have been used to classify gapped phases of 1D systems \cite{chen_classification_2011,turner_topological_2011,schuch_classifying_2011}. For this reason  a strong effort has been devoted to incorporate the appropriate exact symmetries even when studying many-body systems with approximate variational ansatz.
In the context of TN, a sufficient condition in order to have symmetric states is that the constituent tensors are symmetric \cite{sanz_matrix_2009,singh_tensor_2010,singh_tensor_2011,perez-garcia_canonical_2010,singh_tensor_2012}.
As an example a  symmetric tensor with respect to the group $G$ with one in-leg and two out-legs obeys the following equation,
\begin{equation}
 T^{a'}_{b',c'}= T^a_{b,c} {\Gamma^{\dagger}_r(g)}_{a}^{a'} {\Gamma_{r''}(g)}^{b}_{b'}{ \Gamma_{r´´´}(g)} ^{c}_{c'} \label{eq:inv_tens},
\end{equation}
with $\Gamma_r(g)$ the matrix of the appropriate unitary representation of the group $G$. This relation is sketched graphically in Fig. \ref{fig:symm}.

Equation \eqref{eq:inv_tens} can be satisfied by non-vanishing tensors only when the tensor product $R=\bar{r}\otimes r' \otimes r''$ contains the trivial representation, that is the representation where all group elements are mapped to the identity. 
It is thus  important to be able to explicitly construct the trivial representation contained in a given tensor product of different representations.
For continuous groups this can be done by diagonalizing the corresponding Casimir operators. Their zero eigenvalues, if present, identify  the trivial factors. The way this is  done in practice is explained in detail in Refs. \cite{singh_tensor_2012,weichselbaum_non-abelian_2012}.

An alternative way, which works both for discrete and continuous groups, is to explicitly build the projector onto the trivial representation as a group sum (or integral in the case of continuous groups). The projector  is given by  
\begin{equation}
 P_0 = \frac{1}{|G|}\sum_g \Gamma_R(g),  \label{eq:proj_triv}
\end{equation}
as can be found i.e. in Ref \cite{serre_linear_1977}. This is the method that we will mostly use. 

A third possibility entails disentangling the symmetry constraints following the ideas of Refs. \cite{aguado_entanglement_2008,tagliacozzo_entanglement_2011,aguado_entanglement_2011}. We postpone the discussion about this until Sec. \ref{sec:proj_tn}, where we will provide an explicit example of this procedure. 

%
%
%
%
%

\section{Hamiltonian Lattice Gauge Theories from the tensor network perspective}
\label{sec:lgt}

Gauge theories originated, at a classical level, in the description of the electromagnetic interactions between charged particles and light. The   physical processes, described by the Maxwell equations, depend only on the electric and the magnetic fields while Maxwell equations  can be written in terms of a vector potential and in this form they show some redundancy. The vector potential can be  modified by  adding to it the gradient of an  arbitrary scalar potential, without affecting the corresponding electric and magnetic field, so giving the same physical results.

At a quantum level the vector potential becomes a full quantum field, and the redundancy appears as a  local symmetry in the action that drives its dynamics. The generalizations of these  ideas to vector potentials describing non-Abelian ``electric and magnetic'' fields, and their success in describing the hadron spectrum, gave rise to the modern gauge theories, and to our understanding of particle physics. 

Actual calculations away from the perturbative regime are most of the times carried out numerically in the framework of LGT by discretizing the space-time on a lattice. In this formalism the vector potential is associated to the links of the lattice, while the charged matter fields live on the sites.

A Hamiltonian version of the system has been obtained by identifying one of the lattice direction as ``time'', fixing the temporal gauge,  and constructing  the Hamiltonian operator  whose  matrix elements coincide, in the time continuum limit, with those of transfer matrix in the time direction \cite{kogut_hamiltonian_1975,creutz_gauge_1977}. 

In the Hamiltonian formulation LGT become  many-body quantum systems, whose constituents are divided in two groups, \emph{gauge bosons}  attached to the links of  an oriented lattice $\Lambda$ and \emph{matter constituents} attached to the sites $s$ of $\Lambda$. Here, we will work on an oriented 2D square lattice but, what follows, can be generalized easily to more complex orientable lattices. 

The original local symmetry of the classical action  is then mapped to a residual local symmetry of the quantum Hamiltonian.
Symmetric Hamiltonians typically have symmetric eigenstates, and thus one can decide to characterize the space of locally symmetric states. 
The residual local symmetry is defined in terms of a set of constraints, 
that the quantum sates and operators should fulfill. As we show in this section, both the operators used to define the symmetry constraints and the local constraints have a natural expression in terms of TN diagrams. In particular in this section we describe the Kogut-Susskind version of the Hamiltonian LGT (KS).

\subsection{Hilbert space of constituents}
\label{sec:lgth}

The Hilbert space for gauge bosons is the group algebra $\mathbb{C}(G)$. In this case, one can associate a state $\ket{g}$ to any group element $g\in G$. States representing different group elements are orthogonal $\braket{g|h}= \delta_g^h$. As a consequence,  the dimension of the local Hilbert space is equal to  $|G|$. In particular, continuous groups require dealing with infinite dimensional Hilbert spaces.

The lattice $\Lambda$ is oriented, and it is made by $L$ links so that the total Hilbert space is $\mathbb{C}(G)^L$. Changing the orientation of one link sends $\ket{g} \to \ket{g^{-1}}$. If we act with the operator ${\cal O}$ on the state $\ket{g}$, this also implies that we need to act with the operator ${\cal O}^{\dagger}$ on the state $\ket{g^{-1}}$, obtained by reversing the link orientation.

A prerequisite for defining the action of the symmetry operators is to be able to define the action of left and right rotations of a state by arbitrary group elements $h$  and $k \in G$. This is achieved by defining  the operators  $L(h^{-1})$ and $R(k)$, which act on  $\ket{g}$ and produce  
\begin{equation}
 L(h^{-1})R(k)\ket{g} \equiv \ket{h^{-1}gk}. \label{eq:left_right}
\end{equation}

\begin{figure}
 \includegraphics[width=\columnwidth]{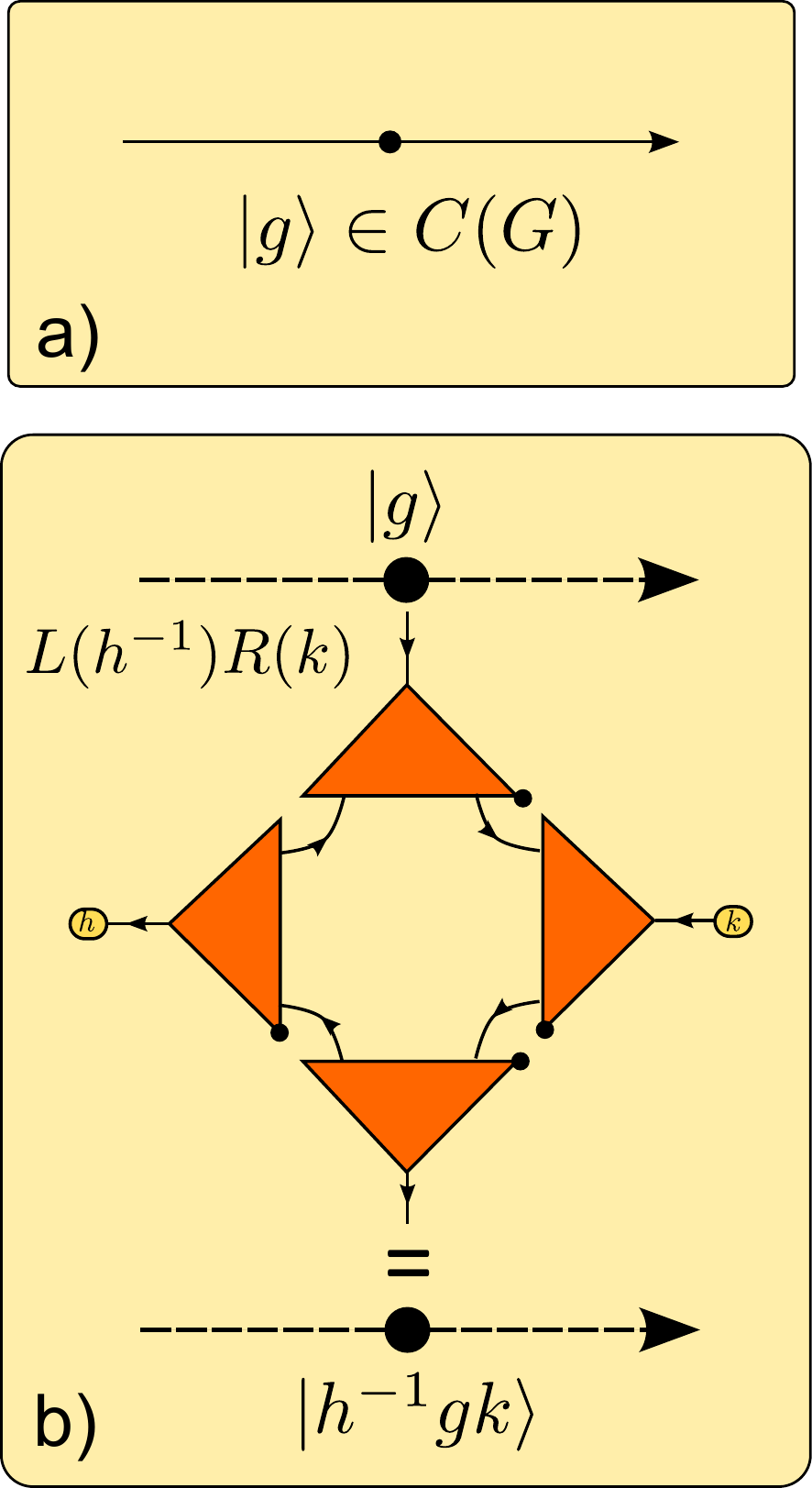}
 \caption{Gauge boson constituents  are defined on the links of an oriented lattice. Links are represented by dashed lines, and constituents are small solid circles along these lines (which should not be confused with tensors, whose legs are solids lines). a) In the  \emph{Kogut-Susskind} LGT, each constituent is described by a state $\ket{g}$ of the group algebra $\mathbb{C}(G)$. 
 b) Left and right rotations $L(h^{-1})$ and $R(k)$ are introduced in Eq. \eqref{eq:left_ks} and Eq. \eqref{eq:right_ks}. They transform the state $\ket{g}$ into $\ket{h^{-1}gk}$. Both operators require  an initial change of basis from $\mathbb{C}(G)$ to $\oplus_r (V_r\otimes \bar{V}_r)$, obtained through $W_G$ of Eq. \eqref{eq:wg} and represented in Fig. \ref{fig:W_G} (the first horizontal triangle). One then applies to each of the legs the corresponding rotation matrix given by $\oplus_r \left[ \Gamma_r(h^{-1})\otimes \Gamma_r(k)\right]$, with the individual  $\Gamma_r$  defined in  panel c)  of Fig. \ref{fig:ortho} and represented here by the two vertical triangles. At last  $W^{\dagger}_G$,  represented by the inverted triangle, allows to go back to  to $\mathbb{C}(G)$ .  \label{fig:hilb_space}}
\end{figure}

 This is done first by using $W_G$ of Eq. \eqref{eq:wg}, to change basis from $\mathbb{C}(G)$ to ${\oplus_r} (V_r\otimes \bar{V}_r)$. At this stage the rotation is performed through the direct sums of the rotation matrices in each representation, and then one rotates back to $\mathbb{C}(G)$ with $W^{\dagger}_G$. This is  expressed graphically in the  panel b) of Fig. \ref{fig:hilb_space}, and reads
\begin{equation}
 L(h^{-1}) = W_G \oplus_r\left[ \Gamma_r(h^{-1}) \otimes 1_r\right], W^{\dagger}_G \label{eq:left_ks}
\end{equation}
\begin{equation}
 R(k) = W_G \oplus_r \left[ 1_r\otimes \Gamma_r(k) \right] W^{\dagger}_G. \label{eq:right_ks}
\end{equation}

\subsection{Gauge transformations}
\label{sec:lgt_gt}
\begin{figure}
 \includegraphics[width=\columnwidth]{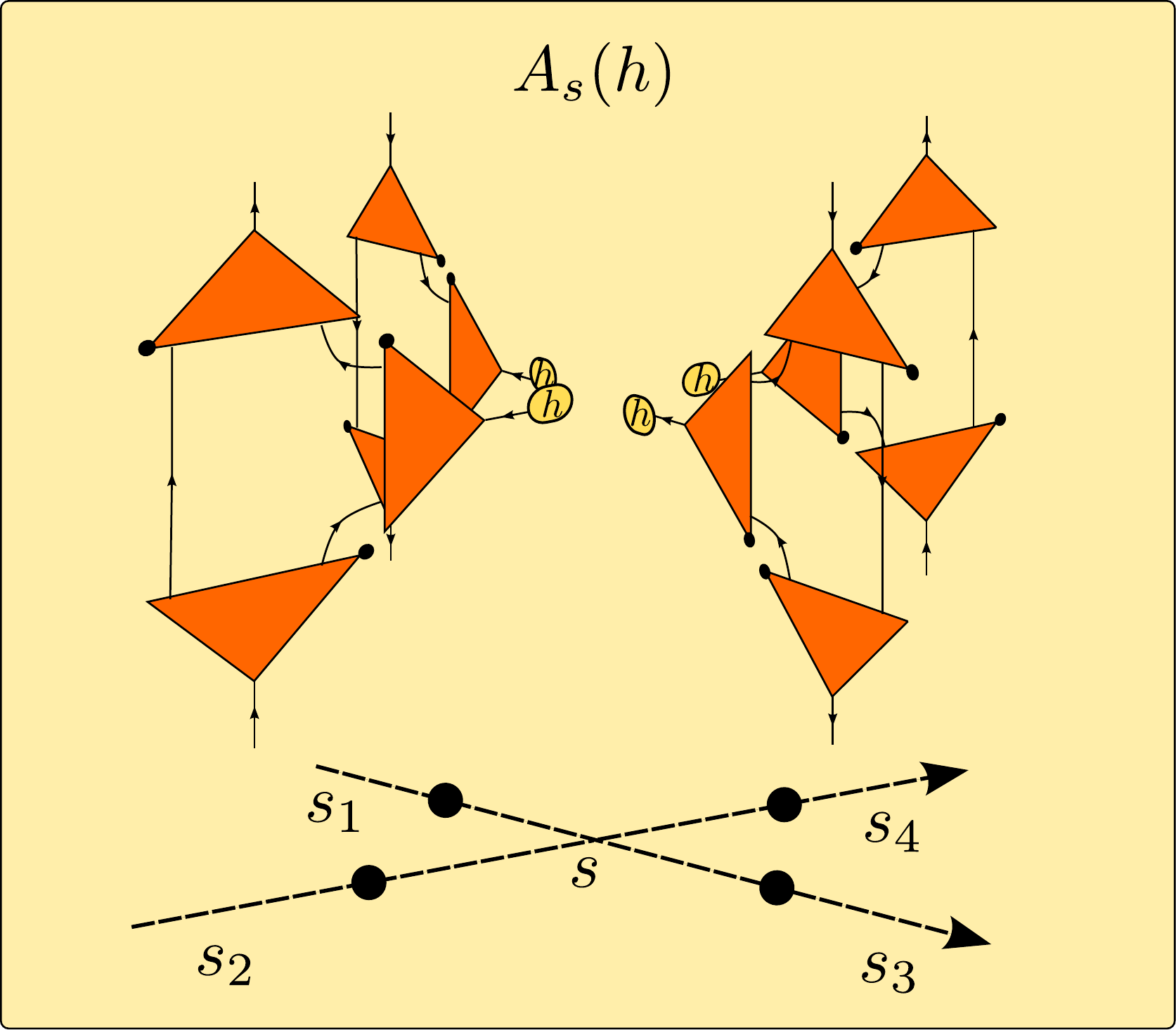}
 \caption{Graphical representation of the operator $A_s(h)$, the building block of gauge transformations in the Kogut-Susskind LGT. On every link entering a site $s$, it either rotates the state of the link through $R(h)$ or $L(h^{-1})$, depending on whether the link enters or leaves the site. The form of $L$ and $R$ are given in panel b) of  Fig. \ref{fig:hilb_space}. \label{fig:a_ks}}
\end{figure}

Having the operators that perform  the  left and right rotations, we are now in the position to define the operators $A_s(h)$, the building blocks of local gauge transformations.
In particular, the local transformation  rotates all the states of the links touching a  site $s$ by an element  $h\in G$. Since the lattice is oriented, the transformation induced by $A_s(h)$ is different for entering $s$ (in-links) and  links leaving $s$ (out-links).  All in-links are rotated through $R(h)$ while all the out-links are rotated on the left by $L(h^{-1})$. 
Concretely, $A_s(h)$ at site $s$, acting on links $s_1$ to $s_4$ ordered counter-clockwise starting from the left, is defined as  
\begin{equation}
 A_s(h) = R(h)_{s_1} \otimes R(h)_{s_2} \otimes L(h^{-1})_{s_3} \otimes   L(h^{-1})_{s_4}, \label{eq:gauge_trans}
\end{equation}
 with $L$ and $R$ defined respectively in Eq. \eqref{eq:left_ks} and \eqref{eq:right_ks}.
 Notice that $[A_s(g), A_s'(h)]= 0$ if $s \neq s'$, as a consequence of the commutation between $L$ and $R$ operators defined on the same links.  They are represented graphically in Fig. \ref{fig:a_ks}.
A generic gauge transformation is then a product of local gauge transformations, where for each site $s$ one chooses a different element $g_s$ to perform the desired rotation. Given a lattice of $L_s$ sites and a choice of $L_s$ elements $h_i, i=1\dots L_s \in G$, we obtain the transformation ${\cal T}$,
 \begin{equation}
  {\cal T}(\set{h_1\dots h_{L_s}}) = \prod_{s=1}^{L_s} A_s(h_s). \label{eq:glob_gauge_transf}
 \end{equation}

\subsection{The gauge invariant Hilbert space}
 \label{sec:lgt_ph}
The set of states in the Hilbert space that are invariant under all  $A_s(g)$, the building blocks of the gauge transformations, defined by Eq. \eqref{eq:gauge_trans},
constitutes the \emph{physical Hilbert space} (or gauge invariant Hilbert space) ${\cal H}_p$,
\begin{align}
{\cal H}_p \equiv \left\{ \ket{\phi} \in \mathbb{C}(G)^L, \right. \nonumber\\ 
\left. A_s(g)\ket{\phi} = \ket{\phi} \quad \forall s \in \Lambda, g\in G \right\}, \label{eq:gauge}
\end{align}
where $s$ are the sites of the lattice $\Lambda$, $L$ is the number of links, and $g$ is an arbitrary group element.

An example of state in  ${\cal H}_p$ is given by
\begin{equation}
 \ket{\phi} = \ket{+}^{\otimes L} =\left( \frac{1}{\sqrt{|G|}}\sum_g \ket{g} \right)^{\otimes L} \label{eq:st_+}
\end{equation}
since $(L(h^{-1})\ket{+}R(k)) =\ket{+}$. 

${\cal H}_p$ is a subspace of the original tensor product Hilbert space $\mathbb{C}(G)^L$, as sketched in Fig. \ref{fig:hilb_emb}. 
By construction all states  in ${\cal H}_p$ are also invariant under any global transformation ${\cal T}$, defined in Eq. \eqref{eq:glob_gauge_transf}.

\subsection{Gauge invariant operators}

\label{sec:g_inv_op}
We are now interested in introducing the dynamics of a LGT, and this is first done by  characterizing gauge invariant operators. 

Gauge invariant operators are operators ${\cal O}$ that commute with all the $A_s(g)$, that is
\begin{equation}
 [{\cal O}, A_s(g)] =0 , \forall g \in G, s \in \Lambda. \label{eq:g_inv_op}
\end{equation}
We would like to find the ``minimal'' gauge invariant operators, that is those that constitute the building blocks for   gauge invariant Hamiltonians, and with as small support as possible. In order to find them we take inspiration from the QED Hamiltonian, whose gauge part can be written as $H_{QED}= E^2 + B^2$, with $E$ and $B$ being the electric and magnetic fields. On the lattice the $E^2$ term is mapped to a link operator (remember that $E = - \nabla V$, with $V$ being a scalar field, so that in a mathematical sense $E$ is naturally a one form, and thus geometrically  attached to links). The term $B^2$ is  mapped to a  plaquette operator, where plaquettes, the smallest possible closed loops made by links,  are  the elementary pieces of the lattice surface (again remember that $B= \nabla \wedge A$, with $A$ being a vector field, and thus $B$ is naturally identified with a two form, discretized on plaquettes).

For this reason we look for a generalization to an arbitrary group $G$ of the $E$ operator of electromagnetism, as a single-link operator,   and of  the $B$ operator, as an operator acting on plaquettes of the lattice. 
We start by the single  link operator. In an  Abelian LGT  any matrix representing the rotation by a group element commutes with all the others, and thus fulfills  Eq. \eqref{eq:g_inv_op}. For this reason we can choose any $\Gamma_{\textrm{reg}}(g) +H.c.$ \footnote{ reg. stands for the regular representation, the representation that acts on the group algebra} as a link operator  \cite{tagliacozzo_optical_2013}. When dealing with non-Abelian LGT,  the only link operator that commutes with an arbitrary rotation, as requested by Eq.  \eqref{eq:g_inv_op}, is an operator proportional to the identity in each of the irrep's, as a consequence of the Schur lemma \cite{tinkham_group_2003}.   A gauge invariant link operator  acting on a link $s_n$ in the KS LGT thus can be expressed as  
\begin{equation}
 {\cal E}^2_{s_n}=  \left[ W_G ^{\dagger}  \oplus_r\left[  c^r\, \mathbb{I}_{r} \otimes \mathbb{I}_{\bar{r}} \right]W_G ^{\dagger}\right]_{s_n}, \label{eq:one_site_op}
\end{equation}
 with $c^r$ arbitrary numbers and $W_G$ defined in Eq. \eqref{eq:wg}. 

\begin{figure}
 \includegraphics[width=\columnwidth]{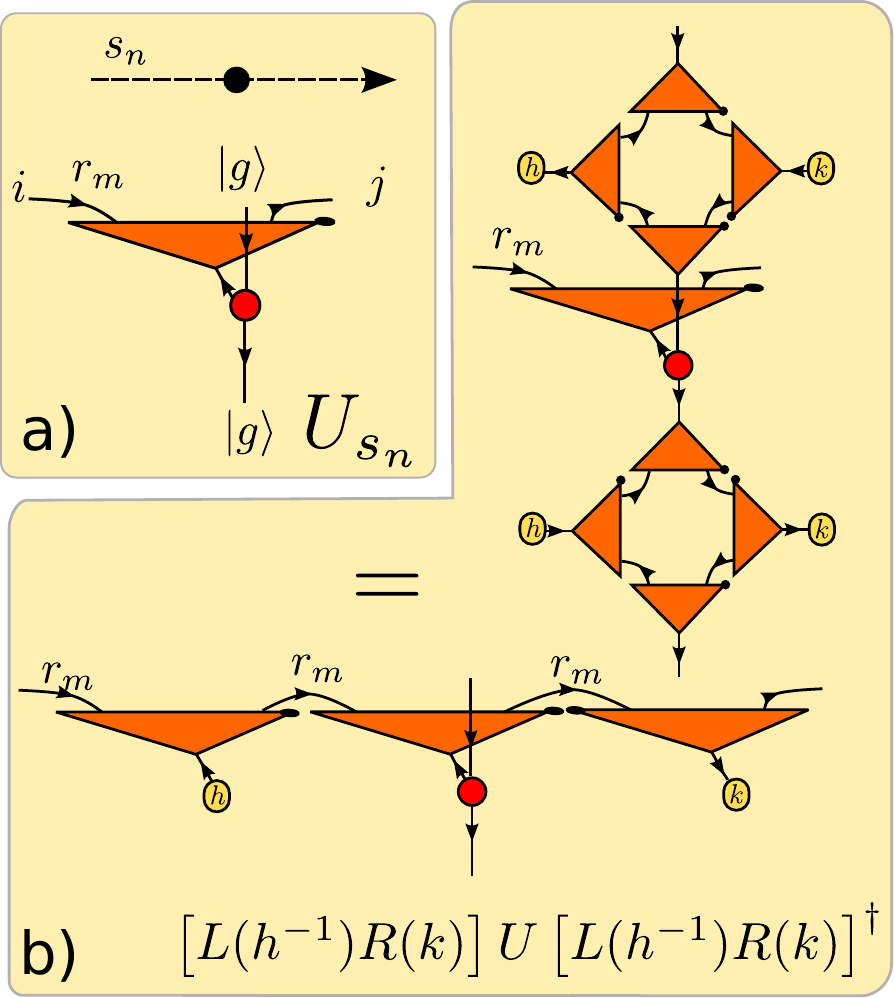}
 \caption{a) The $U_{s_n}$ operator of \eqref{eq:u_ks} in the KS LGT. It acts on the tensor product of the physical space ${\mathbb C}(G)$ of the link $s_n$  and of an auxiliary space that defines the  irrep $r_m$, ${\mathbb C}(G)\otimes V_{r_m}$. It is built from the contraction of a copy tensor ${\cal C}\equiv\proj{g}\otimes\bra{g}$ (red circle), and the corresponding rotation matrices $\Gamma_{r_m}(g)^{i}_{j}$ (defined in Eq. \eqref{eq:gamma_g}).
 b) The above definition implies that the $U_{s_n}$ transmits the rotation onto the physical index, as induced by conjugation by $L$ and $R$, to the auxiliary indexes. This property allows to use $U_{s_n}$ as building block of the gauge invariant plaquette operator in Eq. \eqref{eq:plaquette}, where, thanks to the trace, the rotations attached to the auxiliary indexes cancel.
 \label{fig:u_ks}}
\end{figure}

Plaquette operators can be defined as  matrix product operators whose elementary tensors $U$ act on the Hilbert space of a single link and an auxiliary Hilbert space. Specifically  $U$  acts on the tensor product of the $\mathbb{C}(G) \otimes V_{r_m}$. $V_{r_m}$ is the defining space of the irrep $r_m$ of $G$  that defines how the matter transforms under gauge transformation  \cite{kogut_hamiltonian_1975,creutz_gauge_1977,creutz_quarks_1983}. $U$ is defined as
\begin{equation}
 U_{s_n} = \sum_g \proj{g}_{s_n} \otimes \Gamma_{r_m}(g)^{i}_{j}. \label{eq:u_ks}
\end{equation}
It acts diagonally on the Hilbert space of the link $s_n$, and  perform a controlled rotation in the auxiliary space, in other words it ``reads'' the state of the link  and rotates accordingly the state on the auxiliary space $r_m$.
The operator $U_{s_n}$ is represented graphically in Fig. \ref{fig:u_ks}.

From the definition we can derive its covariance properties under left and right rotations.
Indeed  after the gauge transformation that sends $\ket{g}$ to $\ket{h^{-1} g k}$, a $U_{s_n}$ becomes 
\begin{align}
U'_{s_n} \equiv\left[ L(h^{-1}) R(k) \right]\  U\ \left[ L(h^{-1}) R(k)\right] ^{\dagger} = \nonumber \\
= \sum_g \proj{h^{-1} g k}_{s_n} \otimes \Gamma_{r_m}(g)^{i}_{j}, \label{eq:phys_rot}
\end{align}
as illustrated in the lower panel of Fig. \ref{fig:u_ks}.
By re-defining $g' = h^{-1} g k$ we obtain 
\begin{equation}
 U'_{s_n} = \sum_{g'} \proj{g'}_{s_n} \otimes \Gamma_{r_m}(h g' k^{-1})^{i}_{j}. \label{eq:u_ks_rotated}
\end{equation}
Eq. \eqref{eq:phys_rot} together with \eqref{eq:u_ks_rotated} show that the $U_{s_n}$ allows to transfer the rotation on the physical Hilbert space of the link   to corresponding rotations on the auxiliary space.

This relation is often considered as the defining relation of a LGT \cite{creutz_quarks_1983}, since $U$ can be thought of the equivalent of the position operator in the group manifold, while  $L, R$  are equivalent to translation  operators on the group manifold.

\begin{figure}
 \includegraphics[width=\columnwidth]{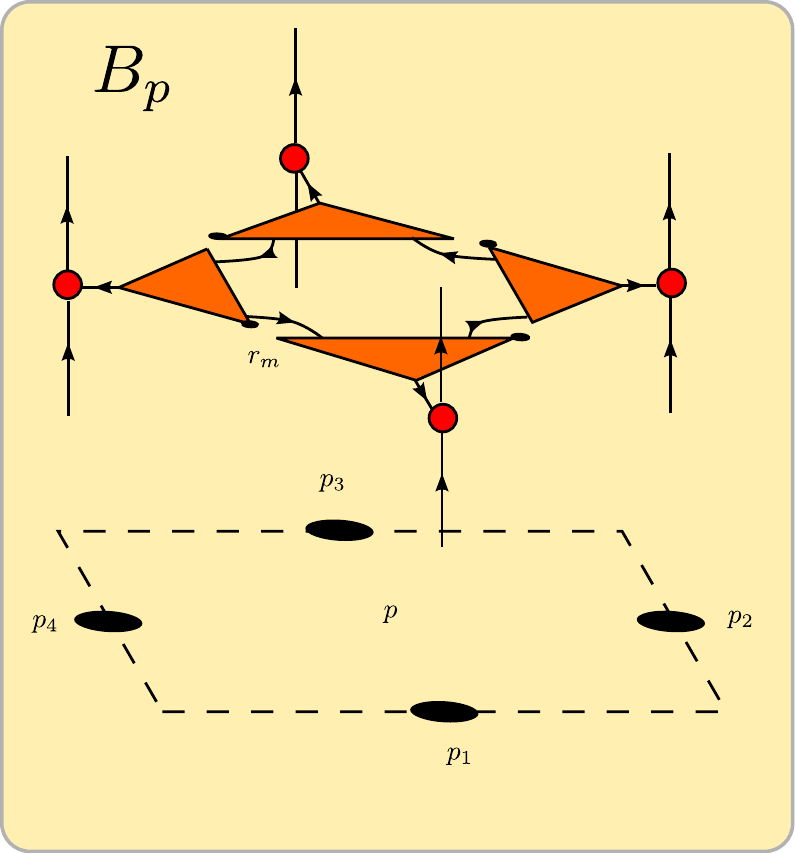}
 \caption{The plaquette operator $B_p$ of Eq. \eqref{eq:plaquette} constructed as a matrix product operator from four $U_{p_n}$'s, acting on the links around a plaquette in the KS LGT. As a consequence of the covariance of the $U_{p_n}$ under conjugation by $L$ and $R$ (see panel b) of Fig. \ref{fig:u_ks}). It transfers the rotation onto the physical legs to rotations onto the auxiliary legs. The operator commutes with the gauge transformations, and, thus, it be used as building block of a gauge invariant Hamiltonian (see Eq. \eqref{eq:h_ks}). \label{fig:plaquette_op_ks}}
\end{figure}

By appropriately building a close path out of $U$ we can get rid of the rotations on the auxiliary legs, and thus obtain a gauge invariant operator.
In particular we can now construct  the simplest closed path that leads to the plaquette operator we are after.
\begin{equation}
 B_p = \textrm{tr}_{r_m} \left( U_{p_1}  U_{p_2}  U^{\dagger}_{p_3}  U^{\dagger}_{p_4} \right), \label{eq:plaquette}
\end{equation}
where we denote by $p$ a plaquette of $\Lambda$,  and  by $p_1
\cdots p_4$ the links around it, ordered counter-clockwise. The dagger is related to the fact that the orientation of the plaquette is in some cases opposite to the natural orientation of the lattice,  and the trace is intended over the auxiliary indexes only.  The equivalence between Eq. \eqref{eq:phys_rot} and \eqref{eq:u_ks_rotated}  guarantees that the plaquette operator, represented in Fig.  \ref{fig:plaquette_op_ks}, commutes with the gauge transformations that have non-trivial overlap with it. 

Having both gauge invariant link-operators and plaquette operators, we can now write the generalization of the $E^2 + B^2$ Hamiltonian of QED for a LGT with gauge group $G$. This reads
\begin{equation}
H_{LGT} = \sum_{l}  {\cal E}^2_{l} + \frac{1}{\alpha^2}\sum_p \left( B_p +  B_p^{\dagger}\right) \label{eq:h_ks}.
\end{equation}
where $\alpha$ is the coupling constant, and the first sum runs over links $l \in  \Lambda$, while the second over plaquettes $p \in \Lambda$.  

\section{Projector onto  ${\cal H}_p$ as a tensor network}
\label{sec:proj_tn}
The gauge invariant Hilbert space  ${\cal H}_p$ defined in Eq. \eqref{eq:gauge}, is made of those states that fulfill all constraints 
$A_s(g) \ket{\phi} = \ket{\phi}$. This can be obtained through a projector ${\cal P}$ 
\begin{equation}
 {\cal P} : \mathbb{C}(G) ^L \to {\cal H}_p,\ {\cal P}^2 = {\cal P}. \label{eq:cal_p}
\end{equation}

\begin{figure}
 \includegraphics[width=\columnwidth]{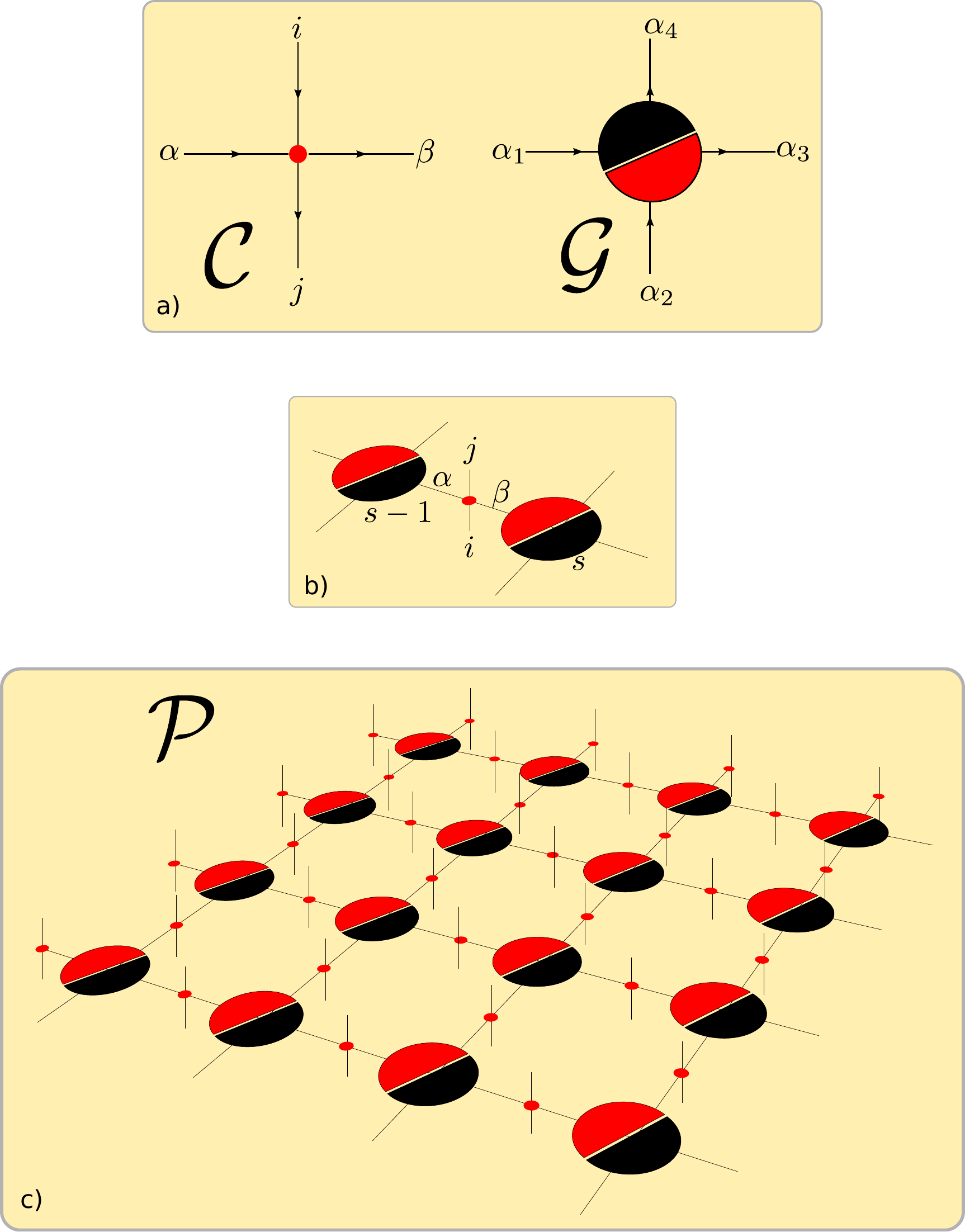}
 \caption{The projector on the gauge invariant states. a) The corresponding TN  is defined through the contraction of several copies of two elementary tensors, ${\cal C}$, which copies the physical Hilbert space onto the auxiliary Hilbert space,  and ${\cal G}$, which selects only configurations fulfilling the gauge invariance condition. b) An example of gauge-constraint decoupling at site $s-1$ and $s$ obtained through the insertion of the copy tensor ${\cal C}$. c) The projector on the gauge invariant states  for  a $4 \times 4 $ square lattice with PBC.\label{fig:proj_gauge}}
\end{figure}

 Here we show that ${\cal P}$ has an exact TN representation. The idea is very general and requires the contraction of  several copies of  two types of elementary  tensors, ${\cal C}$ tensors and ${\cal G}$ tensors. The two have different roles as shown in Fig. \ref{fig:proj_gauge}. The  ${\cal C}$ tensors copy  the states from the \emph{physical} legs to the  \emph{auxiliary} legs so that gauge constraints are decoupled and are imposed  individually by  ${\cal G}$ tensors. There is a    ${\cal C}^{\alpha, j}_{i, \beta}$,  for every link. They have all elements zero but  those corresponding to  $\alpha =  i =   \beta = j$ \cite{swingle_topological_2010,biamonte_categorical_2011,denny_algebraically_2012}. For every site, there is a  ${\cal G}^{\alpha_1 \alpha_2}_{ \alpha_3 \alpha_4}$ that only possesses auxiliary indexes (all of its indexes are contracted) \footnote{In general, ${\cal G}$ has as many legs as the coordination number of the lattice, i.e. six for a triangular lattice, three for an hexagonal lattice...}.   It selects, among all states of the tensor product of the four auxiliary constituents around a site, only those  that fulfill the gauge-symmetry requirements in Eq. \eqref{eq:gauge}.

Concretely, consider an horizontal  link ($s_1$ using the notation of Fig. \ref{fig:a_ks}) connecting  sites $s-1$ and $s$. Its state is copied, through the corresponding ${\cal C}$, to two auxiliary states, one located close to $s-1$,  $\alpha$ and the other close to $s$, $\beta$ (see panel b) of Fig. \ref{fig:proj_gauge}).  
In this way we can treat the   gauge constraint defined at the site $s-1$ as acting on $\alpha$ rather than on the link $s_1$, analogously the gauge constraint at the site $s$ can be imposed on $\beta$ rather than on $s_1$.
In this way we have been able to  completely decouple the gauge-symmetry constraints acting on site $s-1$ and $s$. Before the copy tensors they were acting on the same link $s_1$, while after it they act on two different auxiliary  sites $\alpha$ and $\beta$. This allows to address each gauge constraint individually.

By contracting as many copies of ${\cal C}$ as there are links on the lattice with as many copies of  ${\cal G}$ tensors as there are sites on the lattice, following the pattern in Fig. \ref{fig:proj_gauge}, we obtain the desired projector ${\cal P}$.

Let us write explicitly the tensors ${\cal C}$ and ${\cal G}$.
The only non zero elements of the copy tensor are $C^{g,g}_{g,g} = 1, \forall g$. Regarding $\cal G$ ,
 there are several ways to obtain them (all providing equivalent tensors).  Here we  follow the one inspired by the known TN expressions for the ground states of quantum doubles \cite{aguado_entanglement_2008,aguado_entanglement_2011,tagliacozzo_entanglement_2011,osborne}. The idea is to disentangle the gauge-symmetry constraint. It originally acts on four auxiliary sites, and we want to design an appropriate unitary transformation that  transforms it to a single auxiliary site operator. In practice, 
  we obtain this by constructing ${\cal G}$ itself as the result of an elementary TN contraction,  whose building blocks are  unitary tensors that act on two constituents, and perform controlled rotations. In particular we define the tensors, $C_R(\alpha, \beta)$ and $C_L(\alpha, \beta)$, 
\begin{eqnarray}
 C_R (\alpha, \beta)=\sum_g \proj{g}_{\alpha} \otimes R(g)_{\beta}; \nonumber \\ 
 C_L(\alpha, \beta) =\sum_g \proj{g}_{\alpha} \otimes L(g)_{\beta}.  \label{eq:cnots}
\end{eqnarray}
where $\alpha$ and $\beta$ specify the location of the constituents in the virtual lattice. The generic properties of these tensors are  independent on the position of the constituents.
They act on the tensor product $\mathbb{C}(G)\otimes\mathbb{C}(G)$ and transform the state $\ket{g,h}$ as $C_R : \ket{g,h} \to \ket{g, hg}$ and $C_L : \ket{g,h} \to \ket{g, gh}$. The two operators have the following properties:
\begin{eqnarray}
 C_L (R(h)\otimes L(h^{-1})) C_L ^{\dagger}= (R(h)\otimes Id), \nonumber \\
 C_R^{\dagger} (R(h)\otimes R(h)) C_R = (R(h)\otimes Id). \label{eq:dis}
\end{eqnarray}
These properties can be used in order to simplify the gauge-symmetry building blocks. Concretely, we now specify one possible arrangement of $C$ operators that allows to simplify the gauge condition. If we define the following unitary operator $I_s =  C_R^{\dagger}({\alpha}_1, \alpha_4) C_L(\alpha_4, \alpha_3) C_L(\alpha_1,\alpha_2)$ by using the properties \eqref{eq:dis} we obtain
\begin{equation}
I_s^{\dagger} A_s(h) I_s = R(h)_{\alpha_1} \equiv A'_s(h),
\end{equation}
where $A_s(h)$ are the building blocks of local gauge transformation defined in Eq. \eqref{eq:gauge_trans}.
The physical interpretation of this transformation is that we have concentrated a four-body constraint  onto a single-body constraint acting on $\alpha_1$ that it is now easy to fulfill. On $\alpha_1$, gauge invariance \eqref{eq:gauge} requires indeed to pick  the only state that is invariant under the rotation for an arbitrary group element. This  is the state $\ket{+}=1/\sqrt{|G|}\sum_g \ket{g}$. 

\begin{figure}
 \includegraphics[width=\columnwidth]{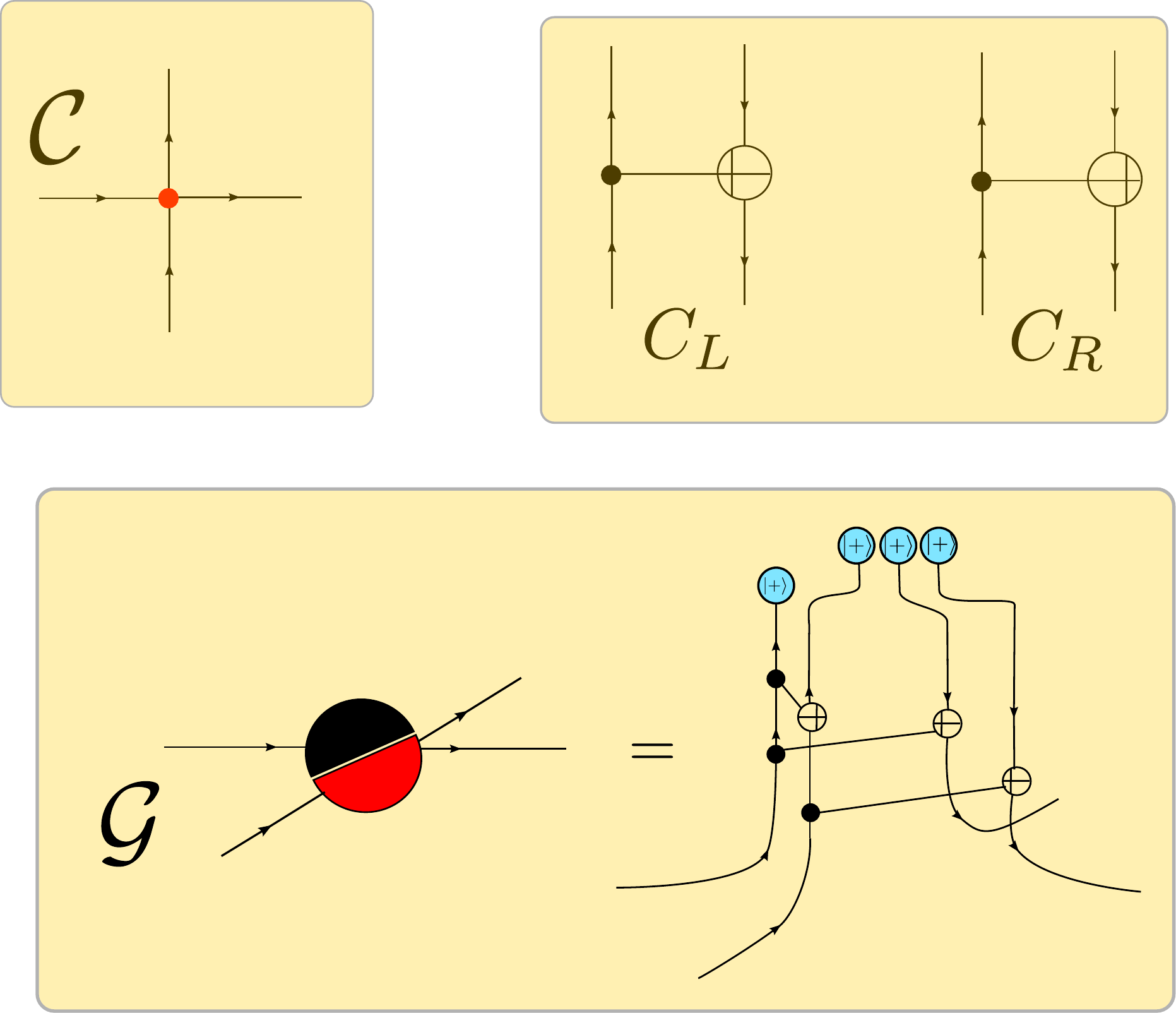}
 \caption{The definition of the tensors ${\cal C}$ and ${\cal G}$ for the Kogut-Susskind LGT. The tensor ${\cal G}$ is obtained by composition of several copies of unitaries $C_L$ and $C_R$ described in the main text. They are followed by the projection onto the states  $\ket{+}$ and $\tilde{\ket{+}}$ defined in the main text.\label{fig:proj_k_s} }
\end{figure}

Every state of the other three auxiliary sites, forming   $\mathbb{C}(G)^ 3$ is by construction gauge invariant, since after the unitary transformation the gauge constraint does not act on those sites.
The projector operator is obtained by the equal superposition of all the gauge invariant states. This can be done by projecting each of the three $\mathbb{C}(G)$ onto $\tilde{\ket{+}}=\sum_g \ket{g}$.
This construction is sketched in Fig. \ref{fig:proj_k_s}.

Before proceeding, let us summarize what we have obtained so far. By re-expressing the KS LGT in the language of TNs we have been able to provide a TN prescription for the projector onto the gauge invariant Hilbert space ${\cal H}_p$. This TN has bond dimension $D$ equal to the number of elements in the group,  $D=|G|$. This means that for discrete groups this TN can be used in actual computations since it has finite bond dimension. Furthermore, the construction can be improved, as discussed in  section \ref{sec:proj_tn_tlgt}, where we explain  an alternative choice of  ${\cal C}$ and ${\cal G}$ that allows  to express ${\cal P}$ with a TN with bond dimension $D=\sum_r n_{r}$ where $n_{r}$ is the dimension of the irrep $r$. This is of the order of  the square root of $|G|$ since $|G|=\sum_r n_{r}^2$.  

In any case, when we study LGT with continuous groups, the bond dimension of the TN  is infinite and thus is not useful for numerical simulations. It provides, nevertheless, an  interesting analytical results since it encodes, in a TN, the exact projector onto the gauge invariant Hilbert space. 

We now generalize the KS LGT to models that are described by finite dimensional local Hilbert spaces  while  invariant under continuous groups. This will allow to generalize our TN construction and obtain a TN expression for ${\cal P}$ with finite bond dimension that can be used in actual numerical calculations.

\section{Truncated LGT}
\label{sec:t_ks}

The constructive approach that we have followed so far  allows to generalize the original KS LGT. Here we are departing from the Hamiltonian LGT  whose Lagrangian formulation  provides the standard Yang-Mills action in the continuum limit. 
However, we would still like to construct models that are related to the original KS LGT through a truncation of the local Hilbert space that commutes with local gauge transformations. In this way, what we have discussed so far applies as a whole to the truncated models \cite{horn_finite_1981}.

Within our formalism the truncation of the Hilbert space we are looking for is very natural. We use  $W_G$ of Eq. \eqref{eq:wg} to pass from $\mathbb{C}(G)$ to ${\oplus_r}W_r$ and to truncate the direct sum to an arbitrary finite set of irrep's. The minimal choice requires keeping at least two  irrep's. $r\oplus r'$ (the reason why we need at least two irrep's will become clearer in the following), so that the projector can be written as 
\begin{equation}
W_T \equiv  W_r \oplus  W_{r'}, \label{eq:wt}
\end{equation}
where the $W_r$ are defined in Eq. \eqref{eq:wr}. 

\begin{figure}
 \includegraphics[width=\columnwidth]{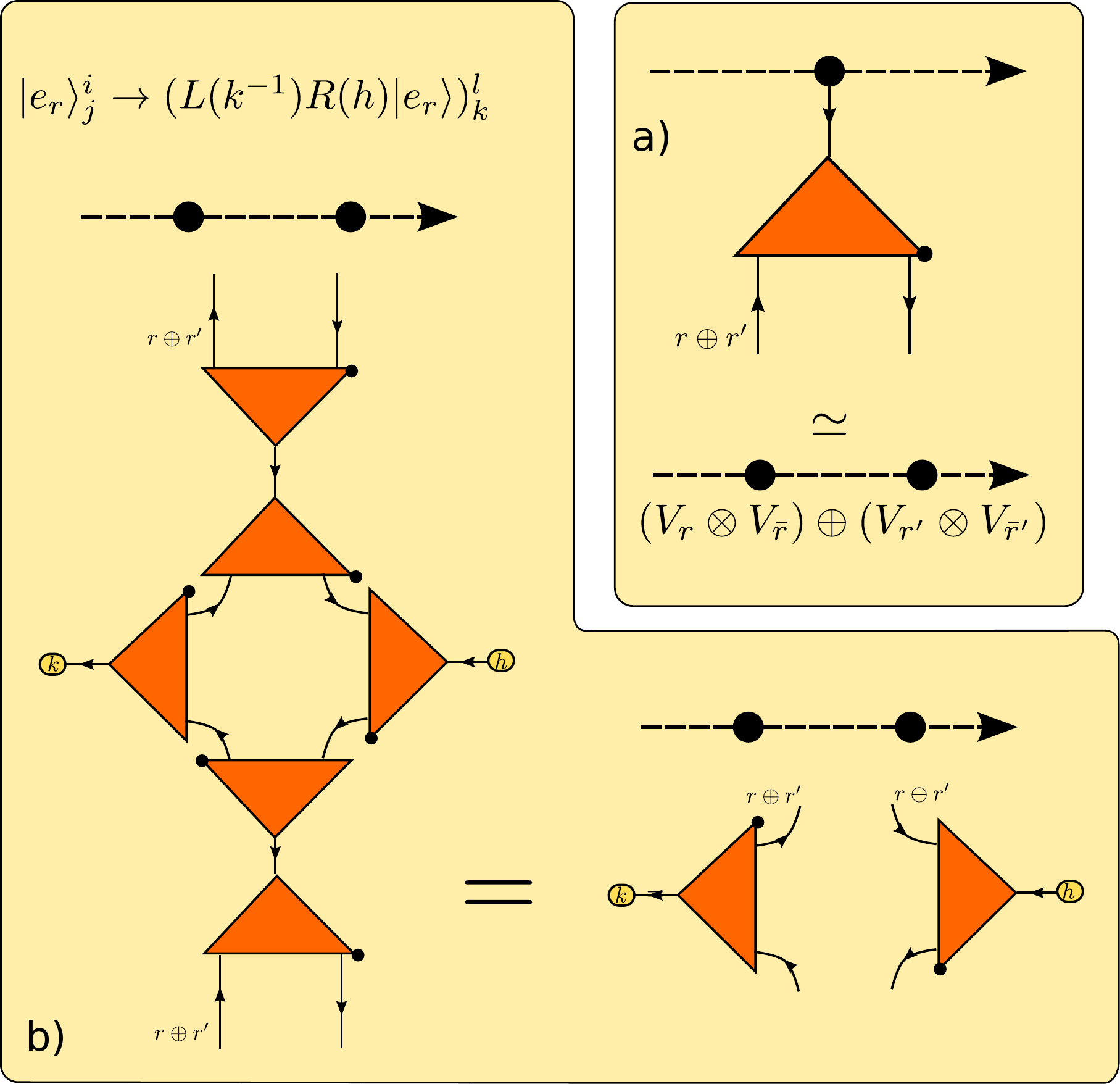}
 \caption{a) The Hilbert space of a gauge boson in the \emph{truncated} KS LGT is obtained by truncating $\mathbb{C}(G)$ with the isometry $W_T$ defined in  Eq. \eqref{eq:wt}. It  is isomorphic to $(V_r \otimes \bar{V}_r)\oplus (V_{r'} \otimes \bar{V}_{r'})$.
 Each term of the direct sum is a tensor product of two constituents so that we represent it by two solid circles on each link.  b) The operators implementing the left and right rotations defined in Eq. \eqref{eq:left_tr_k_s} and \eqref{eq:rigth_tr_k_s} are obtained from those of  Fig. \ref{fig:hilb_space} after conjugation with  $W_T$.  \label{fig:hilb_space_t_lgt}}
\end{figure}

After the projection, if the group $G$ is compact,   the Hilbert space on a link  becomes finite dimensional.  It still preserves the property of the group algebra that we have used extensively; namely in each block it is the tensor product of $V_r\otimes V_{\bar{r}}$. This immediately allows to write  left and right rotations $L$ and $R$ as
\begin{align}
  L(h^{-1}) =  \left( \Gamma_r(h^{-1}) \otimes 1_r \right) \oplus \left( \Gamma_{r'}(h^{-1}) \otimes 1_{r'} \right)\label{eq:left_tr_k_s}\\
  R(k) = \left( 1_r \otimes \Gamma_r(k)^{l}_{k} \right)\oplus \left( 1_{r'} \otimes \Gamma_{r'}(k)^{l}_{k} \right) \label{eq:rigth_tr_k_s}
\end{align}
as represented graphically in  panel b) of  Fig. \ref{fig:hilb_space_t_lgt}.

As already mentioned, there is a lot of freedom on the choice of $r$ and $r'$.
A legitimate choice  is the one that minimizes the dimension of the local Hilbert space, since this dimension takes part in the computational cost of TN algorithms.  In this case one should choose $r'$ as the trivial representation and  $r$ as the smallest faithful irrep of the group. 
For example, for $SU(N)$ groups, the dimension of the local Hilbert space with such choice is $N^2+1$, where $N$ is indeed the dimension of the fundamental representation.

\begin{figure}
  \includegraphics[width=\columnwidth]{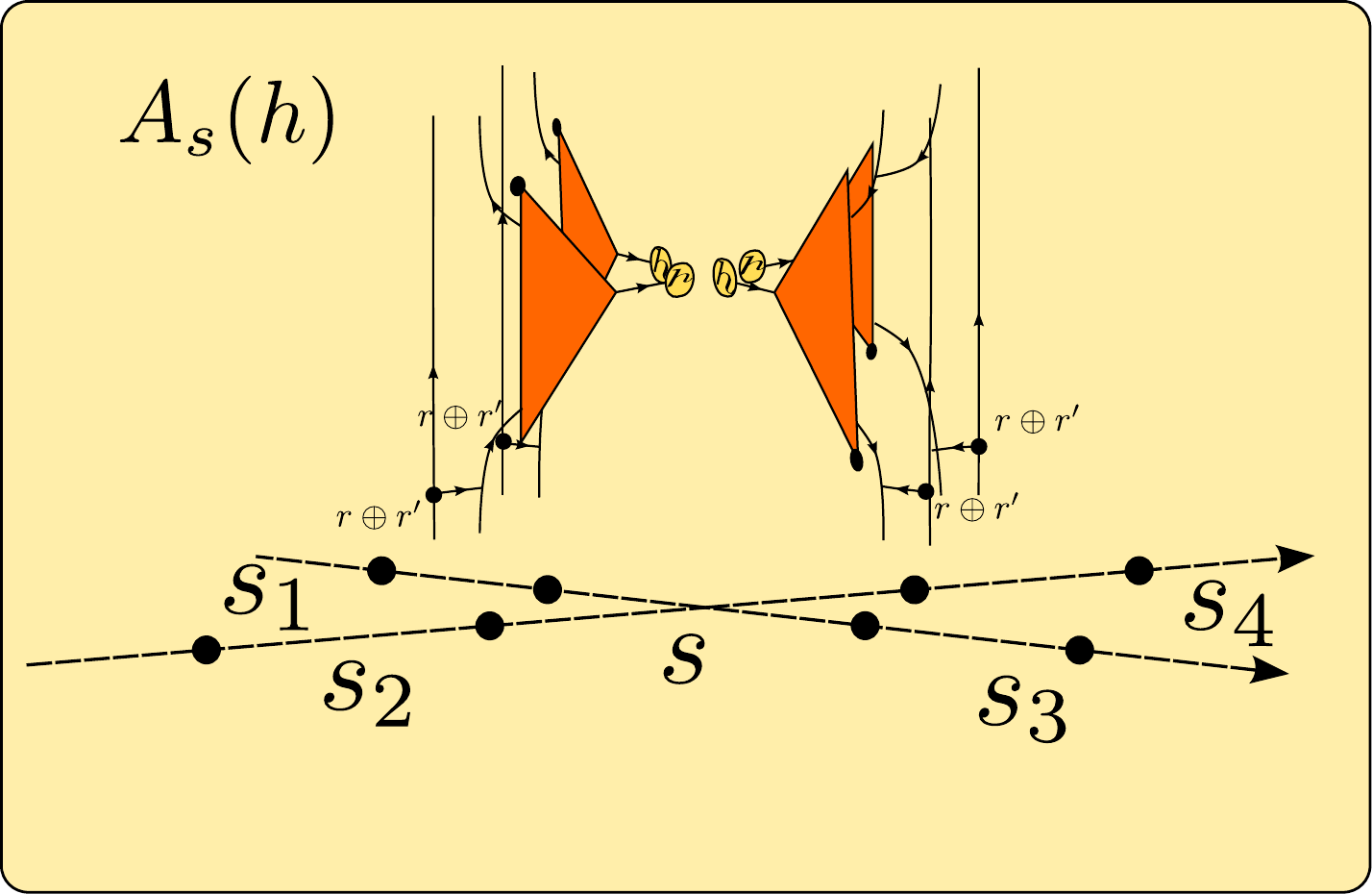}
 \caption{The operator $A_s(h)$ that generates the gauge transformations in the truncated  KS LGT.
 It is the result of the truncation 
 of the operator $A_s(h)$ of Fig. \ref{fig:a_ks} with $W_T$ defined in Eq. \eqref{eq:wt}. 
On each link, the operator acts on the constituent closer to $s$ through a rotation, $\Gamma_r(h)$ or $\Gamma_{r'}(h)$, which is controlled by the irrep of the neighboring constituent on the same link.
 It thus performs controlled operations in between the two constituents of a link, as suggested  by a dot on the controller leg and by an arrow that joins it with the controlled leg. \label{fig:a_t_ks}}
\end{figure}

The building blocks of gauge transformations $A_s(h)$  are still defined by Eq. \eqref{eq:gauge_trans} with $L$ and $R$ given by Eq. \eqref{eq:left_tr_k_s} and \eqref{eq:rigth_tr_k_s} as represented graphically in Fig. \ref{fig:a_t_ks}. 

They allow to define the gauge invariant Hilbert space ${\cal H}_p$ by using Eq. \eqref{eq:gauge}.

Gauge invariant operators have to commute with all the $A_s(g)$ defined above.  Both single-link operators (electric-like) and plaquette operators (magnetic-like) are just the truncation with $W_T$ of the corresponding operators in the KS LGT. The single-link operator is the truncation of Eq. \eqref{eq:one_site_op},
\begin{align}
 {{\cal E}_T^2}_{s_n}=  \left[ W_T ^{\dagger}{\oplus_r}\left[ c^r \left ( \mathbb{I}_{r} \otimes \mathbb{I}_{\bar{r}}\right) \right]W_T\right]_{s_n} \nonumber \\
 = \left[ c^r \left ( \mathbb{I}_{r} \otimes \mathbb{I}_{\bar{r}}\right)\oplus c^{r'} \left ( \mathbb{I}_{r'} \otimes \mathbb{I}_{\bar{r'}}\right) \right]_{s_n} \label{eq:one_site_op_t}
\end{align}
and depends only on the two free parameters $c_r$ and $c_{r'}$.
Here is where it becomes clear that we need to keep in $W_T$ at least two irrep's.
 The truncation of  Eq.  \eqref{eq:one_site_op} to a single irrep is indeed proportional to the identity.
 
\begin{figure}
 \includegraphics[width=\columnwidth]{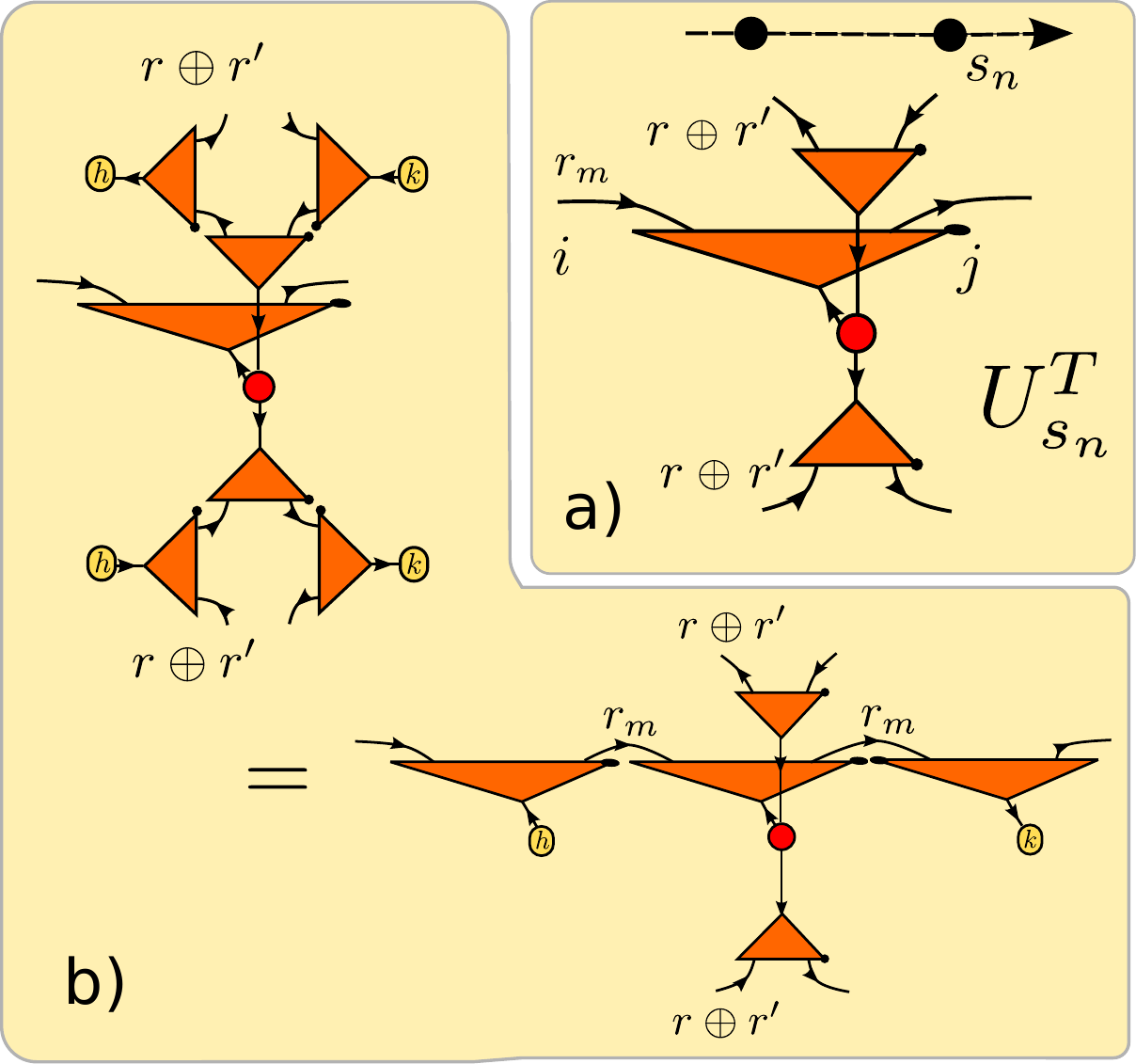}
 \caption{a) The ${U_T}_{s_n}$ operator in the truncated KS LGT as defined in Eq. \ref{eq:u_t}.  b) It has by construction  the desired covariance properties, meaning that rotations on the physical legs (upper part of the panel)  are transmitted to rotations on the auxiliary legs (lower part of the panel), as described by Eq. \eqref{eq:phys_rot} and \eqref{eq:u_ks_rotated}. \label{fig:u_t_ks}}
\end{figure}

 The truncated plaquette operator is built from $U_{s_n}$ in Eq. \eqref{eq:u_ks},
 \begin{equation}
  {U_T}_{s_n} = W_T^{\dagger} U_{s_n} W_T, \label{eq:u_t}
 \end{equation}
and is represented graphically  in Fig. \ref{fig:u_t_ks}.

Depending on the choice of $r$ and $r'$ inside $W_T$, ${U_T}_{s_n}$ could  vanish. In order to see this we have to remind that   in the  KS LGT  $V_{r_m}$,  is  chosen as the smallest faithful irrep (for $SU(N)$ this is the fundamental irrep of dimension $N$). 
 From the above definition of $U_T$ in Eq. \eqref{eq:u_t}, and the definition of $U_{s_n}$ in Eq. \eqref{eq:u_ks}, 
we see that $U_T$ entails terms of the kind $\sum_g \Gamma_r (g) \Gamma_{r'}(g^{-1}) \Gamma_{r_m}(g)$. These terms are proportional to the projector onto the trivial representation defined in Eq. \eqref{eq:proj_triv}, where now the $R$ in Eq. \eqref{eq:proj_triv} is given by $R= V_r \otimes V_{r'}\otimes V_{r_m}$. This implies that the above terms will  be non-vanishing only if  the decomposition of $R$ in direct sum of irrep's contains the trivial representation.

This consideration implies that in order to have a non-trivial model, a certain care should be taken  when choosing $r,r'$ and that their choice depends on the choice of  $r_m$.
In particular, the minimal prescription provides a valid truncation scheme with non-trivial dynamics since in that case $r_m=r$ and $r'$ is by itself the trivial representation.

Once the $U_T$ is non trivial, it automatically fulfills the  desired commutation relations with the $L$ and $R$ operators defined in Eq. \eqref{eq:phys_rot} and \eqref{eq:u_ks_rotated}, as illustrated once more in the panel b) of Fig. \ref{fig:u_t_ks}. 
In this way it can be used to construct the desired plaquette operators, using the same formula of Eq. \eqref{eq:plaquette} where the $U$ are substituted with the $U_T$ just defined. 

The Hamiltonian of the truncated LGT has thus the same form as the one of the KS LGT defined in Eq. \eqref{eq:h_ks}, with the appropriate substitution of  ${{\cal E}^2}$ by  ${{\cal E}_T^2}$ and $U$ inside the plaquettes by the corresponding  $U_T$ operators.

\subsection{${\cal P}$ in the truncated LGT}
\label{sec:proj_tn_tlgt}
In the truncated LGT we can also consider the projector on to the Hilbert space of gauge invariant states, 
\begin{equation}
 {\cal P}_T : \left[ \left(V_r \otimes V_{\bar{r}} \right) \oplus \left( V_{r'} \otimes V_{\bar{r'}} \right) \right] ^{\otimes L} \to  {\cal H}_p,
\end{equation}
with ${\cal H}_p$ defined as in Eq. \eqref{eq:gauge} with the $A_s(g)$ of \eqref{eq:gauge_trans} containing the operators $L$ and $R$ of Eq. \eqref{eq:left_tr_k_s} and \eqref{eq:rigth_tr_k_s}.
Even in this case this projector can be written as an exact TN.

The construction is very  similar to the one used for the KS LGT in Sect. \ref{sec:proj_tn}. There, however, the specific form of ${\cal G}$ has been derived by exploiting that, in $\mathbb{C}(G)$, we are able to disentangle the symmetry requirements. In general we are unable to  do this explicitly, so that here  we introduce a generic approach (which can also be used  for the full KS LGT and provides, indeed  a TN with lower bond dimension).
The TN structure of ${\cal P}_T$ is the same as the one of ${\cal P}$ in the KS.
It consists of the contraction of various copies of ${\cal C}$ tensors (one per link) and ${\cal G}$ tensors (one per site) following the pattern in Fig. \ref{fig:proj_gauge}.

In particular, the four-leg tensor ${\cal C}$ copies the physical states to the auxiliary states, while the gauge  fixing tensor ${\cal G}$ selects among the auxiliary states only those that fulfill the gauge invariance condition, $A_s(g) \ket{\varphi} =\ket{\varphi}$. 

Chosen a basis $\ket{i_r,j_r }, \ket{k_{r'} l_{r'}} $, with $\set{i_r, j_r}= 1,\dots,d_r$ and $\set{k_{r'}, l_{r'}} = d_r+1,\dots,d_r + d_{r'} $ of the Hilbert space of a link $\left[ \left(V_r \otimes V_{\bar{r}} \right) \oplus \left( V_{r'} \otimes V_{\bar{r'}} \right) \right]$, the ${\cal C}$ tensor copies the left constituent to the left and the right constituent to the right:
\begin{align}
 {\cal C}= \ket{i_r j_r}_{s_n} \bra{i_r j_r}_{s_n} \otimes \ket{i_r}_{\alpha} \bra{j_r}_{\beta} \nonumber \\
 +\ket{k_{r'} l_{r'}}_{s_n} \bra{k_{r'} l_{r'}}_{s_n} \otimes \ket{k_{r'}}_{\alpha} \bra{l_{r'}}_{\beta}.
\end{align}

\begin{figure}
 \includegraphics[width=\columnwidth]{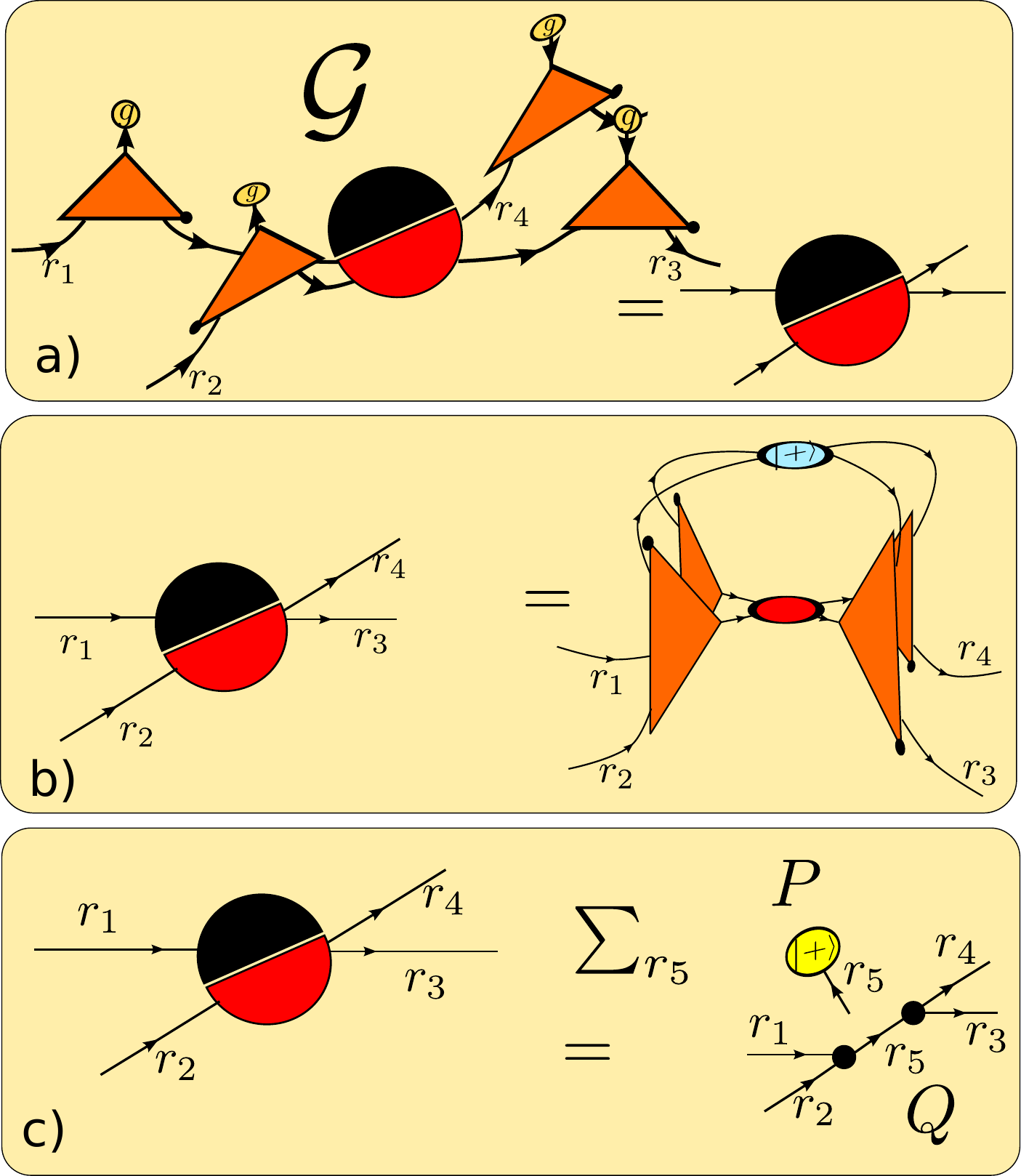}
    \caption{a) The tensor ${\cal G}$ that is used to build the projector onto the physical Hilbert space is an invariant  tensor under the action of $G$ as defined in Eq. \eqref{eq:inv_tens}. The legs of ${\cal G}$ carry an index of the irrep, since the Hilbert space on which they act has a direct sum structure labeled by the irrep $r$.
    b)  ${\cal G}$ can be obtained by using the explicit form of the projector onto the trivial irrep \eqref{eq:proj_triv}. The red circle is the copy tensor in the group algebra. After acting with the projector, we have to take an equal superposition of all vectors in the trivial-irrep space, so to obtain ${\cal P}_T$. This implies finding the image of the projector, taking an equal superposition of vectors spanning such image, state that we represent by the cyan $\ket{+}$, and  contracting on it   all open legs.
    c) Alternatively one can use the construction introduced in Ref. \cite{singh_tensor_2012} for building symmetric tensors. The tensor is then divided into a piece $Q$ that takes care of correctly matching the various irrep's (made by the Clebsch-Gordan coefficients), and a degeneracy tensor $P$, made of free parameters, that assigns a four-leg tensor to each of the irrep's which enter in composition of representations. In the example we draw,  $P$  depends on the irrep's $r_5$ that are contained into the tensor product of  $r_1$ and $r_2$. Once more, the projector ${\cal P}_T$ is obtained by taking a uniform superposition of the possible values of $r_5$, that is setting $P= \ket{+}$.\label{fig:proj_k_s_t}}
 \end{figure}

${\cal G}$ acts then   non trivially only on  $V_G \simeq \left(V_r \oplus V_{r'}\right)^{\otimes 2}\otimes \left(V_{\bar{r}} \oplus V_{\bar{r}'}\right)^{\otimes 2} $. On this space, the requirement of gauge invariance is equivalent to asking that  ${\cal G}$ is a symmetric tensor with respect to rotations of elements in the group $G$, that is it fulfills the requirements in Eq. $\eqref{eq:inv_tens}$ as explicitly shown in panel a) of Fig. \ref{fig:proj_k_s_t}.

Furthermore, since we are interested in building the projector ${\cal P}_T$, we need to build ${\cal G}$ out of the equal superposition of all symmetric tensors acting on the above space. 

As explained in Sec. \ref{sec:symm}, there are two possible ways of building a symmetric tensor. The first one consists of applying the projector onto the trivial irrep defined in \eqref{eq:proj_triv} to the space $V_G$. The projector acts separately in each block of irrep's and involves terms of the type $\frac{1}{|G|}\sum_g \Gamma_{r_1}(g) \otimes \Gamma_{r_2}(g) \otimes \Gamma_{r_3}(g^{-1}) \Gamma_{r_4}(g^{-1})$ with $\set{r_i} = r, r'$. It is important to notice that not all of the blocks contain a trivial irrep, and the projector will then give zero when acting on those blocks without it.  After the action of the projector, we take the equal superposition of all symmetric tensors with equal weight. It is important at this point to find the rank of the previous projector, and take an equal superposition of all possible normalized states on its support. This is obtained by using the tensor $\ket{+}=\ket{\underbrace{1\dots1}_{d_0}}$,  where $d_0$ is the number of copies  of the trivial irrep (the rank of the projector) as is represented in panel b) of Fig. \ref{fig:proj_k_s_t}. 

Alternatively, one can use the technology developed in Ref. \cite{singh_tensor_2012} for constructing invariant tensors. The idea is to decompose the tensor ${\cal G}$ into a $Q$ part (the part that is fully dictated by the symmetry constraints) and a $P$ part, the part that contains the variational parameters. The $Q$ part, in our case, is built out from the Clebsch-Gordan coefficients that  decompose the tensor product of the two irrep's $r_1, r_2$ into the direct sum of irrep's $r_5$. Then every $r_5$ is again decomposed into the tensor product of two irrep's $r_3, r_4$. In general,  this implies that the tensor $P$ has an extra index corresponding to $r_5$. The projector is obtained by taking the equal superposition of all the $Q$ tensors with different $r_5$, that is $P = \ket{+}$ . This is represented in panel c) of Fig. \ref{fig:proj_k_s_t}.

For concreteness we discuss how to construct ${\cal G}$ for the KS LGT with $G=SU(2)$ truncated to the sum of the trivial irrep, plus  the product of two $J=1/2$ irrep's.
${\cal C}$  copies the trivial irrep on both sides, while it copies one of the two $J=1/2$ to the left and the other to the right. 
${\cal G}$ is then used to project onto the invariant states. Each auxiliary site then lives on $V_{J=0}\oplus V_{J=1/2}$  that has dimension $d=3$. The possible blocks of the four fold tensor product $\left( V_{J=0}\oplus V_{J=1/2}\right) ^{\otimes 4}$ that contain the trivial irrep are those with an even number of $J=1/2$ factors. This implies that there is one block   $V_{J=0}^{\otimes 4}$ six blocks with $V_{J=0}^{\otimes 2}\otimes V_{J=1/2}^{\otimes 2}$ and one block $V_{J=1/2}^{\otimes 4}$. 
This last block furthermore leads to two values of $r_5$, $r_5=0$ and $r_5=1$ that need to be equally taken into account.
The equal superposition of all these blocks gives the ${\cal G}$ necessary to build the projector ${\cal P}_T$. The explicit form of ${\cal G}$ is given in the appendix \ref{app:tensors}.

We conclude this section by discussing the bond dimension of the TN ${\cal P}_T$. It clearly depends on the choice of irrep's that one decides to consider in $W_T$ of Eq. \eqref{eq:wt}. In the minimal case for a $SU(N)$ LGT, it is $D=N+1$.

\section{Gauge magnets and quantum link models.}
\label{sec:gm}
The truncated LGT we have just presented is not the only  LGT with continuous groups, defined on a finite dimensional Hilbert space. In particular a set of models have been proposed in the literature that have the same features and are known as  gauge magnets (GM) \cite{orland_lattice_1990,orland_exact_1992}. Here we briefly recall their construction using the tools that we have described in the previous sections.

In gauge magnets the local Hilbert space is the direct  sum $V_r\oplus V_{r}$. By rewriting the direct sum into a tensor product of $\mathbb{C}^2\otimes V_r$ we obtain a natural basis  spanned by $\set{\ket{0}, \ket{1}}\otimes \set{\ket{v}}$ in terms of a position qubit times a spin vector in $V_{r}$ (more details are given in Sect. \ref{sect:p_gm}).  In this case the constituents live in a space of dimension $d_{GM}=2n_r$, where $n_r$  the dimension of the irrep $r$. 
This is,  in general,  smaller than $d_{tKS}=n_r^ 2+1$ the Hilbert space of the KS LGT truncated to the same irrep $r$. 
For this reason GM, in their simplest version, can be considered  the ``minimal LGTs'', that is the LGT with the smallest local Hilbert space. 

\begin{figure}
 \includegraphics[width=\columnwidth]{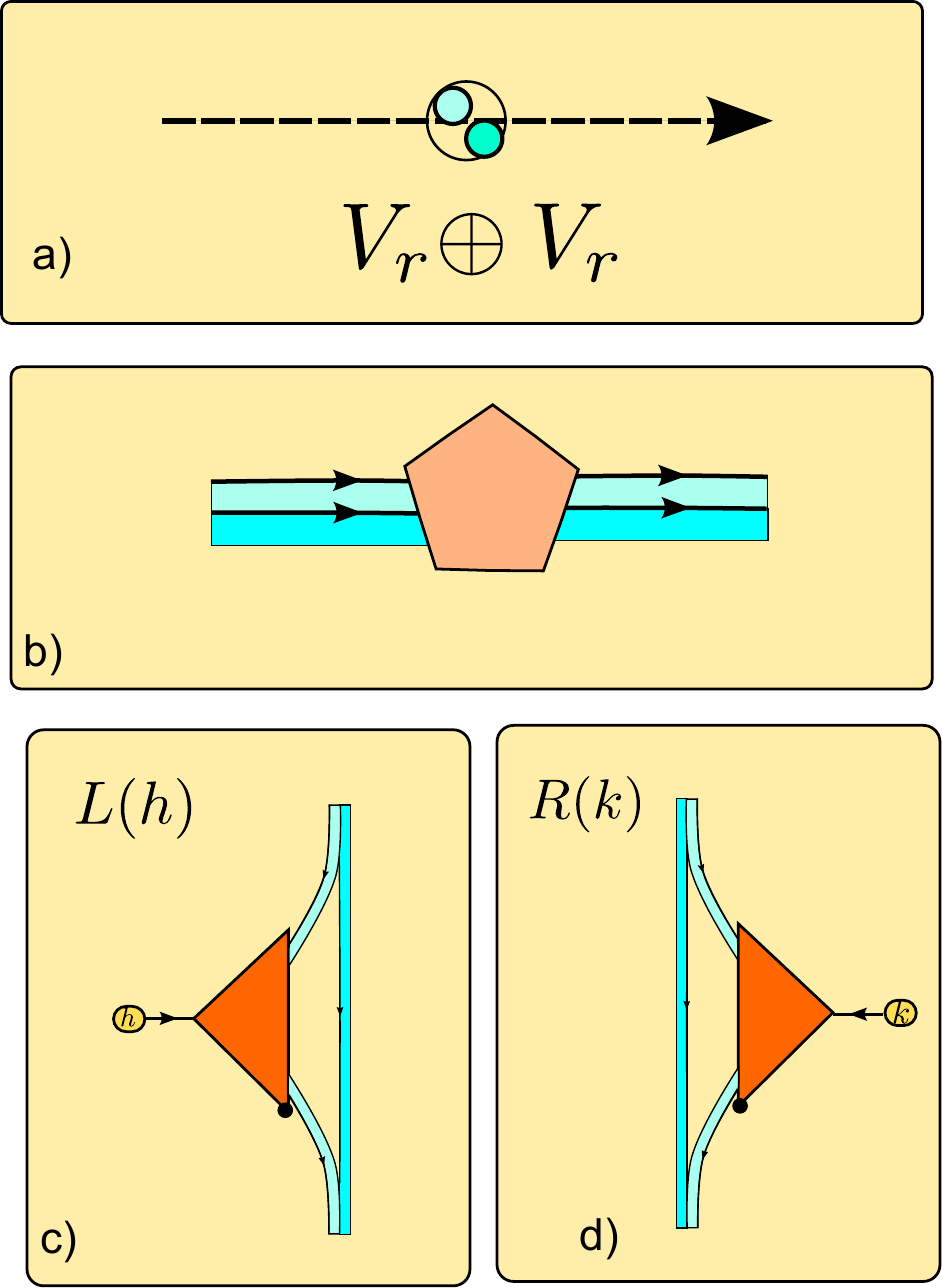}
 \caption{a) Gauge boson constituents in \emph{gauge-magnets} are states of the direct sum of two irrep's, $V_r \oplus V_r$, which can be identified as the left and the right constituents at the ends of each link. Graphically, we represent this local Hilbert space as a single circle embracing the two copies of $V_r$, each of them being a smaller circle identified by a different color and position inside the bigger circle. b) The legs of the tensors acting on this Hilbert space are represented by bands rather than lines so that we can specify operators that only act on one sector as acting on half of the band. Each sub-sector is colored differently. c)-d) The left and right rotations in the gauge magnets only act on half of the direct sum as depicted here and discussed in Eq. \eqref{eq:left_rigth_g_m}. \label{fig:hilb_space_gm}}
\end{figure}

However, the two terms in the direct sum still allow to define left and right rotations of the state of a link for an arbitrary element of the group that, as we have seen,  is the prerequisite for being able to  define gauge transformations. The left and right rotation for  elements $h,k$ in $G$ are defined through  
\begin{equation}
 L(h)R(k) \equiv (\Gamma_r(h) \oplus  1 )(1 \oplus \Gamma_r(k)) =  \Gamma_r(h)\oplus \Gamma_r(k).\label{eq:left_rigth_g_m}
 \end{equation}
 The above equation is represented graphically in panels c)-d) of  Fig. \ref{fig:hilb_space_gm} and  can be rewritten in the tensor product basis as $ L(h) =\proj{0} \otimes \Gamma_r(h) +\proj{1} \otimes 1$ and $R(k) = \proj{0} \otimes 1+ \proj{1} \otimes \Gamma_r(k)$. In this notation it is clear that $\ket{0}$ represents the left end of the link while $\ket{1}$ represents the right end of the link.
The above  equation  allows to identify the gauge boson with a  boson that can occupy one of the  two extremes of  the link (for a physical implementation of this ideas with cold-atoms please see \cite{banerjee_atomic_2012,tagliacozzo_simulation_2013}).   

\begin{figure}
  \includegraphics[width=\columnwidth]{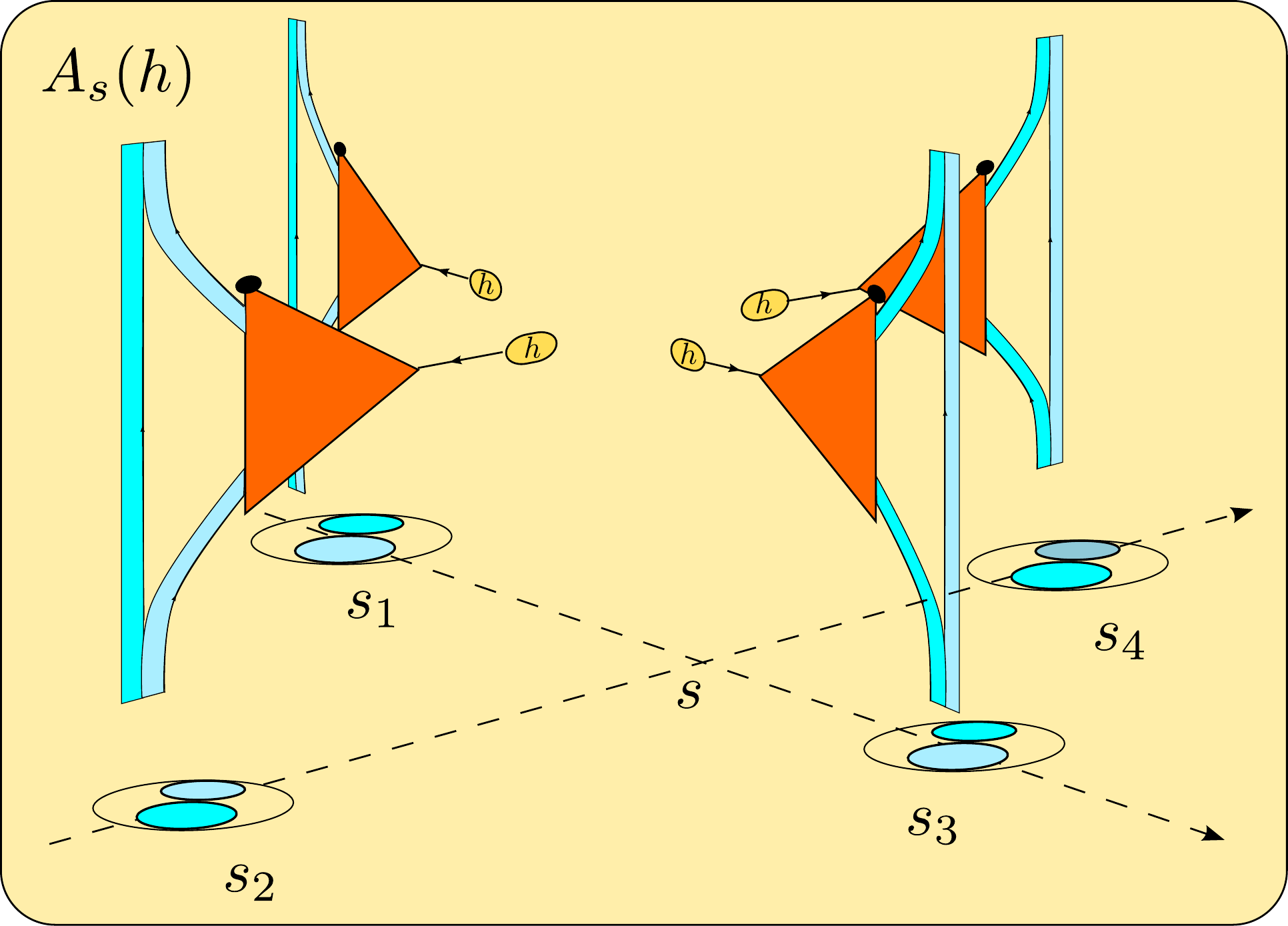}
 \caption{The operator that generates gauge transformations in the gauge-magnet LGT defined in Eq. \eqref{eq:left_rigth_g_m}.\label{fig:a_gm} It rotates both incoming and outgoing links by the same group elements. On every link it only acts on the constituent  that is closer to the site.}
\end{figure}

 The definition of local gauge transformations at a site slightly differs from the one in KS LGT of Eq. \eqref{eq:gauge_trans}, 
 \begin{equation}
  A_s(h) = R(h)_{s_1} \otimes R(h)_{s_2} \otimes L(h)_{s_3} \otimes   L(h)_{s_4}, \label{eq:gauge_trans_gm}
 \end{equation}
 since it rotates all the links by the element $h$, independently if they are entering or leaving the site $s$.  In terms of left and right constituents $A_s(h)$  only acts non trivially on those constituents of the links that are located close to $s$ as illustrated in Fig.  \ref{fig:a_gm}.

Once the building blocks of gauge transformations are defined, the discussion parallels the one for the others LGT, and in particular we can use the $A_s$'s in order to define the physical Hilbert space of gauge invariant states ${\cal H}_p$.
Furthermore we use the new $A_s(h)$ in order to define gauge invariant operators as from  Eq. \eqref{eq:g_inv_op}.
As a result all  gauge invariant link operators  are defined as  ${{\cal E}^2_G}_{s_n}=\left ( c_0\mathbb{I}_l\oplus c_1  \mathbb{I}_r\right)_{s_n}$ with $c_0$ and $c_1$ arbitrary numbers. 
The equivalent of the $U_{s_n}$ operator is used to build gauge invariant plaquette operators. In the literature it is possible to find the specific form for the $U_{s_n}$ operators for $SU(N), SP(N)$ and $G_2$ groups \cite{brower_qcd_1999,schlittgen_low-energy_2001,baer_quantum_2001}. Here we provide a recipe to generalize it to an arbitrary group $G$, either discrete or continuous. 

We use a similar construction than the one used for the KS LGT and define an operator that acts on $(V_r \oplus V_r) \otimes V_{r_m}$,  
 \begin{equation}
 U^G_{s_n}= \sum_g \left[ \left (\ket{0}\bra{1} \otimes \Gamma_r(g) +\ket{1}\bra{0}\Gamma_r(g^{-1}) \right) \otimes \Gamma_{r_m}(g) \right]. \label{eq:U_gm}
\end{equation}
 If we now study how  $U^G_{s_n}$ changes under a left rotation, we immediately see that 
 \begin{align}
 &L(h) U^G_{s_n} L(h^{-1}) = \nonumber \\
  &\sum_g \left[ \left (\ket{0}\bra{1} \otimes \Gamma_r(g) +\ket{1}\bra{0}\Gamma_r(g^{-1}) \right) \otimes \Gamma_{r_m}(h^{-1} g) \right], \label{eq:u_left_gm}
 \end{align}
 and that under a right rotation 
 \begin{align}
& R(k) U^G_{s_n} R(k^{-1}) = \nonumber \\
 & \sum_g \left[ \left (\ket{0}\bra{1} \otimes \Gamma_r(g)) +\ket{1}\bra{0}\Gamma_r(g^{-1}) \right) \otimes \Gamma_{r_m}( g k) \right].\label{eq:u_right_gm}
 \end{align}
\begin{figure}
 \includegraphics[width=\columnwidth]{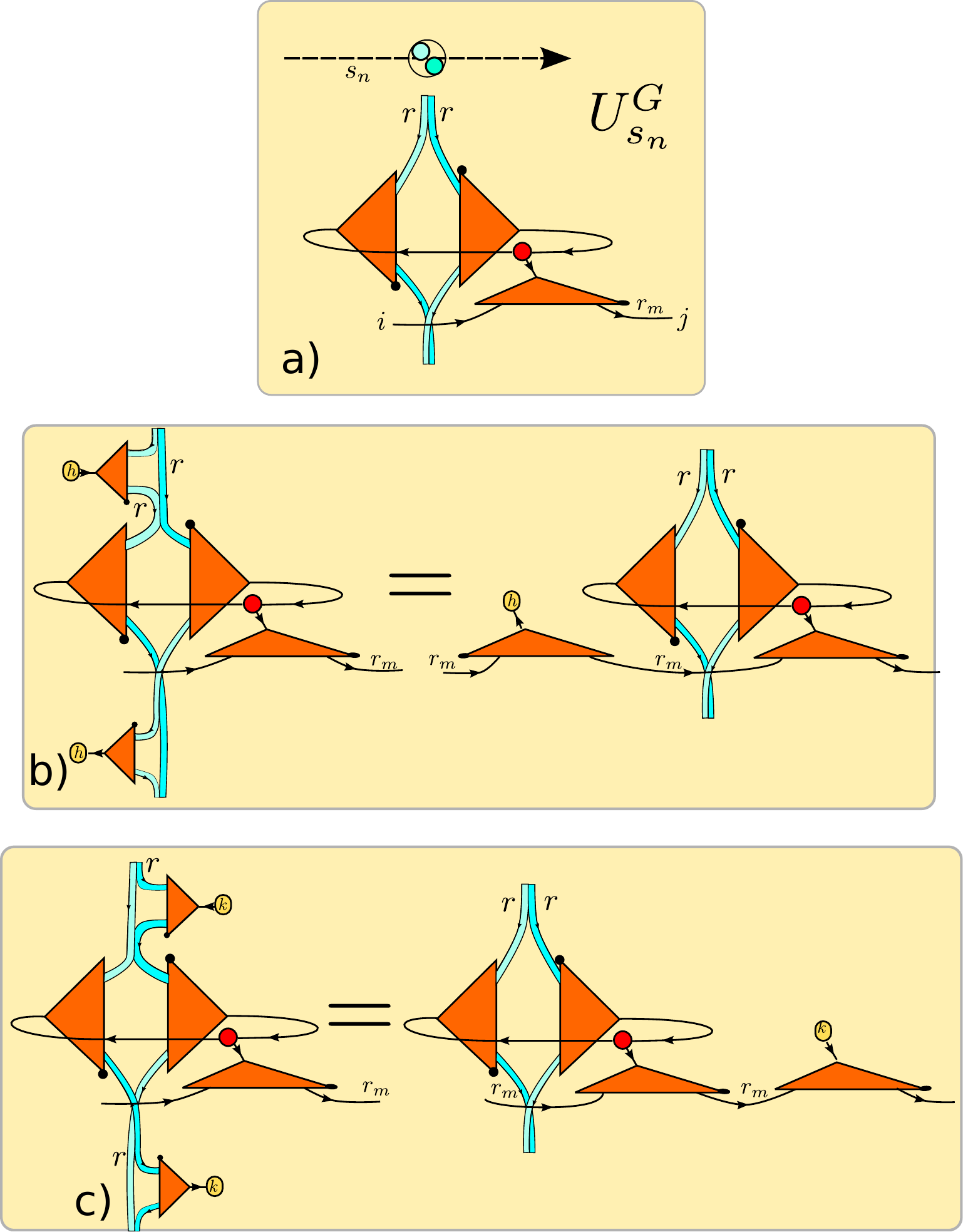}
 \caption{a) The $U^G_{s_n}$ operator in the gauge magnet LGT defined in Eq. \eqref{eq:U_gm}. It can be defined, once more, as the contraction of three $\Gamma_r(g)$'s, and a copy tensor in $\mathbb{C}(G)$ (red circle). b)-c) From the definition, one can check that it allows to pass the $L$ and $R$ rotations  on the physical legs, induced by the gauge transformation of Fig. \ref{fig:a_gm}, to analogous  rotations onto the auxiliary legs. This is exactly the covariance property that allows to define a gauge invariant plaquette operator, defined as in Eq. \eqref{eq:plaquette}, where the rotation on the auxiliary legs cancel by taking the trace onto the auxiliary legs. \label{fig:U_gm}}
\end{figure}
Once more, $U^G_{s_n}$ allows to transmit the rotations on the link to rotations onto the auxiliary space.
This allows to use it as the building block for a plaquette operator analogous to the one of the KS LGT in Eq. \eqref{eq:plaquette}.
 
 The Hamiltonian of a gauge magnet has thus the same structure as the Hamiltonian for the KS LGT defined in Eq. \eqref{eq:h_ks}, where this time  ${\cal E}^2$ is substituted by ${\cal E}^2_G$  and plaquette operators are built from the $U^G_{s_n}$ operators just described.
As a final comment, it is important to notice that in general for non-Abelian groups the gauge magnets and the truncated LGT are different models. Indeed, there is no consistent truncation at the level of a single link that allows to map the KS LGT onto the GM. This would require to decompose $\mathbb{C}(g)$ into a direct sum of irrep's and keep only two of them.  Unfortunately such decomposition cannot be performed in an invariant way. 
This is a consequence  of the  fact that $\mathbb{C}(G)$ can be decomposed in an invariant way only as a direct sum of tensor products of irrep's (as we have done through $W_G$ of \eqref{eq:wg} in the previous sections). Any further decomposition of each of the terms $V_r\otimes V_{\bar{r}}$ into a direct sum of irrep's requires a choice of basis in one of the two factors, and thus cannot be invariant under rotations by elements of $G$ \cite{serre_linear_1977}.

There is an exception to  this rule in the case of Abelian LGT where  $n_r=1$,  and consequently  $d_{GM}=d_{tKS}$. In Ref. \cite{tagliacozzo_optical_2013} we have shown how to construct Abelian  GM as a specific  truncation of the Abelian KS LGT.

Gauge magnets have been independently re-formulated in Ref. \cite{chandrasekharan_quantum_1996} as \emph{quantum link models}. The same authors generalized  them to arbitrary groups and representations \cite{brower_qcd_1999,schlittgen_low-energy_2001,baer_quantum_2001} introducing the concept of rishons. Both the truncated LGT and the gauge magnets,  can be understood as specific quantum link models constructions, where the GM corresponds to a quantum link model with a single rishon per link while the truncated LGT corresponds to a quantum link model with two rishons per link \footnote{We thank U. Wiese for such clarification}. It is however interesting to point out that some of the link models can be obtained as a consistent truncation of the KS LGT and others not.

\subsection{${\cal P}_{GM}$ for gauge magnets}
 \label{sect:p_gm}

 We now discuss how to obtain the projector ${\cal P}_{GM}$ onto the physical Hilbert space ${\cal H}_P$ for gauge magnets as a TN.
 The construction is similar to the one used for the other LGT. In particular ${\cal P}_{GM}$ is obtained by contracting as many copies of tensors ${\cal C}$ as there are links on the lattice  and as many copies of tensors ${\cal G}$ as there are sites, following the patterns of Fig. \ref{fig:proj_gauge}.
 
\begin{figure}
 \includegraphics[width=\columnwidth]{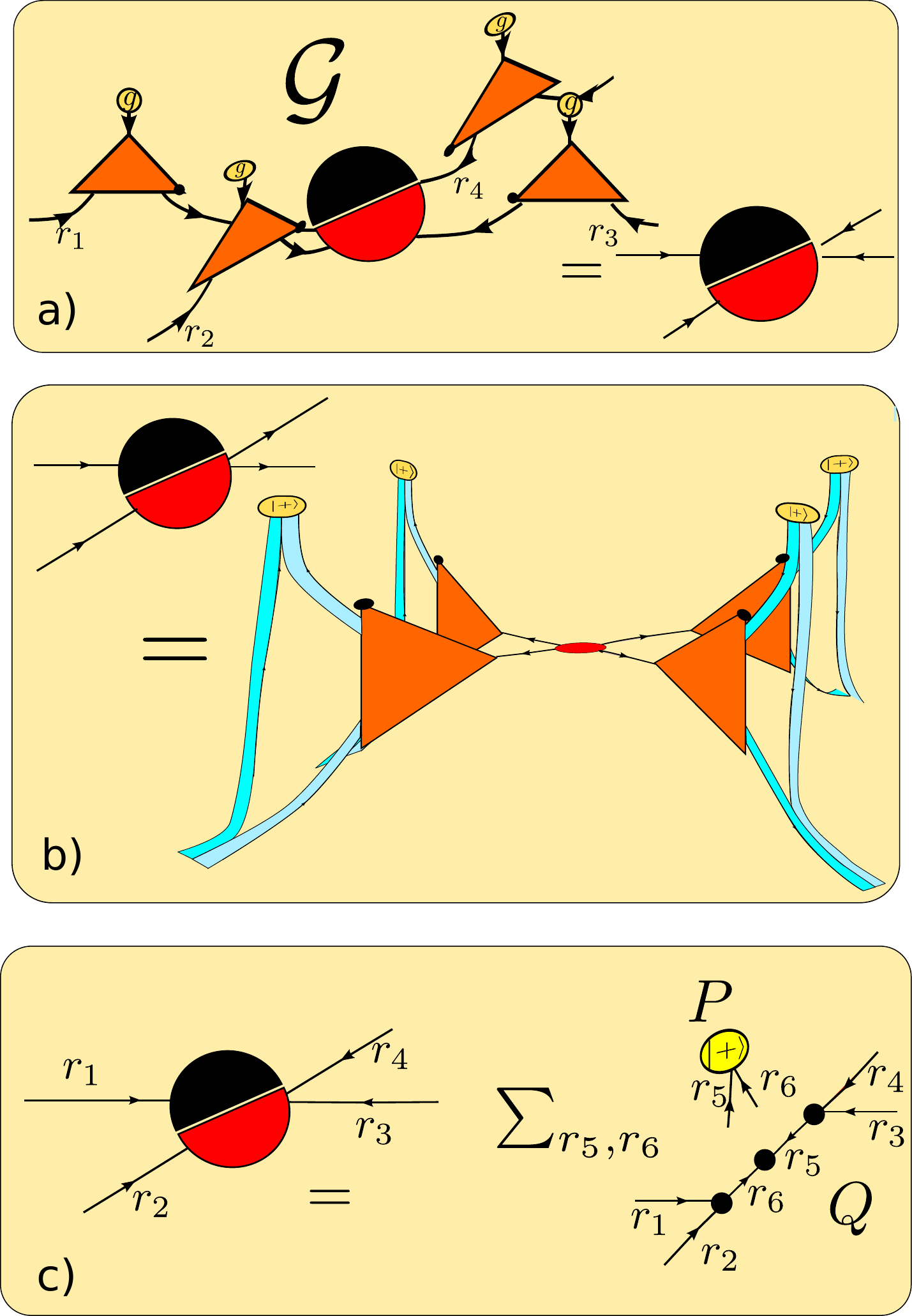}
   \caption{a)  In gauge magnet LGT, the tensor ${\cal G}$ is an invariant $G$-tensor, i.e., it is left invariant under multiplying all its legs by $\Gamma_r(g)$.
   b) ${\cal G}$  is obtained, for a finite group, by summing all the representation matrices through the projector on the trivial representation. After projecting,  the  uniform superposition of all the states  of the trivial representation is obtained by closing the free legs with the product of $\ket{+}$. c) For continuous gauge groups, the projector can either be written as an integral over the group elements with the appropriate invariant measure, or it can be obtained by using the Clebsch-Gordan coefficients, which transform the product of the four representations into a direct sum of irrep's, and, then, by projecting onto to the trivial irrep.\label{fig:proj_gm}}
\end{figure}

 The tensor ${\cal C}$ is a four-leg tensor that copies the state of physical Hilbert space half into the left auxiliary space and half into the right auxiliary space. Notice that both auxiliary spaces need to be extended so that $V_r$ is embedded in a larger vector space that has at least one extra orthogonal direction that we call $\ket{\phi}$.
 Concretely chosen a basis $\ket{l}\ket{v}$ with $\ket{l} =\left\{ \ket{0}, \ket{1} \right\}$ and $\ket{v} \in V_r$, ${\cal C}$ has elements
 \begin{equation}
  {\cal C} = \proj{0\ v}_{ij} \otimes \ket{v}_{\alpha} \bra{\phi}_{\beta} +\proj{1\ v}_{ij} \otimes \ket{\phi}_{\alpha} \bra{v}_{\beta},
 \end{equation}
where $\ket{v}\in V_r$ (remember that  $\braket{\phi| v}=0$ for all $\ket{v}\in V_r$). This means that the bond dimension of the ${\cal C}$ tensor is $n_r +1$ with $n_r$ the dimension of $V_r$.
We can now define the operator ${\cal G}$ that acts on such Hilbert space. Once more $\cal{G}$ needs to be an invariant tensor, that is it needs to fulfill Eq. \eqref{eq:inv_tens}. 
Interestingly the auxiliary Hilbert space is isomorphic to auxiliary Hilbert space of the truncated LGT TN. Nevertheless since  $A_s(h)$ is defined differently in the two models, ${\cal G}$ is different. The symmetry requirements induced by the gauge transformation on $A_s{h}$ are shown in panel a) of Fig. \ref{fig:proj_gm}.

The explicit form of ${\cal G}$ can be obtained  once more using different techniques, either by using the projector onto the trivial irrep of Eq. \eqref{eq:proj_triv} as done in the panel b) of Fig.  \ref{fig:proj_gm}, or by using the standard recipes for constructing symmetric tensors of Ref. \cite{singh_tensor_2012}, as shown in panel c) of Fig. \ref{fig:proj_gm}. In the latter case, one first fuses the irrep's $r_1$ and $r_2$ to $r_5$, then $r_3$ and $r_4$ to $r_6$ and then fuse $r_5$ and $r_6$ to the trivial irrep. In this way the $Q$ part of the tensor is well defined, while the $P$ part depends explicitly on all $r_6$ and $r_5$ compatible with the incoming irrep. Furthermore it should be chosen such to provide a uniform superposition of all symmetric tensors. 

In the appendix \ref{app:tensors} we provide the explicit tensors for both a $U(1)$ and $SU(2)$ gauge magnets.

\section{Tensor network variational ansatz for gauge invariant states}
\label{sec:var_anst}

So far we have discussed how to construct the projector onto the gauge invariant Hilbert space ${\cal H }_P$ defined in Eq. \eqref{eq:gauge} as a TN. 
This means that any state of the gauge invariant Hilbert space can be constructed by acting with ${\cal P}$ on  a generic state $\ket{\phi} \in {\cal H}$ . The challenging problem is still how to express $\ket{\phi}$, since in general it is a state of an exponentially large Hilbert space. One possibility would be to express $\ket{\phi}$ itself as a TN, and then project it with ${\cal P}$. However, even in the best scenario, in which the two states share the same structure as a TN, this rapidly becomes computationally intractable since the bond dimension of the combined TN would be the sum of the bond dimensions for $\ket{\phi}$ and $\ket{P}$. 

The alternative is to construct all states in ${\cal H}_P$ directly from a symmetric variational TN ansatz. This has the big advantage of being a sparse TN so that computations are much cheaper than in full TNs having the same bond dimension. 

The idea is represented schematically in Fig. \ref{fig:hilb_emb}, where we show that ${\cal H }_P$ is embedded in ${\cal H}$ by drawing it as a membrane inside a 3D box.  States described by the  variational ansatz belong to  ${\cal H}_P$ and are represented as orange ovals on it. Their size increases by increasing the bond dimension of the elementary tensors, as represented by the label $D \uparrow$.

The simplest gauge invariant states are the ground states of the Hamiltonian \eqref{eq:h_ks} in the strong coupling limit, which typically are product states. As we increase slightly the complexity of the ground-state wave function we find  another simple class of gauge invariant  states. They consist of uniform superpositions of all gauge invariant states and are obtained by acting with ${\cal P}$ onto a product state $\ket{\varphi}=\ket{+} ^{\otimes L}$, 
\begin{equation}
 \ket{\phi} =  1/\sqrt{Z} {\cal P} \left(\prod _i^{\otimes L} \ket{+}_i \right),\label{eq:RK}
\end{equation}
where  $Z$ is a normalization constant such that $\braket{\phi|\phi}=1$ and $\ket{+}=1/\sqrt{d}(\ket{0}+\cdots +\ket{d})$. These states are an example of generalized Rokshar-Kivelson (RK) states for arbitrary gauge groups \cite{rokhsar_superconductivity_1988,castelnovo_quantum_2005}, and are described by TNs with the same bond dimension as  ${\cal P}$. 


Slight generalizations allow to describe a larger family or RK states by still a TN with the same bond-dimension. This can be done for example by projecting different sates than $\ket{+}$ with ${\cal P}$ with the net-effect of moving away from the equal superposition of all gauge invariant states by  changing the matrix elements inside ${\cal C}$. Alternatively, one could change the weights in the linear superposition  of gauge invariant states by changing the tensors ${\cal G}$ (We will discuss the operators that allow to deform RK state by acting onto the physical Hilbert space  in Sec. \ref{sec:vertex_op}). At this point we have exhausted all possible of gauge invariant states that can be obtained without extending the TN.

%

%


For this reason, in order to describe all other states in ${\cal H}_P$  we have to introduce a more complex  variational ansatz. This is done through a TN ansatz that consists in a superposition of  spin networks \cite{rovelli_spin_1995}.
It is again formed by two types of tensors, ${\cal C}$ and ${\cal G}$, and as before, every bond  of the TN is decorated by an irrep index. 
There is one  ${\cal C}$ tensor for each link $s_n$ of the lattice, and one ${\cal G}$ tensor for each site of the lattice .
The elementary  tensors are contracted following the usual pattern of Fig. \ref{fig:proj_gauge}.  The structure imposed by gauge invariance is exactly the same as the one discussed for the projector ${\cal P}$;  we will refer to it as the ``symmetry part'' of the TN. 
Now, however, we add to each tensor a ``degeneracy part''. We  promote every element of the elementary tensors in ${\cal P}$ to a full tensor acting on the appropriate degeneracy space. 
This implies that to every irrep $r$, we attach a degeneracy space, unconstrained by the symmetry, with  dimension $D_r$ (that can be chosen independently for every irrep r) that contains the variational parameters of the tensor. The various blocks $r$ are patched correctly thanks to the symmetry part of the TN.

\begin{figure}
 \includegraphics[width=\columnwidth]{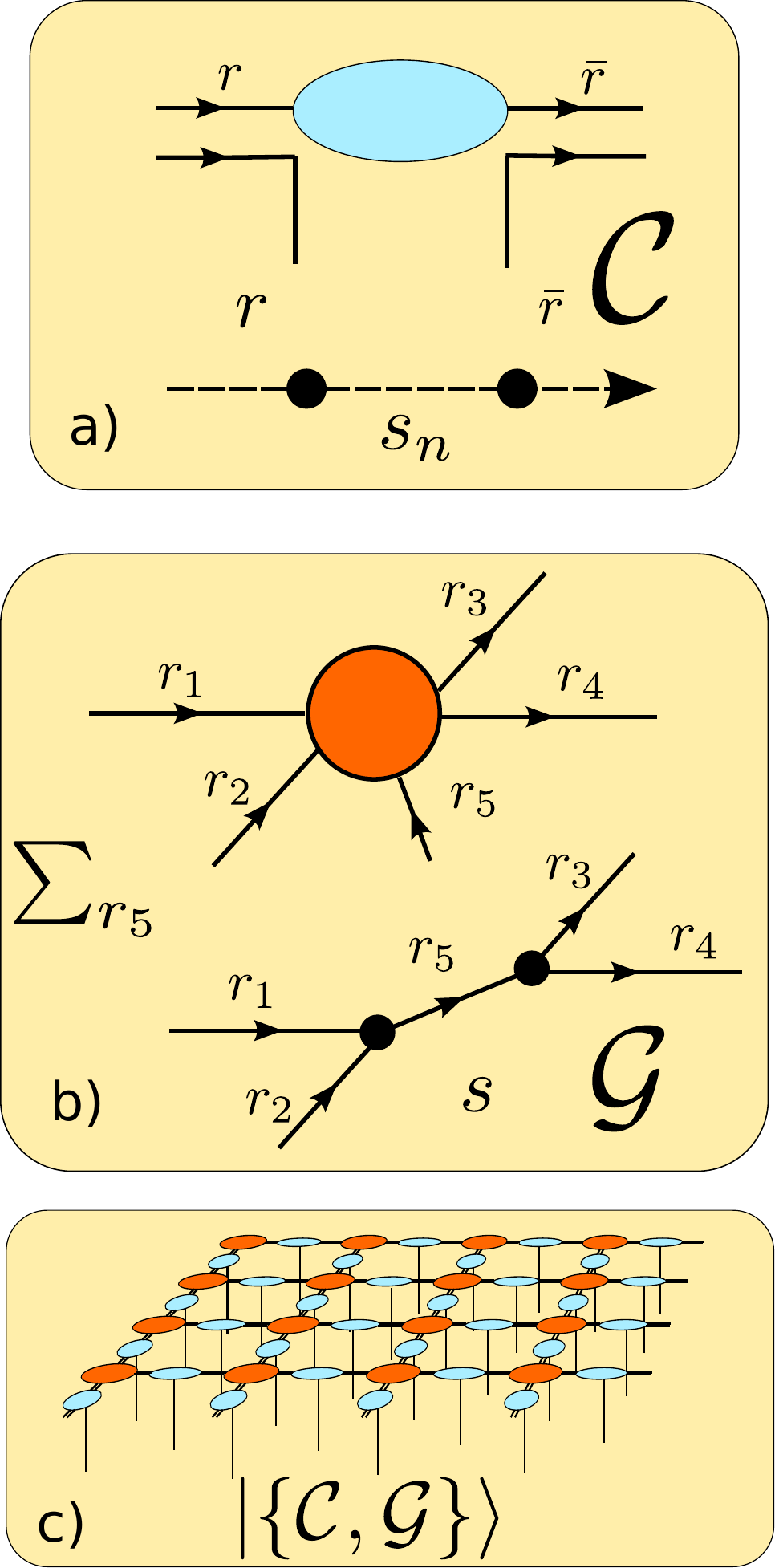}
 \caption{a) The ${\cal C}$ tensor used to build a variational ansatz for a gauge invariant state. The tensor acts on a link $s_n$ and embeds its state onto the auxiliary space. It is composed by two parts, a ``symmetry part'' that does not contain any variational parameter (the two lower lines), and a degeneracy part that contains the variational parameters (the cyan tensors). The tensor has several blocks labeled by the irrep $r$. 
 b) The ${\cal G}$ tensor only acts on the auxiliary space, and it has again a block structure. It is divided in two pieces. The first is responsible of the correct symmetry properties of ${\cal G}$ (see panel a) of Fig. \ref{fig:proj_k_s_t}).
 This part does not contain any free parameters and is given in terms of the Clebsch-Gordan coefficients of the group (small black dots). The second part is  a degeneracy part that is formed by a sum of several tensors (one for each allowed value of $r_5$)  acting on $D_{r_1} \otimes D_{r_2} \otimes D_{r_3} \otimes D_{r_4}$. These tensors store the variational parameters of the ${\cal G}$ tensor in the appropriate symmetry blocks.
 c) Variational ansatz for  gauge invariant states on a lattice of $4 \times 4$ sites and periodic boundary conditions. The network contains  one ${\cal C}$ per link of the lattice, and one ${\cal G}$ per site. The double lines connecting the tensors are used  to remind that each of the elementary tensors  has a double structure, one part dictated by the symmetry and the other one containing  the actual variational parameters. In the figure we are assuming the sum over all the irrep's $r$ on every bond of the TN, so that the specific irrep label is omitted.\label{fig:gen_ans}}
\end{figure}

We illustrate the construction for the KS LGT, but this is applicable to all the other cases we have discussed so far. 
On each link, the Hilbert space is isomorphic to  $\sum_r V_r \otimes V_{\bar{r}}$. The tensor ${\cal C}$ is a three-leg tensor composed by a symmetry part and a degeneracy part. This is illustrated  in panel a) of Fig. \ref{fig:gen_ans}, where the symmetry part is just made of lines representing identity matrices,  while the degeneracy part is made by a tensor of free parameters.
Its symmetry part is indeed the same as the ${\cal C}$ tensor of an RK state  discussed above. 
Inside each block defined by the irrep $r$, it copies  the states in $V_r$ to the left and those in  $V_{\bar{r}}$ to the right.
Its degeneracy  part is novel with respect to the previous examples of RK states. It is   made of matrices of size $D_r \times D_r$. These matrices (one per block) are shown in cyan in panel a) of Fig. \ref{fig:gen_ans} and contain the variational parameters. 
Turning to the tensor ${\cal G}$ in panel b) of Fig. \ref{fig:gen_ans}, it is also  made of two pieces, a symmetry part and a degeneracy part.
The symmetry part (once more the lower piece in the graphical representation of the tensor  in panel b) of Fig. \ref{fig:gen_ans}) does not contain any free variational parameter but simply takes cares of correctly matching the irrep's in such a way that the obtained tensor is symmetric, that is it fulfills Eq. \eqref{eq:inv_tens}. In the graphical representation, this is done by the Clebsch-Gordan tensors represented by the small black circles. To every irrep label, one associates now a degeneracy tensor (shown in orange) in panel b) of Fig. \ref{fig:gen_ans}, which is populated by the variational parameters. Importantly, the degeneracy tensor is obtained as a sum over the irrep's corresponding to the internal lines of the symmetric part ($r_5$ in the figure).
The variational state is thus expressed by the contraction of the various ${\cal C}$ and ${\cal G}$ given by panel c) of Fig. \ref{fig:gen_ans}, where it is represented for the specific case of a  lattice made by $4 \times 4$ sites and periodic boundary conditions are assumed in both directions.


As a second example, we can consider the $U(1)$ GM, whose ${\cal P}$ is defined in appendix \ref{app:tensors}. We start with the ${\cal C}$ tensor for  a generic gauge invariant state. In this  case the physical Hilbert space of a link is two dimensional and  involves only  two blocks labeled by the irreducible representations $0,1$. On the other hand the auxiliary Hilbert space has two blocks with arbitrary  dimension $D_0$ and $D_1$ so that the auxiliary space is  ${\cal H}_{\textrm{aux}}={\cal H}_0 \oplus {\cal H}_1$.
The non-zero elements of ${\cal C}$ are ${\cal C}^{ \alpha_0,\beta_0,0}$, a  full matrix that acts on the sub-block ${\cal H}_0 \otimes {\cal H}_0$ of the  Hilbert space ${\cal H}_{\textrm{aux}}\otimes {\cal H}_{\textrm{aux}}$
\begin{equation}
 {\cal C}^{ \alpha_0,\beta_0,0}:   {\cal H}_0 \to {\cal H}_0.
\end{equation}
Similarly ${\cal C}^{ \alpha_1\beta_1,1}$  is an independent matrix that acts on a different block of  ${\cal H}_{\textrm{aux}}\otimes {\cal H}_{\textrm{aux}}$,
\begin{equation}
{\cal C}^{ \alpha_1,\beta_1,1}:  {\cal H}_1 \to {\cal H}_1.
\end{equation}
The same idea applies to ${ \cal G}$  that now can be thought as a collection of six tensors each of them acting on one of the only six blocks of the four fold tensor product $\prod_{i =1}^{\otimes 4}\left( {\cal H}_0 \oplus {\cal H}_1\right)_i$ (this has 16 blocks) allowed by the symmetry constraints. 
Just to give an example, one of the allowed blocks is ${\cal H}_0\otimes {\cal H}_0\otimes{\cal H}_1\otimes {\cal H}_1$, where  ${ \cal G}$  has entries ${ \cal G}^{\alpha_0, \beta_0,\gamma_1,\delta_1}$.  

Before continuing we summarize the results of this section.
We have proposed a variational ansatz for states of the physical Hilbert space ${\cal H}_P$. The ansatz involves the contraction of several copies of two families of tensors, ${\cal C}$ tensors (one per link) and ${\cal G}$ tensors (one per site). Each of those tensors has two components, one completely determined by the requirements of gauge symmetry and a second one that contains the free variational parameters.
The resulting TN has bond dimension 
\begin{equation}
 D=\sum_r n_r D_r, \label{eq:bond_d}
\end{equation}
where $n_r$ is the dimension of the irrep $r$, $D_r$ is the dimension of the degeneracy space associated to the irrep $r$ (which is a free parameter), and the sum over $r$ extends to all the irrep's  one needs to consider. 

There are several  advantages in dealing with such symmetric ansatz. On one side, the ansatz can be manipulated with a cost smaller than the cost involved in manipulating a non-symmetric ansatz with the same $D$, since one can work in each block separately \cite{singh_tensor_2012}. Also, the ansatz can be used to obtain approximations of  interesting gauge invariant states (such as eigenstates  of gauge invariant Hamiltonians), while exactly preserving the gauge symmetry.
Furthermore, the ansatz allows to target not only the invariant states, but also co-variant states belonging to separate symmetry sectors.
A typical application of this scenario is the  characterization  of the effects  of background charges on the physics of the  gauge bosons.

\section{Gauge invariant vertex operators}
\label{sec:vertex_op}
We have just discussed how one can change, given the projector ${\cal P}$ on ${\cal H}_P$ as a TN,  the parameters defining $\cal{ C}$ and ${\cal G}$ and thus one can define a family of RK states. Here we want to address the question about what are the operators that allow to modify the RK wave function by acting on the physical Hilbert space. Those operators, in principle  can be added to the standard LGT Hamiltonian of Eq. \eqref{eq:h_ks} so to extend it and allow  to explore  extended phase diagrams.
The operators that allow to change the entries of ${\cal C}$ are just the ${\cal E}^2$ operators already discussed. On the other hand,  the operators that allow to modify the entries of ${\cal G}$ act on crosses, and thus are are co-diagonal with ${\cal G}$. 

\begin{figure}
 \includegraphics[width=\columnwidth]{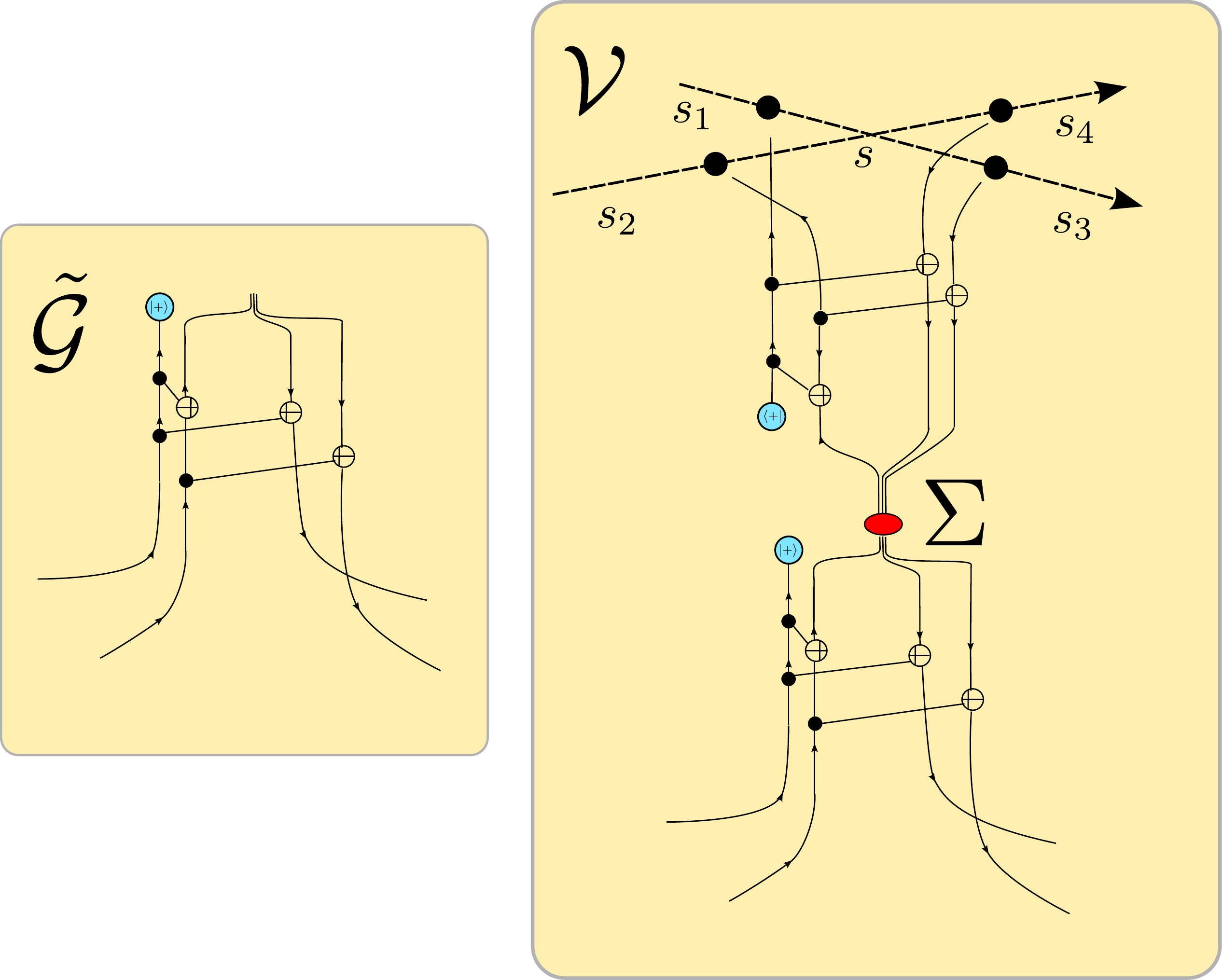}
 \caption{ Left) The isometry  $\tilde{{\cal G}}$ can be used to go to the gauge invariant states. Now a diagonal operator in this space commutes with the gauge transformations, and is thus gauge invariant. Right) The vertex operator described in Eq. \eqref{eq:v_op} is built by concatenating the tensor $\tilde{{\cal G}}$,  a diagonal tensor $\Sigma$ (a red circle) (acting on ${\cal C}(G)^3$), and then $\tilde{{\cal G}}^{\dagger}$. It acts as a potential for different gauge invariant configurations and allows to favor one with respect to another. \label{fig:vertex_op}}
\end{figure}

In particular, ${\cal G}$  consists in an isometric tensor followed by a uniform projection onto the state $\prod_{\otimes }\ket{+}$, that is ${\cal G}\equiv\tilde{{\cal G}}\prod_{\otimes }\ket{+}$ as explicit in Fig. \ref{fig:proj_k_s}. We can now use $\tilde{G}^{\dagger}$ to get to the correct basis of the gauge invariant configurations and weight each of them differently through a diagonal tensor $\Sigma$, 
\begin{equation}
 {\cal V} =  \tilde{G} \Sigma \tilde{G}^{\dagger}. \label{eq:v_op}
\end{equation}
In  the KS construction of ${\cal G}$ given in Sect. \ref{sec:proj_tn}, $\Sigma$ has $|G|^3$ elements that can be chosen arbitrarily.
Both the isometry $\tilde{{\cal G}}$ and the vertex operator ${\cal V}$ are represented graphically in Fig. \ref{fig:vertex_op}.

In particular we will use these operators, in oder to characterize the transition from the eight-vertex to the six-vertex model in the next section.

\section{Benchmark numerical results}
\label{sec:num_res}

In this section we benchmark our proposal against known analytic results. 
We start by describing  the RK wave function of the $\mathbb{Z}_2$ LGT. This is the exact ground state of the  Hamiltonian \eqref{eq:h_ks} for $\alpha=0$. In this case the $U$ operator is just the standard $\sigma^z$ Pauli matrix, while  ${\cal E}^2$ is the $\sigma_x$ Pauli matrix. The Hamiltonian reads
\begin{equation}
 H_{z2} = \sum_l \sigma^x_{l} + \frac{1}{\alpha} \sum_p \sigma^z_{p_1}  \sigma^z_{p_2} \sigma^z_{p_3} \sigma^z_{p_4},
\end{equation}
where $p$ are the plaquettes and $l$ the links of the lattice $\Lambda$.
In the limit  $\alpha \to \infty$ the ground state becomes the RK state compatible with the symmetry constraints
\begin{equation}
  \sigma^x_{s_1}  \sigma^x_{s_2} \sigma^x_{s_3} \sigma^x_{s_4} \ket{\psi} =\ket{\psi}, \forall s\in \Lambda,
\end{equation}
where as usual $s_1,\dots,s_4$ are the links around a site $s$.
The above RK state is obtained by contracting a TN with $D=2$ of the form of the one in Fig. \ref{fig:proj_gauge}, where 
${\cal C}^{0,0,0}=1$, ${\cal C}^{1,1,1}=1$, and ${\cal G}^{0,0,0,0}={\cal G}^{1,1,1,1}=1$, 
${\cal G}^{0,1,1,0}={\cal G}^{1,0,0,1}=1$, ${\cal G}^{1,0,1,0}={\cal G}^{0,1,0,1}=1$, ${\cal G}^{0,0,1,1}={\cal G}^{1,1,0,0}=1$, where as always we denote the indexes by $s_1, \cdots, s_4$, following the pattern of Fig. \ref{fig:a_ks}.

As written explicitly in the appendix \ref{app:tensors_u1}, the ${\cal P}$ on ${\cal H}_P$ for the $U(1)$ gauge magnet has a very similar tensor structure. In that case however,  the tensor ${\cal G}$ misses the last two entries since ${\cal G}^{0,0,1,1}={\cal G}^{1,1,0,0}=0$.  The $U(1)$ RK state  is the ground state of the following Hamiltonian
\begin{align}
 H_{GM} = \sum_p \left[ \left(a_{p_1} a_{p_2}a^{\dagger}_{p_3} a^{\dagger}_{p_4} +H.c.\right)\right. \nonumber \\
 \left.- \left(a_{p_1} a_{p_2}a^{\dagger}_{p_3} a^{\dagger}_{p_4} +H.c.\right)^2\right],
\end{align}
with $a=\ket{0}\bra{1}$, that is it is the ground state of the gauge magnet Hamiltonian at its RK point \cite{moessner_short-ranged_2001,shannon_cyclic_2004,banerjee_2_2013}.

By applying the vertex operators defined in Eq. \eqref{eq:v_op}  to the $\mathbb{Z}_2$ RK state, we can switch off the two extra elements in ${\cal G}$, thus effectively interpolating between the $\mathbb{Z}_2$ RK state and the $U(1)$ RK state. In particular, in order to study the transition between the two models we parameterize the elements  ${\cal G}^{0,0,1,1}={\cal G}^{1,1,0,0}=\cos(\theta)$, with $0\le \theta \le \frac{\pi}{2}$. At $\theta=0$ we have the  $\mathbb{Z}_2$ RK state while at $\theta=\frac{\pi}{2}$ we have the $U(1)$ RK state.

\begin{figure}
 \includegraphics[width=\columnwidth]{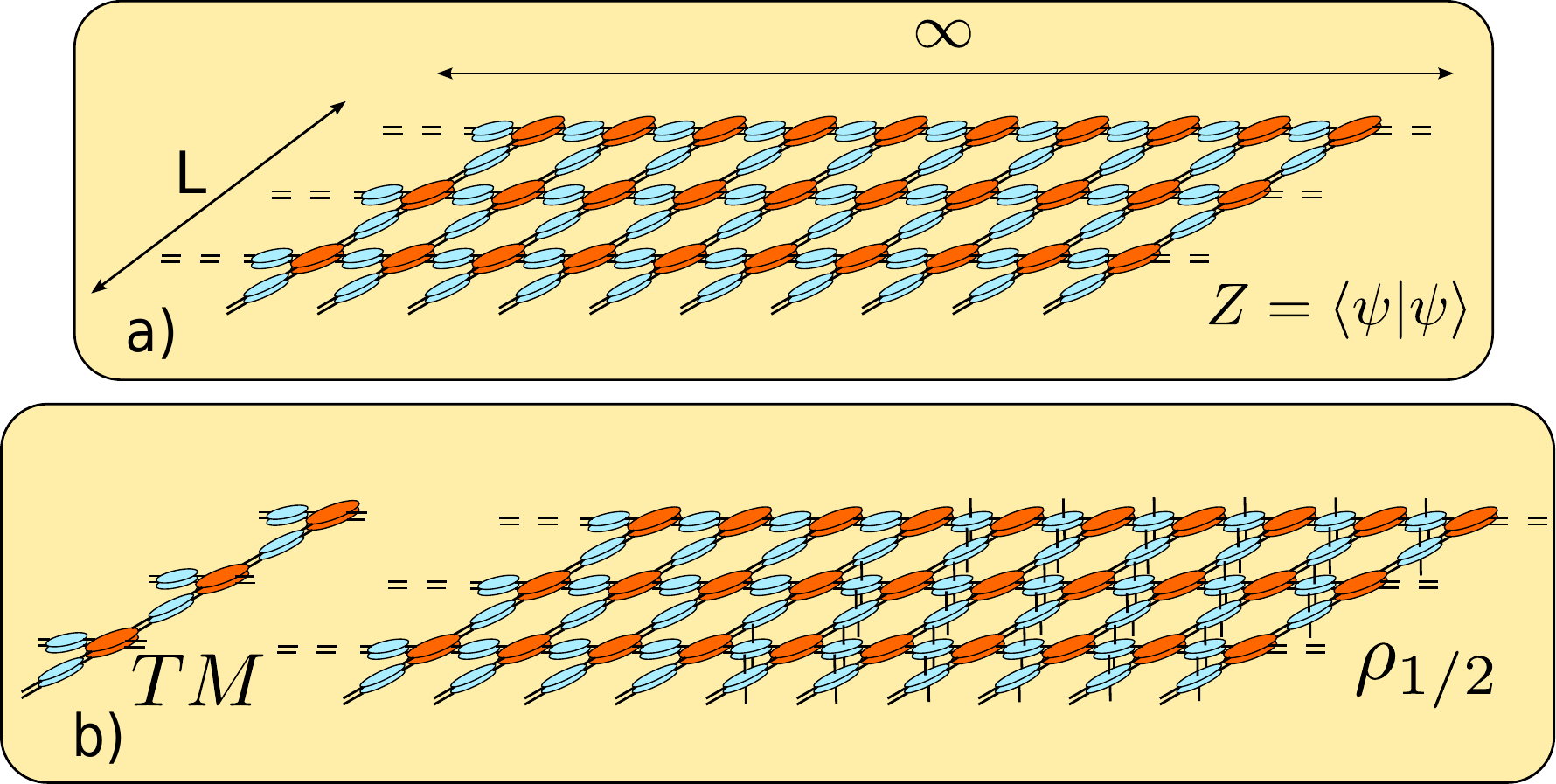}
 \caption{a) The setup used in our numerical calculations. Each of the elementary tensors ${\cal C}$ (cyan) and ${\cal G}$ (orange) has bond dimension $D=2$ and its matrix elements are those described in the main text. The elementary tensors are contracted so to provide the quantum states $\ket{\psi}$ of a 2D infinitely large cylinder of circumference $L$. In the drawing one has to assume PBC along the vertical direction. Here we represent the norm of the state that defines the partition function $Z$ of a 2D classical model. b) We compute the spectrum of the transfer matrix $TM$ across the cylinder (sketched on the left), which characterizes the decay of the correlations addressed in Sect. \ref{sec:corr}, and the spectrum of the reduced density matrix of half of the cylinder $\rho_{1/2}$ (shown on the right), which gives access to the entanglement characterization of the states addressed in Sect. \ref{sec:ent}. Both calculations are performed via sparse exact diagonalization and their cost increase exponentially with $L$ so that we can address at most systems with $L=20$. \label{fig:num}  }
\end{figure}

We characterize the corresponding  RK wave functions for 2D infinite cylinders with circumference $L$ as sketched in the panel a) of Fig. \ref{fig:num}.  

For each value of $\theta$ the norm of the state obtained from the above tensors gives the partition function of the eight-vertex model, whose phase diagram was uncovered by Baxter \cite{2008_exactly}. In particular in the language of the eight-vertex model we follow the line at $a=b=c=1$  and vary $d$ in the range $0\le d\le 1$. Such line has also been studied in Fig. 5 of Ref. \cite{ardonne_topological_2004}.

The interest in this specific line stands in the fact that along it, the model approaches a transition between two topological phases. At $d=1$  the eight-vertex is in a  $\mathbb{Z}_2$ deconfined phase. This is the paradigm of a $\mathbb{Z}_2$ gapped spin liquid. At $d=0$, there is the transition between the eight- and six-vertex model so that the system enters in an algebraic spin-liquid phase. Here we analyze  how to characterize the two phases and the transition between them by using our numerical ansatz.

\subsection{Decay of correlations}
\label{sec:corr}
The correlations across the cylinder are mediated by the transverse transfer matrix (TM), made by the contraction of all ${\cal C}$s and ${ \cal G}$s along a transverse  slice of the cylinder as sketched in panel b) of Fig. \ref{fig:num}. In particular at the $\mathbb{Z}_2$ LGT point $\theta=0$ the TM has only two degenerate non-vanishing eigenvalues $t_1$ and $t_2$. The first gap $\Delta_1 =-\log(t_2/t_1)=0$, while all others are infinite. The model thus has zero correlation length. As we start departing from $\theta = 0$, the two degenerate eigenvalues start to split so that $\Delta_1$ starts to diverge as $\Delta_1 \propto \exp(L)$. A family of new eigenvalues starts to  appear and the model acquire a non-zero correlation length. The decay of correlation functions is thus exponential in all this region. The new eigenvalues tend to  approach $t_1$ . The bigger of them $t_3$ is separated from $t_1$ by a gap $\Delta_2 = -\log(t_3/t_1)$. This is the gap that closes to zero when approaching $d=0$ (that is at $\theta=\pi/2$) as $\Delta_2 \propto 1/L$. The model thus develops algebraic correlations at $\theta=\pi/2$. Our benchmark numerical results agree with this exact picture. All the eigenvalues are computed by exact sparse diagonalization of the TM with a cost that increases exponentially with $L$. These results give a first check that our numerical technique works and we can now apply it to characterize the phase transition in terms of two of the proposed order parameters based on the scaling of the entanglement entropy: the topological entropy and the Schmidt gap. Both quantities could be relevant in understanding the phase diagram of gauge theories.
Gauge theories indeed present phases that cannot be distinguished by using a local order parameter. A legitimate question is whether the scaling of entanglement allows to discern those eluding phases.

\begin{figure}
 \includegraphics[width=\columnwidth]{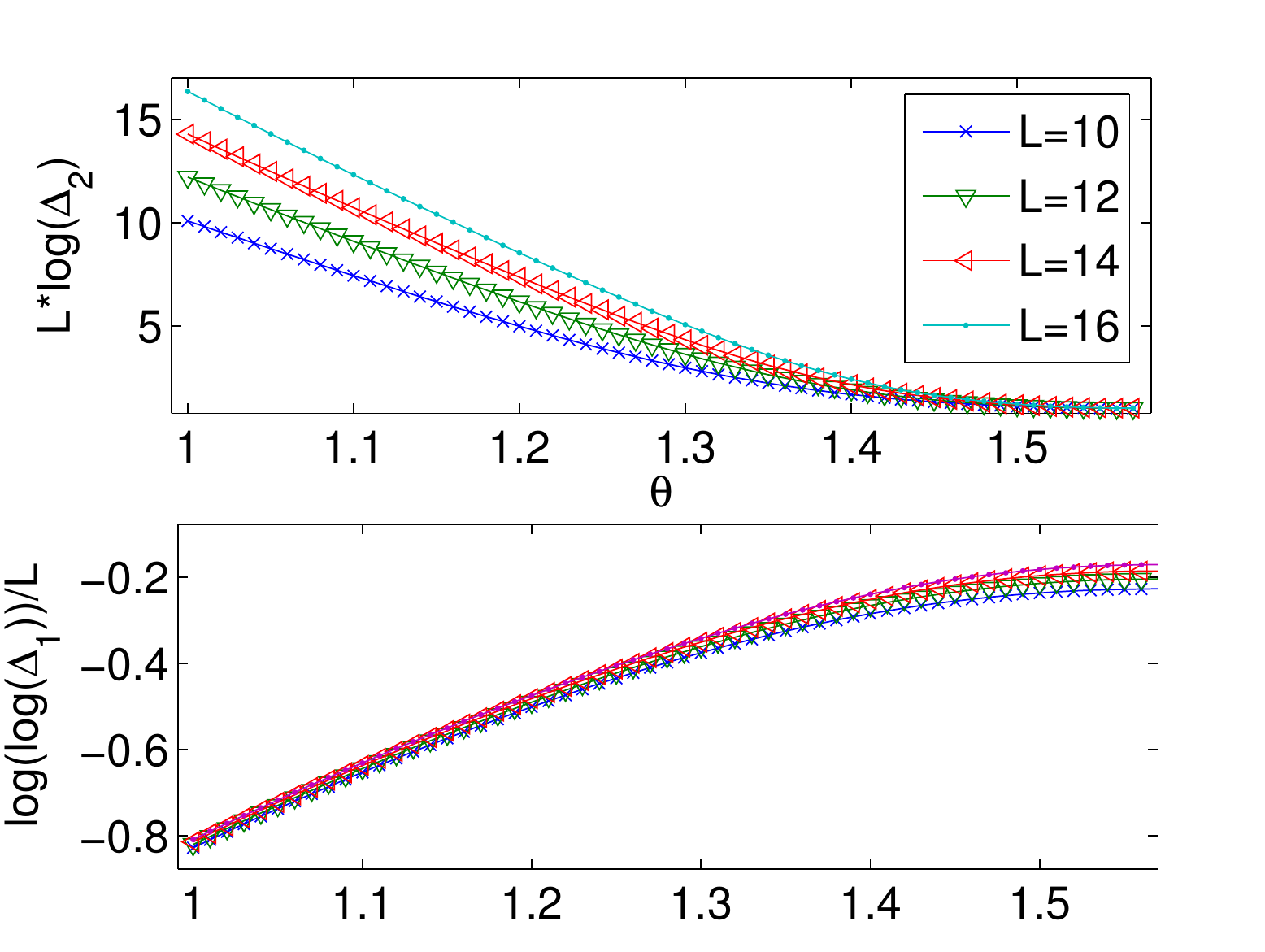}
 \caption{Upper panel) The second gap of the TM represented in panel b) of Fig.  \ref{fig:num} in the $\mathbb{Z}_2$ spin-liquid phase closes as $\theta$ approaches $\pi/2$, where there is the transition from the  $\mathbb{Z}_2$ spin-liquid phase to the $U(1)$ algebraic spin-liquid phase.  The collapse of the data for the $\Delta_2 L$ close to $\pi/2$, for the values of $L$ in the range $L=10, \dots ,16$, confirms that the gap at the transition closes as $1/L$ as expected. Thus, what we are studying is a transition from a gapped phase with exponential decay of correlations to a gapless phase governed by algebraic decay of correlation. Lower panel) On the other hand, the first gap of the TM $\Delta_1$,  representing the gap between the two different topological sectors appearing on the cylinder, opens exponentially with $L$ as $\theta$ tends to $\pi/2$. This is again confirmed by the collapse of our numerical data for $\log(\Delta_1)/L$, with $L$ in the  range $L=10,\dots,16$. \label{fig:num_corr}}
\end{figure}

\subsection{Order parameters based on entanglement}
\label{sec:ent}
 
We compute the  entanglement entropy of the reduced density matrix of half of the infinite cylinder $\rho_{1/2}$, sketched in panel b) of Fig. \ref{fig:num}. The spectrum $\set{\lambda_n}$, $n=1\cdots D^{L}$ of $\rho_{1/2}$ is  computed using sparse exact diagonalization of the eigenvectors of the TM as explained in detail in Ref. \cite{cirac_entanglement_2011}. The cost of the computation increases exponentially with $L$.
We are interested in characterizing the scaling of the entanglement entropy as a function of $L$ since we want to extract the topological entropy $\gamma_T$.
In a gapped spin liquid the entanglement entropy $S_A$ of a region $A$ with boundaries of length $L$ scales as 
\begin{equation}
 S = c_1*L +\gamma_T +c_2/L + \cdots, \label{eq:ent_scaling}
\end{equation}
where the dots stand for the omission of higher order corrections starting with $(1/L)^2$.
The constant $\gamma_T$, in the topological phases, is negative and universal, and encodes the topological entropy \cite{levin_detecting_2006,kitaev_topological_2006}. In the specific case of a $\mathbb{Z}_2$ spin liquid it is known to be 
\begin{equation}
\gamma_T = -\log(2). \label{eq:gamma_z2}
\end{equation}
Equation \eqref{eq:ent_scaling} holds  also for the gapless spin-liquid phase described by the six-vertex model. In a series of seminal works St\' ephan and collaborators \cite{stephan_shannon_2009,stephan_renyi_2012} have shown that it is indeed possible to get an exact expression for $\gamma_T$ for the whole phase diagram of the six-vertex model. 
In particular they have expressed it in the language of the $XXZ$ spin chain defined on a space-like section of the cylinder.  The eigenvectors of the Hamiltonian of the XXZ model \eqref{eq:six_vertex} are indeed equal to those of the transfer matrix of the six-vertex model \cite{2008_exactly}.
The entanglement entropy of half of the infinite cylinder  corresponds, in the $XXZ$ model, to the Shannon entropy of the ground-state wave function  of a chain with periodic boundary conditions and length $L$.  The $XXZ$ Hamiltonian is given by
\begin{equation}
 H = \sum_i \sigma^x_i \sigma^x_{i+1} +\sigma^y_i \sigma^y_{i+1}+\Delta \sigma^z_i \sigma^z_{i+1}. \label{eq:six_vertex}
\end{equation}
In the range $-1 < \Delta \le 1$ the Hamiltonian is critical and the low-energy physics is described by a CFT with $c=1$ describing a free boson compactified on a circle with radius $R = \sqrt{2- \frac{2}{\pi}\textrm{arccos} (\Delta)}$. In the whole phase the topological  entropy is given by \cite{stephan_shannon_2009},
\begin{equation}
 \gamma_T = \log(R) - \frac{1}{2}. \label{eq:gamma_u1}
\end{equation}
The specific point we are studying, called the ice-point of the six-vertex model corresponds to $\Delta = -1/2$.

\begin{figure}
 \includegraphics[width=\columnwidth]{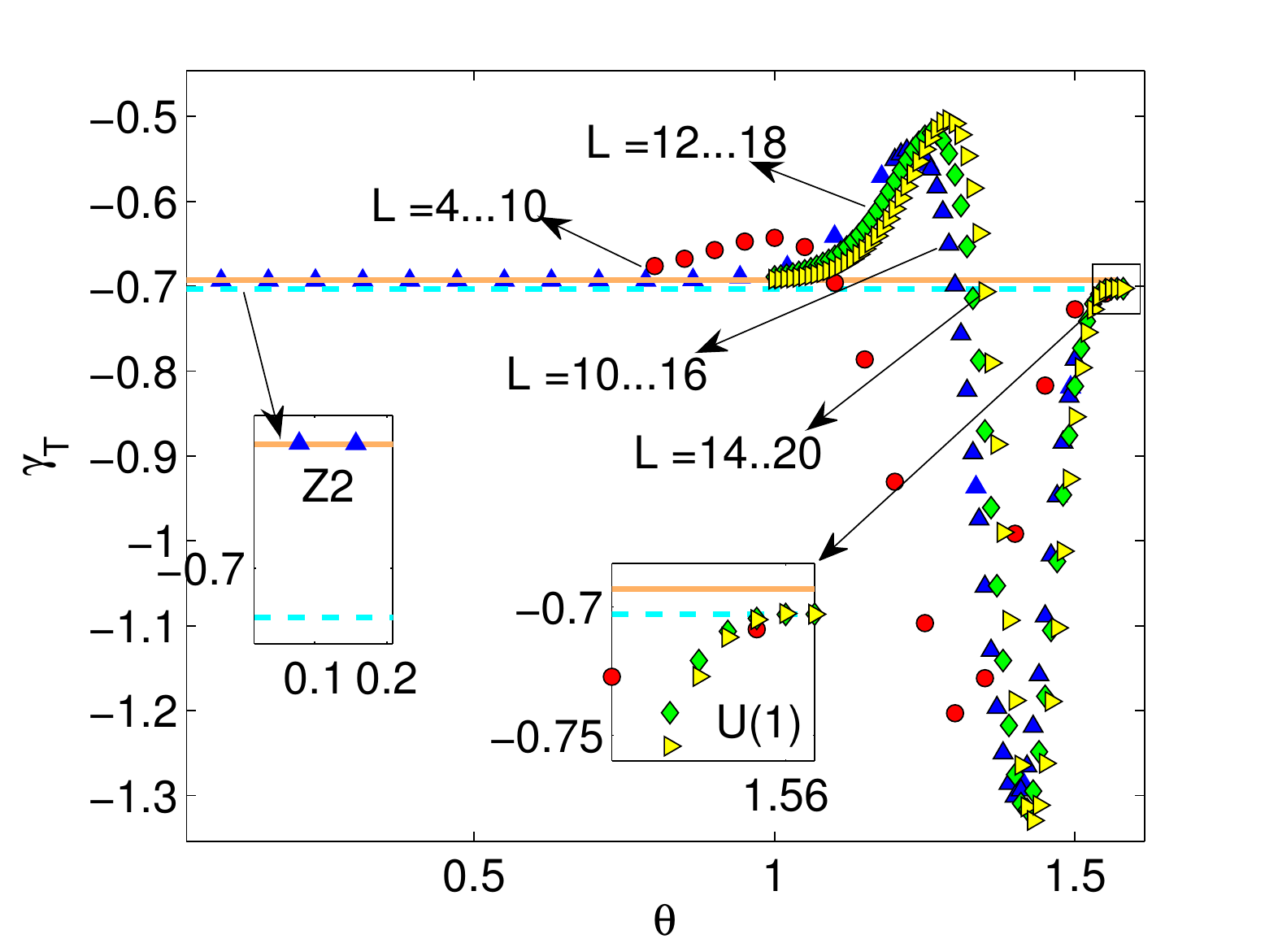}
\caption{The topological entropy $\gamma_T$ defined in Eq. \eqref{eq:ent_scaling}, and extracted from the scaling of the entropy of half infinite cylinders with respect to their circumference $L$. The red dots are obtained by considering the scaling of the entropy in the interval  $L=4,\dots 10$,  blue up-facing triangles  $L=10,\dots,16$, green squares  $L=12,\dots,18$, and the yellow triangles pointing to the right $L=14,\dots,20$. The red solid orange line represents the exact value of $\gamma_T$ for the  $\mathbb{Z}_2$ spin liquid in Eq. \eqref{eq:gamma_z2}, while the cyan dashed  line represents its value for the $U(1)$ spin liquid from  Eq. \eqref{eq:gamma_u1} with $R$ associated to $\Delta =-1/2$. We see that, for small $\theta$, $\gamma_T$ coincides, independently of the size of the cylinder considered,  with the expected exact value $\mathbb{Z}_2$ (left inset). As we move towards the transition $\gamma_T$ shows a transient oscillation that tends to become sharper, deeper and move towards $\pi/2$ for larger $L$. At $\pi/2$ it attains again the expected analytical value with very small corrections induced by considering the two different set of data (right inset). In the main text we provide a discussion of these results.\label{fig:topo_ent}}
 \end{figure}

The numerical results we have obtained are presented in Fig. \ref{fig:topo_ent}.  We have only access to modest sizes in the transverse direction  $L=4,\dots,20$. $\gamma_T$ is easy to extract close to $\theta=0$ and $\theta= \pi/2$ where we are able to recover, from our numerics, its exact analytical value.
As a cross-check we have further reduced the size of the system and considered only the smaller cylinders from  $L=4,\dots,10; L=10,\dots,16;  L= 12,\dots,18; L=14, \cdots, 20$. The results,  obtained  with those sets of data (represented by different symbols in Fig. \ref{fig:topo_ent}), still provide estimates of $\gamma_T$ in agreement with the theory close to $\theta=0$,  where finite size corrections are completely negligible,  and  close to $\theta =\pi/2$ where they are reasonably  small (see the rightmost  inset of  Fig. \ref{fig:topo_ent}). 

The situation is very different for intermediate values of $\theta$. Especially in the region between $0.7\le \theta \le 1.5$, we observe strong cross-over effects. The value of  $\gamma_T$  extracted from different series of $L$ do not agree as shown by the fact that curves made by different symbols are distinct.  This effect could be related to the sub-leading corrections that become more  important when we approach the transition.  
We also observe that for  larger systems,  the cross-over region shrinks and move towards the phase transition at  $\theta =\pi/2$. 
It looks like that, if we were able to reach  the thermodynamic limit, $\gamma_T$ would present  a very sharp jump between the two asymptotic values.

These results give further evidence  that, provided one is able to address large enough systems,  $\gamma_T$ can be used as an order parameter, away from a relatively small cross-over region (which further-more shrinks with increasing system size). In our specific case indeed it allows to discern the gapped $\mathbb{Z}_2$ spin-liquid phase  from the algebraic $U(1)$ spin-liquid phase  at $\theta=\pi/2$.

We now analyze another possible order parameter based on the scaling of the entanglement. Li and Haldane in Ref.  \cite{li_entanglement_2008} suggested that phases could be easier to identify by  considering  the scaling of the full entanglement spectrum  rather than focusing on a single number as  the  topological entropy.
The entanglement spectrum is the collection of  the logarithm of the eigenvalues of the reduced density matrix $\log(\lambda_n)$.

In particular,  numerical studies of  1D systems have provided  a precise characterization of  the scaling of the lowest part of the entanglement spectrum, the one associated to the largest eigenvalues of the reduced density matrix \cite{de_chiara_entanglement_2012,lauchli_operator_2013}. In many cases, the lowest gap of the entanglement spectrum, called the Schmidt gap, vanishes when approaching a quantum phase transition following universal scaling laws. The authors thus have proposed to use  the Schmidt gap as an order parameter. This idea is further supported by the recent results that show that for conformal invariant critical points, several gaps in the entanglement spectrum close in a way that allows to identify the critical exponents of the underlying CFT \cite{lauchli_operator_2013}. It is still unclear how general these results are. 

For this reason we have decided to analyze the behavior of the lowest part of the entanglement spectrum  of $\rho_{1/2}$ of half cylinders, when approaching the transition at $\theta = \pi/2$. In particular we address the cross-over region between $1\le \theta \le \pi/2$, where the analysis of the topological entropy is unreliable.  

\begin{figure}
 \includegraphics[width=\columnwidth]{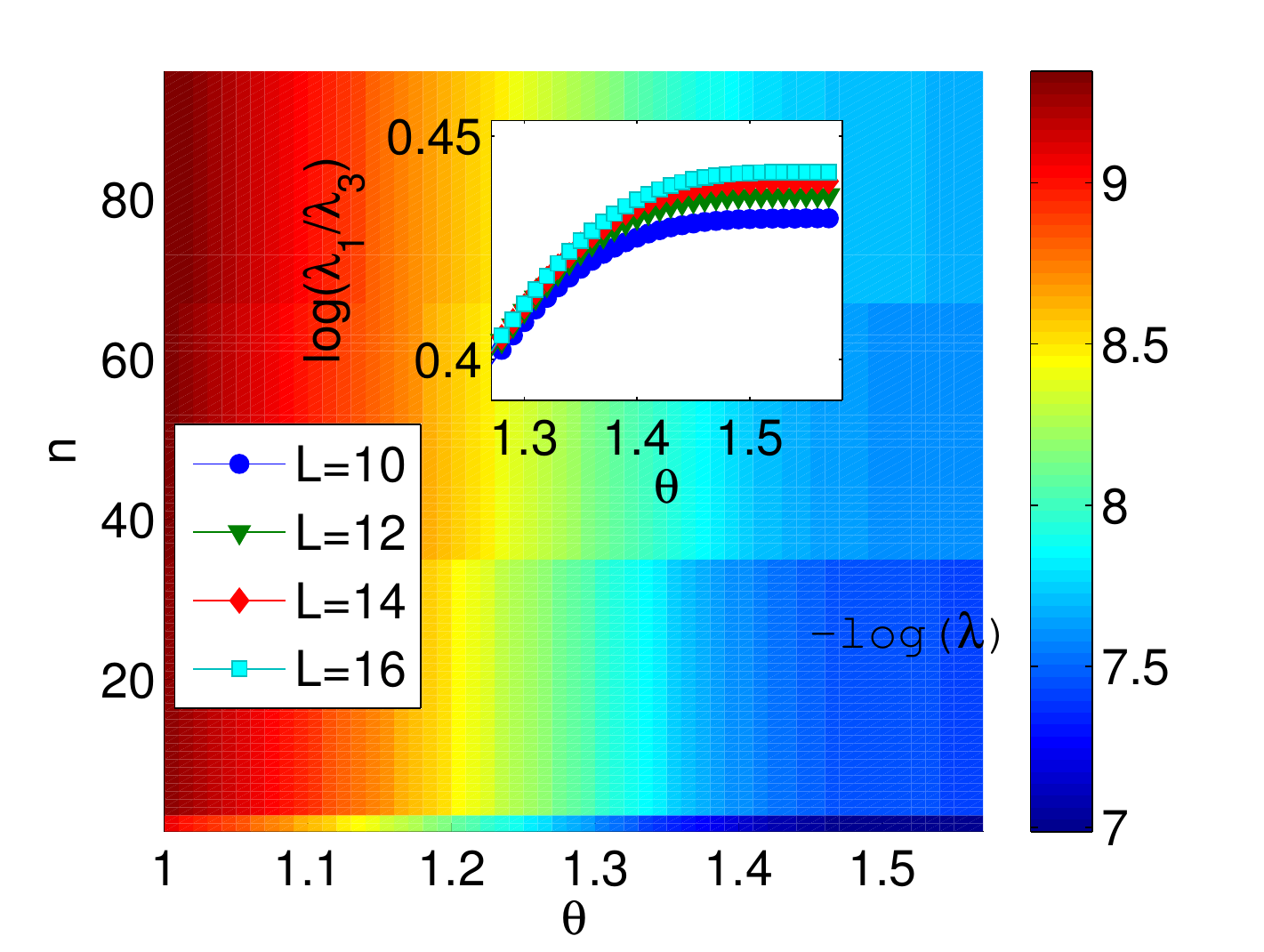}
 \caption{The behavior of first 100  eigenvalues  $\lambda_n$ ($n$ is on the y-axis) of the reduced density matrix of  half infinite cylinders $\rho_{1/2}$ defined in the panel b) of Fig. \ref{fig:num} with $L=16$. We plot them as as a function of $\theta$ (x-axis). $\theta$ varies in the cross-over range $1\le \theta\le \pi/2$ identified during  the analysis of the topological entropy in Fig. \ref{fig:topo_ent}. The entanglement spectrum presents clear plateaux, footprint of  RK wavefunctions. Nevertheless, its lower part does not seem to detect the phases transition. The first Schmidt gap is identically zero everywhere (there are two degenerate eigenvalues), and the second Schmidt gap increases while approaching the transition (inset in the Figure) for all the $L=10,\dots,16$.   \label{fig:num_sp} }
 \end{figure}

The results are presented in Fig. \ref{fig:num_sp}. The main panel shows the first 100 values of the entanglement spectrum $\log(\lambda_n)$ as a function of $\theta$ in the cross-over region. Surprisingly nothing strange seems to happen. The spectrum presents the plateaux structure characteristic of the RK wave functions. The structure of the first plateaux seems quite stable, and the only effect of increasing $\theta$ towards $\pi/2$ is to shift the relative height of the  plateaux so to accommodate the appearance of new ones in the tails. In particular the first two eigenvalues are degenerate for all the interval considered. The Schmidt gap is indeed constantly zero and does not detect the transition. Even from the plot of the second Schmidt  gap, which increases monotonically with $\theta \to \pi/2$, we are unable to appreciate that we are approaching a phase transition as shown in the inset of Fig. \ref{fig:num_sp} for several values of $L$. We thus conclude that the low-energy part of the entanglement spectrum seems to fail to detect the phase transition between the the gapped $\mathbb{Z}_2$ spin-liquid phase and the algebraic $U(1)$ spin liquid  occurring at $\theta=\pi/2$.

%
%
%

\section{Previous work on the subject}
\label{sec:about}
Here we try to  list the works that have contributed to our understanding of various topics and to the final formulation of our proposal as we have outlined.

There are several good references for group theory in a diagrammatic notation. In particular the book  \cite{cvitanovic_group_2008} deals with continuous groups. Continuous groups and the specific case of $SU(2)$ have been extensively studied in the literature of TNs \cite{mcculloch_non-abelian_2002,mcculloch_density-matrix_2007,singh_tensor_2010,singh_tensor_2011,singh_tensor_2012}. The reader not familiar with the elementary concepts in the theory of group  representations would benefit from studying the first few chapters of  some of the  standard text-books \cite{serre_linear_1977,tinkham_group_2003}.  A nice summary of relevant material can also be found in the appendices of \cite{bombin_family_2008}.

A nice introduction to LGT in the Lagrangian formulation can be found in \cite{kogut_introduction_1979} and in any of the standard textbook on the subject, i.e.  \cite{creutz_quarks_1983,montvay_quantum_1997}. The standard Hamiltonian formulation was obtained by Kogut and Susskind,  and by Creutz \cite{kogut_hamiltonian_1975,creutz_gauge_1977}.

Truncated LGT where discussed independently by many authors \cite{horn_finite_1981,orland_lattice_1990,chandrasekharan_quantum_1996} that have called them gauge magnet or quantum link models. Horn originally introduced a model similar to the one described here in Sect. \ref{sec:t_ks}, which we generalize here to arbitrary continuous and discrete groups.
Later in Ref. \cite{brower_qcd_1999}, the quantum link models  where generalized to several continuous gauge groups with a different strategy from the one presented here. In the same work, they where also given an  interpretation  in terms of rishons. In that language, the truncated LGT we have discussed in Sec. \ref{sec:t_ks} when dealing with continuous groups,  is a specific quantum link model with two rishons per link, while the one in Sect. \ref{sec:gm} is a quantum link model with one rishon per link \footnote{We acknowledge U. J. Wiese for pointing this out to us}. Still the results we present here can be applied also to discrete groups, and thus provide a further generalization of quantum link models. Also they allow to easily distinguish between those quantum link models that can be obtained as a consistent truncation of the KS LGT and those that cannot.

The study of lattice gauge theories with Matrix Product States, the simplest 1D TN structure, has by now a quite long tradition, \cite{byrnes_density_2002,sugihara_matrix_2005,banuls_mass_2013,banuls_matrix_2013,buyens_matrix_2013,rico_tensor_2013,silvi_lattice_2014}
2D LGT with TNs on the other hand have been less studied. LGT with discrete groups  have been addressed with entanglement renormalization in \cite{tagliacozzo_entanglement_2011}, some aspects of them have  been studied in \cite{denny_algebraically_2012} with categorical TNs, a construction that has strong connections to the present proposal. There however the emphasis is on Abelian discrete groups and the LGT are addressed  at the exactly solvable point obtained at $g=0$ while here we discuss a construction that allows to tackle generic groups  both Abelian and non-Abelian, discrete and continuous at any point in the phase diagram. LGT have also recently been addressed by using tensor renormalization schemes \cite{dittrich_coarse_2011, dittrich_coarse_2013, liu_exact_2013, he_modular_2014}.

There have been several proposal on how to obtain the gauge invariant Hilbert space. The original proposal by KS was to act on a reference state with all possible gauge invariant operators. However, there the states generated are not necessarily orthogonal and thus special care need to be taken. T. Osborne has pointed out  Refs. \cite{baez_spin_1996,ligterink_toward_2000} to us that is  related (implicitly)  to our work, and we also found the discussion in Ref.  \cite{burgio_basis_2000} very illuminating. T. Osborne himself is working along similar ideas \cite{osborne} and in particular he has independently worked out operators similar to the ones we have presented in Eq. \eqref{eq:u_ks} and \eqref{eq:cnots}.

In the context of characterizing a family of states (such as the RK states we have discussed), T. Osborne has pointed out Refs. \cite{greensite_calculation_1979,feynman_qualitative_1981}. In condensed matter, in particular, the recent results presented in Ref.  \cite{wang_constructing_2013,poilblanc_rvb_2014} have been obtained by applying similar ideas to the characterization  of singlet states.

At last we have also used the available literature about TN and topological order in order, whenever possible to make connections between our ideas and the one presented in that context. In particular we have found particularly useful the Ref. \cite{bombin_family_2008,swingle_topological_2010,schuch_peps_2010}.

To our knowledge the only previous mention to vertex operators in the context of LGT is the one of Ref. \cite{ardonne_topological_2004} however there the analysis is limited to $Z_2$ and $U(1)$ LGT while here we give a prescription for arbitrary groups.

It is also worth mentioning that there is an alternative connection between LGT and TN through a  map of the low-energy physics of QCD in the chiral limit to the physics of specific spin chains as pursued in  \cite{perez-garcia_heisenberg_2013,perez-garcia_chern-simons_2014}.

Finally, several groups have recently addressed the experimental implementation of truncated LGT  \cite{weimer_rydberg_2010,zohar_simulating_2012,tagliacozzo_optical_2013,tagliacozzo_simulation_2013,zohar_cold-atom_2013,banerjee_atomic_2012,banerjee_2_2013,hauke_quantum_2013,stannigel_constrained_2014,kosior_simulation_2014}.

With respect to the absence of the closure of the Schmidt gap across a phase transition, a similar observation was made in Ref. \cite{stephan_renyi_2012} and while we were preparing the final version of our manuscript in Ref. \cite{alet} the authors have provided a plausible argument to understand that this is a quite general phenomenon for phase transitions between different RK states. 
\section{Conclusions}
\label{sec:conc}
In this paper we have defined a TN  framework for  studying  LGT. 
It allows to use TNs as both a LGT model-building tool (and as such we have used to construct the minimal  consistent-truncation scheme for the KS LGT) and as a  practical tool to numerically explore LGT, their phase diagrams, and their emerging properties. 

The ansatz we have proposed follows the same spirit as the one proposed in Ref. \cite{tagliacozzo_entanglement_2011}. The TN indeed has a symmetric part that allows to exactly encode the constraints imposed by the gauge-symmetry conditions, and a variational part, which can be optimized numerically in order to characterize interesting physical states such as the low-energy states of gauge invariant Hamiltonians.

The new framework is also very powerful from the theoretical point of view. In this paper we have been indeed able to derive through it a  consistent truncation of local Hilbert space of the KS LGT with continuous group to finite dimensional Hilbert spaces. We have obtained also an explicit alternative construction of gauge magnets and of their $U$-operators for arbitrary gauge groups [Eq. \eqref{eq:U_gm}] that is also applicable to discrete groups, and the construction of  gauge invariant vertex operators for arbitrary gauge groups [Eq. \eqref{eq:v_op}]. We have also been able to show that, differently from the Abelian case, the  non-Abelian gauge magnets cannot be obtained as a consistent local truncation of the KS LGT.  For this reason they stand as an alternative  microscopic formulation of LGT.
This result  does not exclude the possibility that both gauge magnets and the KS LGT can encode  the same emergent physics. It excludes however that they are locally (where by locally we mean at the level of a single link) unitarily equivalent.
Furthermore, the distinct form of their projectors onto ${\cal H}_P$ in terms of TN (given explicitly in App. \ref{app:tensors_su2} and \ref{app:tensors_su2gm}) also points to the fact that their RK states are probably different (as we will analyze in a subsequent paper). This is not particularly surprising since the relation between the low-energy physics of quantum link models and of standard LGT was already discussed in Ref. \cite{schlittgen_low-energy_2001} and required the use of dimensional reduction arguments (so that $D+1$ dimensional quantum link models are expected in some limit to be equivalent to $D$ dimensional standard LGT).

The tools that we have developed here can be used to analyze the entanglement content of interesting LGT states. The entanglement, studied in a basis of states belonging to the original Hilbert space, tensor product of the constituents, does not have a direct physical meaning (since the only measurable operators in a real GT are gauge invariant operators, c.f. the recent discussion in \cite{casini_remarks_2014}). However,  it still provides an estimate of the computational cost of simulating such states using a TN (c.f.  the related  discussion in \cite{tagliacozzo_entanglement_2011}).

As a benchmark, we have considered the transition between the eight-vertex and the six-vertex model in terms of the RK wave function of the corresponding LGT. We have shown that, while the transition is correctly detected by the behavior of the entanglement entropy, it is hard to detect it by observing the behavior of the Schmidt gap and the lower part of the entanglement spectrum.

We envisage that the tools that  we have developed will play an important role in the characterization of the real time dynamics in LGT, and in the quest for finding model displaying  stable topological phases even at finite temperature. All the recent developments about the characterization of topological phases in terms of 2D TN such as the ones of Ref. \cite{swingle_topological_2010,schuch_peps_2010} can be easily applied to our construction as we plan to in the near future.

Recently, gauge magnets have received a lot of attention from the AMO community,  due to the possibility of implementing them in experiments based on the emerging new quantum technologies such as cold-atoms, trapped ions etc \cite{weimer_rydberg_2010,zohar_simulating_2012,tagliacozzo_optical_2013,tagliacozzo_simulation_2013,zohar_cold-atom_2013,banerjee_atomic_2012,banerjee_2_2013,hauke_quantum_2013,stannigel_constrained_2014,kosior_simulation_2014}. There is still a large room for improvement on these first proposals  and  the tools we have designed will help in this task. We have indeed just become aware that a new proposal for simulating $SU(2)$ LGT along the lines of our discussion has already been independently designed \cite{cirac}. 

Furthermore, our analysis is just the starting point in the development of a TN approach to LGT. In particular, questions relevant for high-energy LGT like, e.g., taking the continuous limit, have not been addressed here and constitute a logic next step to be done. Some of such questions are subject of an ambitious collaborative project, coordinated by Tobias Osborne, that is open to contributions and available on-line \cite{osborne}.

\section{Acknowledgments}
We would like to acknowledge I. Cirac, A. Ferris, F. Gliozzi, G. Misguich, F. Verstraete, G. Vidal for stimulating discussions on the topics presented. We also acknowledge the correspondence with T. Obsorne and U. J. Wiese who pointed to us to many relevant references.  The results of the paper have been already presented at several conferences and the feedback from the audience allowed us to improve them. In particular the slides of the talk given at the  CECAM conference Networking TNs hosted by  the ETH Zurich in  May 2013 are available on-line since then \cite{talk_zur}.  A large part of this work was developed while LT was financed by the project FP7-PEOPLE-2010-IIF ENGAGES 273524. We also acknowledge financial support by the ERC AdG OSYRIS, EU IP SIQS, and EUSTREP EQUAM.
While we were completing this manuscript we become aware of the preprint arXiv:1312.3127 (Ref. \cite{rico_tensor_2013,silvi_lattice_2014}) that discusses the formulation of  quantum link models (with emphasis on 1D systems) in the context of TN.  During the review process another related work appeared on the arXiv \cite{haegeman_gauging_2014}.
\bibliographystyle{apsrev4-1} 

%

\renewcommand{\theequation}{S\arabic{equation}}
\setcounter{equation}{0}
\renewcommand{\thefigure}{S\arabic{figure}}
\setcounter{figure}{0}

\appendix
\section{The explicit form of the tensors for RK states for $U(1)$ and $SU(2)$ LGT}
\label{app:tensors}
\subsection{$U(1)$}
\label{app:tensors_u1}
In the case of $U(1)$ both the local Hilbert space and the auxiliary space are two dimensional and 
the tensor ${\cal C}$ is given in the computational basis as 
\begin{equation}
 {\cal C} = \ket{0} \proj{0} + \ket{1}\proj{1} \label{eq:c_u1}
\end{equation}
while the tensor ${\cal G}$ is given in the computational basis, 
\begin{align}
 {\cal G}&= \ket{0}_{s_1}\ket{0}_{s_2}\bra{0}_{s_3}\bra{0}_{s_4}+ \nonumber \\
  &+ \ket{1}_{s_1}\ket{1}_{s_2}\bra{1}_{s_3}\bra{1}_{s_4}+ \nonumber  \\
  &+\ket{1}_{s_1}\ket{1}_{s_2}\bra{1}_{s_3}\bra{1}_{s_4}+ \nonumber  \\
  &+\ket{1}_{s_1}\ket{0}_{s_2}\bra{1}_{s_3}\bra{0}_{s_4}+ \nonumber   \\
  &+\ket{0}_{s_1}\ket{1}_{s_2}\bra{0}_{s_3}\bra{1}_{s_4}+ \nonumber  \\
  &+\ket{1}_{s_1}\ket{0}_{s_2}\bra{0}_{s_3}\bra{1}_{s_4}+ \nonumber  \\
  &+\ket{0}_{s_1}\ket{1}_{s_2}\bra{1}_{s_3}\bra{0}_{s_4}\label{eq:g_u1}
\end{align}
\subsection{Truncated $SU(2)$ LGT}
\label{app:tensors_su2}
The truncated $SU(2)$ LGT has Hilbert space of dimension five. In it we call the vectors as $\ket{0},\ket{11} ,\ket{12},\ket{21}, \ket{22}$, in order to remember that we have two blocks  the irrep $j=0$ and the irrep $j=1/2$, which are direct sum, one  of dimension 1 and the other of dimension $4$ that is tensor product of two $2$ dimensional spaces. The TN that encodes the RK states can be highly simplified by noting that only a part of the Hilbert space needs to be copied on the left and another part on the right. In particular we can write the  ${\cal C}$ tensor as 
\begin{align}
 {\cal C} = \ket{0} \proj{2} + \nonumber  \\
 \ket{11}\ket{0}\bra{0} + \ket{12}\ket{0}\bra{1}  +\ket{21}\ket{1}\bra{0}+ \ket{22}\ket{1}\bra{1}.\label{eq:c_su2}
\end{align}
From the second line we immediately recognize that ${\cal C}$ in the four dimensional  block  that is tensor product copies the left factor to the left and the right factor  to the right. In this way the auxiliary dimension is only $D=3$.
We now need to select gauge invariant configurations on the auxiliary links by using ${\cal G}$. We give the experssion of the blocks individually
\begin{align}
 {\cal G}_{1/2,1/2,1/2,1/2} &= 1/2 \left(\ket{0}_{s_1}\ket{0}_{s_2}\bra{0}_{s_3}\bra{0}_{s_4}+ \right.\nonumber \\
  &+\ket{1}_{s_1}\ket{1}_{s_2}\bra{1}_{s_3}\bra{1}_{s_4}+ \nonumber \\
  &+\ket{0}_{s_1}\ket{1}_{s_2}\bra{0}_{s_3}\bra{1}_{s_4}+ \nonumber \\
  &\left.+\ket{1}_{s_1}\ket{0}_{s_2}\bra{1}_{s_3}\bra{0}_{s_4}\right)+ \nonumber \\
  &+1/(2\sqrt{3})\left( \ket{0}_{s_1}\ket{1}_{s_2}\bra{0}_{s_3}\bra{1}_{s_4}+ \right. \nonumber \\
   &+\ket{1}_{s_1}\ket{0}_{s_2}\bra{1}_{s_3}\bra{0}_{s_4}+ \nonumber \\
  &-\ket{0}_{s_1}\ket{0}_{s_2}\bra{0}_{s_3}\bra{0}_{s_4}+ \nonumber \\
   &\left.-\ket{1}_{s_1}\ket{1}_{s_2}\bra{1}_{s_3}\bra{1}_{s_4}\right)+ \nonumber \\
  &-1/(\sqrt{3})\left( \ket{1}_{s_1}\ket{0}_{s_2}\bra{0}_{s_3}\bra{1}_{s_4}+\right. \nonumber \\
   &\left. +\ket{0}_{s_1}\ket{1}_{s_2}\bra{1}_{s_3}\bra{0}_{s_4}\right).
\end{align}

\begin{align}
{\cal G}_{0,0,0,0} &= 1/2 \left(\ket{2}_{s_1}\ket{2}_{s_2}\bra{2}_{s_3}\bra{2}_{s_4} \right).\\
  {\cal G}_{0,0,1/2,1/2} &= 1/\sqrt{2} \left(\ket{2}_{s_1}\ket{2}_{s_2}\bra{0}_{s_3}\bra{1}_{s_4}+ \right. \nonumber \\
 &-\left.\ket{2}_{s_1}\ket{2}_{s_2}\bra{1}_{s_3}\bra{0}_{s_4} \right).\\
{\cal G}_{0,1/2,0,1/2} &= 1/\sqrt{2} \left(\ket{2}_{s_1}\ket{0}_{s_2}\bra{2}_{s_3}\bra{0}_{s_4}+ \right. \nonumber \\
 &+\left.\ket{2}_{s_1}\ket{1}_{s_2}\bra{2}_{s_3}\bra{1}_{s_4} \right).\\
  {\cal G}_{0,1/2,1/2,0} &= 1/\sqrt{2} \left(\ket{2}_{s_1}\ket{0}_{s_2}\bra{0}_{s_3}\bra{2}_{s_4}+ \right. \nonumber \\
 &+\left.\ket{2}_{s_1}\ket{1}_{s_2}\bra{1}_{s_3}\bra{2}_{s_4} \right).\\
  {\cal G}_{1/2,1/2,0,0} &= 1/\sqrt{2} \left(\ket{0}_{s_1}\ket{1}_{s_2}\bra{2}_{s_3}\bra{2}_{s_4}+ \right. \nonumber \\
 &-\left.\ket{1}_{s_1}\ket{0}_{s_2}\bra{2}_{s_3}\bra{2}_{s_4} \right).\\
 {\cal G}_{1/2,0,0,1/2} &= 1/\sqrt{2} \left(\ket{0}_{s_1}\ket{2}_{s_2}\bra{2}_{s_3}\bra{0}_{s_4}+ \right. \nonumber \\
 &+\left.\ket{1}_{s_1}\ket{2}_{s_2}\bra{2}_{s_3}\bra{1}_{s_4} \right).\\
 {\cal G}_{1/2,0,1/2,0} & = 1/\sqrt{2} \left(\ket{0}_{s_1}\ket{2}_{s_2}\bra{0}_{s_3}\bra{2}_{s_4}+ \right. \nonumber \\
 &+\left.\ket{1}_{s_1}\ket{2}_{s_2}\bra{1}_{s_3}\bra{2}_{s_4} \right).
\end{align}
\subsection{The $SU(2)$ gauge magnet}
\label{app:tensors_su2gm}
Similarly to the truncated $SU(2)$, ${\cal P}$ for the $SU(2)$ gauge magnet is written as a TN with $D=3$.
The local Hilbert space has dimension $4$ tensor ${\cal C}$ reads
\begin{align}
 {\cal C} = \sum_{j=0,1} \ket{0,j}\ket{j}\bra{2} +\ket{1,j}\ket{2}\bra{j}
\end{align}

The tensor ${\cal G}$ on the other hand reads
\begin{align}
  {\cal G} &= 1/\sqrt{2} \left(\ket{2}_{s_1}\ket{2}_{s_2}\bra{0}_{s_3}\bra{1}_{s_4}+ \right. \nonumber \\
 &-\left.\ket{2}_{s_1}\ket{2}_{s_2}\bra{1}_{s_3}\bra{0}_{s_4} \right) +\nonumber \\
  &+ 1/\sqrt{2} \left(\ket{2}_{s_1}\ket{0}_{s_2}\bra{2}_{s_3}\bra{1}_{s_4}+ \right. \nonumber \\
 &-\left.\ket{2}_{s_1}\ket{1}_{s_2}\bra{2}_{s_3}\bra{0}_{s_4} \right)+ \nonumber \\
 &+ 1/\sqrt{2} \left(\ket{2}_{s_1}\ket{0}_{s_2}\bra{1}_{s_3}\bra{2}_{s_4}+ \right. \nonumber \\
 &-\left.\ket{2}_{s_1}\ket{1}_{s_2}\bra{0}_{s_3}\bra{2}_{s_4} \right)+ \nonumber\\
 &+ 1/\sqrt{2} \left(\ket{0}_{s_1}\ket{1}_{s_2}\bra{2}_{s_3}\bra{2}_{s_4}+ \right. \nonumber \\
 &-\left.\ket{1}_{s_1}\ket{0}_{s_2}\bra{2}_{s_3}\bra{2}_{s_4} \right)+ \nonumber \\
  &+ 1/\sqrt{2} \left(\ket{0}_{s_1}\ket{2}_{s_2}\bra{2}_{s_3}\bra{1}_{s_4}+ \right. \nonumber \\
 &-\left.\ket{1}_{s_1}\ket{2}_{s_2}\bra{2}_{s_3}\bra{0}_{s_4} \right)+ \nonumber \\
 & + 1/\sqrt{2} \left(\ket{0}_{s_1}\ket{2}_{s_2}\bra{1}_{s_3}\bra{2}_{s_4}+ \right. \nonumber \\
 &-\left.\ket{1}_{s_1}\ket{2}_{s_2}\bra{0}_{s_3}\bra{2}_{s_4} \right)+\nonumber \\
 &+1/2 \left(\ket{1}_{s_1}\ket{0}_{s_2}\bra{1}_{s_3}\bra{0}_{s_4}+ \right.\nonumber \\
  &+\ket{0}_{s_1}\ket{1}_{s_2}\bra{0}_{s_3}\bra{1}_{s_4}+ \nonumber \\
  &-\ket{1}_{s_1}\ket{0}_{s_2}\bra{0}_{s_3}\bra{1}_{s_4}+ \nonumber \\
  &-\left.\ket{0}_{s_1}\ket{1}_{s_2}\bra{1}_{s_3}\bra{0}_{s_4}\right)+ \nonumber \\
  &+1/(2\sqrt{3})\left( \ket{0}_{s_1}\ket{1}_{s_2}\bra{0}_{s_3}\bra{1}_{s_4}+ \right. \nonumber \\
  &+\ket{1}_{s_1}\ket{0}_{s_2}\bra{1}_{s_3}\bra{0}_{s_4}+ \nonumber \\
   &\left.+\ket{1}_{s_1}\ket{1}_{s_2}\bra{0}_{s_3}\bra{0}_{s_4}\right.+ \nonumber \\
    &\left.+\ket{0}_{s_1}\ket{0}_{s_2}\bra{1}_{s_3}\bra{1}_{s_4}\right)+ \nonumber \\
  &-1/(\sqrt{3})\left( \ket{0}_{s_1}\ket{1}_{s_2}\bra{1}_{s_3}\bra{0}_{s_4}+\right. \nonumber \\
   &\left. +\ket{1}_{s_1}\ket{0}_{s_2}\bra{0}_{s_3}\bra{1}_{s_4}\right).
 \end{align}

\end{document}